\documentclass[a4paper,10pt]{report}
\usepackage[utf8]{inputenc}
\usepackage{url}
\usepackage[hidelinks]{hyperref}
\usepackage{rotating}
\usepackage{subfigure}
\usepackage{eurosym}
\usepackage{tablefootnote}
\usepackage[colorinlistoftodos,prependcaption]{todonotes}
\usepackage[toc,page]{appendix}
\date{}

\title{Evaluating the Impact of AbuseHUB on Botnet Mitigation\\
\vspace{2mm}
Final  Report 
\\  \vspace{2cm}
\texttt{PUBLIC VERSION}}

\author{Michel van Eeten, Qasim Lone,\\Giovane Moura, Hadi Asghari, and Maciej Korczyński \\ 
\vspace{0.5cm} Economics of Cybersecurity Group\\Faculty of Technology, Policy, and Management\\Delft University of Technology\\
 \url{M.J.G.vanEeten@tudelft.nl}
\\ \url{http://www.tbm.tudelft.nl/econsec}\\
\\ \\ May 25, 2016
}

\begin{document}
\maketitle

\tableofcontents

\listoffigures
\listoftables

%

\newpage\null\thispagestyle{empty}\newpage
\chapter*{Introduction}

This documents presents the final report of a two-year project to evaluate the impact of AbuseHUB, a Dutch clearinghouse for acquiring and processing abuse data on infected machines ~\cite{abusehub}. The report was commissioned by the Netherlands Ministry of Economic Affairs, a co-funder of the development of AbuseHUB. AbuseHUB is the initiative of 9 Internet Service Providers, SIDN (the registry for the .nl top-level domain) and Surfnet (the national research and education network operator). The key objective of AbuseHUB is to improve the mitigation of botnets by its members. 

We set out to assess whether this objective is being reached by analyzing malware infection levels in the networks of AbuseHUB members and comparing them to those of other Internet Service Providers (ISPs). Since AbuseHUB members together comprise over 90 percent of the broadband market in the Netherlands, it also makes sense to compare how the country as a whole has performed compared to other countries. 

This report complements the baseline measurement report produced in December 2013 and the interim report from March 2015. We are using the same data sources as in the interim report, which is an expanded set compared to the earlier baseline report and to our 2011 study into botnet mitigation in the Netherlands~\cite{michel-dutch-market}. 

The document is organized as follows: in Chapter~\ref{sec:methodology}, we present the methodology used in this research -- the data sets and the mapping of that data to the relevant networks of ISPs. In Chapter~\ref{sec:q0}, we compare bot infection levels in the Netherlands against several other countries. In  Chapter 3, we focus on ISPs specifically and compare Dutch ISPs against ISPs in other countries. In Chapter~\ref{sec:q1}, we shift the focus to comparing infection levels within the Netherlands, comparing AbuseHUB members with non-members. Then, in Chapter~\ref{sec:q2}, we compare the performance of AbuseHUB members among each other. Finally, in Chapter~\ref{sec:summary}, we summarize the main findings of the study and discuss whether AbuseHUB has improved the intelligence and practice of botnet mitigation for its members. We conclude by exploring recommendations for further improving botnet mitigation in the Netherlands.
\chapter{Questions and methodology}
\label{sec:methodology}

The basic methodology employed in this report consists of collecting and analyzing Internet measurement data on infected machines. We then interpret these measurements by connecting them to the properties of the operators of the networks containing the infected machines, such as the country in which they are located and the number of subscribers of the operators. This way we can develop comparative metrics to determine the performance in botnet mitigation of Dutch ISPs, members and non-members of AbuseHUB, to each other and to ISPs in other countries.

We evaluate the infection rates of Dutch ISPs over different times frames, depending on the data source. Three sources cover the period from January 2011 to December 2015. Other sources cover only 2015 and parts of 2014.

Please note that we focus on comparing Dutch ISPs. This points to a limitation of the study. The members of AbuseHUB include non-ISPs as well, most notably SIDN and Surfnet. Their networks are quite different from those of the ISPs. We are not in a position to generate reliable comparative metrics that take into account these differences in the nature of these networks and of the subscriber populations. To deal with these issues, we undertake comparisons both at the country level (where we can include all networks) and at the ISP levels (where we include only the ISP networks). This approach is outlined in the next section.

In the remainder of this chapter, we first present the research questions of the study. Next, we turn to the Dutch Internet Service Providers and other network operators that are included in the analysis (Section \ref{sec:dutch-ips}). Next, in Section \ref{sec:datasets}, we present the infection-related datasets that we have analyzed. Last, in Section \ref{sec:mapping}, we explain how the datasets are mapped back to the ISPs covered in Section \ref{sec:dutch-ips}.



\section{Research questions}
\label{sec:questions}

The main goal of the study is to assess the impact of the AbuseHUB initiative on infection rates in the networks of Dutch ISPs. To achieve this goal, four questions we articulated:
\begin{enumerate}
  \item How do member ISPs compare to non-member ISPs? 
  \item How do member ISPs compare among themselves?
  \item Do member ISPs have better data on the presence of bots in their networks?
  \item What recommendations can be identified to improve botnet mitigation in the Netherlands?
\end{enumerate}

\section{AbuseHUB members and Dutch Internet Service Providers}
\label{sec:dutch-ips}

The bulk of the report focuses on the first two questions, which are empirical in nature: How do member ISPs compare to non-member ISPs? And: How do member ISPs compare among themselves? After they are answered, and we have a data-driven understanding of botnet mitigation in the Netherlands, we can evaluate questions 3 and 4. These are discussed in the final part of the report.

Given that AbuseHUB has ISPs as well as non-ISPs among their membership, there are certain limitations in comparing the members among themselves, as well as comparing members with non-members. The number of infections in a network is, to a significant extent, a function of the size of the user population in that network, as well as the type of users. To put it crudely, networks with highly different user populations cannot be meaningfully compared.

For ISPs, we have found that dividing the number of infections by the number of subscribers gives reliable comparative metrics. This works well for a subset of the AbuseHUB members. It does not work for all networks served by AbuseHUB, however. The Autonomous System (AS) of SIDN is basically a type of corporate network and the machines in the network are owned by the organization itself. Surfnet is different from all the others also. It receives abuse data for 16 highly heterogeneous ASes. A similar complication holds for some ISPs, who are also receiving abuse data for other ASes than the one with their broadband customers. All of this means that there is no meaningful way to generate comparative metrics for all 35 ASes covered by AbuseHUB.

We deal with the heterogeneity within AbuseHUB in three ways. First, we generate comparative metrics for those ASes of AbuseHUB that provided broadband access, predominantly to consumers. For those ASes, the number of subscribers that reside in the network is a good basis to take the size of the network into account. The data on the number of subscribers we got from the TeleGeography's GlobalComms database. We disaggregated them further using data supplied to us by the members of AbuseHub. In short, we compare oranges to oranges by including a subset (albeit a large subset) of AbuseHUB ASes. This is continuation of the approach in our earlier reports. 

To include the non-broadband access ASes of AbuseHUB in the study, we have generated additional metrics and units of analysis. These can deal with the heterogeneity, but at the expense of being less comparable across networks. We deploy two workarounds.  

The first workaround is is to look at the Netherlands as a whole. To see whether AbuseHUB has led to a faster improvement in mitigation compared to ISPs in other countries, we will also analyze the data at the country level. Since AbuseHUB covers a large part of the Dutch broadband user population, we will take the Netherlands as a whole as a proxy for AbuseHUB and look at what has happened to infection rates for the country as a whole. At that level, we can generate relative metrics by dividing by the number of Internet users in each country. 

The second workaround is to look at the relative speed of improvement over time. We take the absolute number of infections in the relevant networks and index these at a certain date to track the rate with which mitigation improves these levels after that date. This way, we do not need to normalize the number of infections by the number of subscribers.

Table \ref{tab:isps-sig} presents the networks evaluated in this report. This list was provided by the AbuseHUB members. The count of the address space size of each member as well as the total Dutch IP address space was performed using BGP data from August, 2015. As can be seen, AbuseHUB covers a little less than two-thirds of the Dutch IP address space.

 \begin{table}[t]
  \begin{center}
   \begin{tabular}{|c|c|c|} \hline
	\textbf{ISP}     & \textbf{ASes} & \textbf{$\sum$ IPv4\tablefootnote{Obtained from BGP tables on August, 2015, combining all the ASes of each ISP. We use Maxmind's~\cite{Maxmind}  geolocation databses to filter out IPs employed in The Netherlands.}} \\ \hline

  \parbox{4.5cm}{\vspace{0.5pt}\center{KPN (incl. Telfort)}}  &\parbox{4.5cm}{\vspace{0.5pt}\center{286, 1134, 1136, 20143, , 8737, 49562, 5615 }}&    7,316,075 \\ \hline 
  
  Tele2   &  \parbox{4.5cm}{\vspace{0.5pt}\center{15670, 34430, 13127, 20507}}  &   1,855,392  
\\ \hline 
 
 
 UPC & 6830 &  2,044,918   
 \\ \hline

 RoutIT  & 28685&  212,581 \\ \hline 
 
 SIDN                   &  \parbox{4.5cm}{\vspace{0.5pt}\center{1140,48283}}  &  4,608 \\ \hline 
 
 SOLCON  & 12414 & 145,152\\ \hline

  SURFnet    &  \parbox{4.5cm}{\center{1101, 1102, 1103,
            1104, 1124, 1125, 1126, 1128, 1132, 1133, 1139, 1145, 1161, 1837, 1888, 25182}}  &    9,878,527   \\ \hline 
 
 XS4ALL & 3265 &  1,188,607\\ \hline 
 
 Ziggo      &9143 &   3,499,647 \\ \hline 
 
 ZeelandNet & 15542 &   174,976   \\ \hline 

   & & \\ \hline
 \textbf{Total AbuseHUB IPs:} &  &\textbf{26,320,483}\\ \hline
 \textbf{Total Dutch IPs}   & &  \textbf{41,217,528}\\ \hline
 \textbf{Ratio AbuseHUB/NL} & &\textbf{63.9\%}\\ \hline

    \end{tabular}
    \end{center}
    \caption{Evaluated Internet Service Providers}
  \label{tab:isps-sig}
  \end{table}

\section{Evaluated botnet infection datasets}
\label{sec:datasets}


As covered in~~\cite{michel-dutch-market}, there is no authoritative data source to identify the overall population of infected machines around the world. Commercial security providers typically use proprietary data and shield their measurement methods from public scrutiny. This makes it all but impossible to correctly interpret the figures they report and to assess their validity.

The data that is available for scientific research in this area relies on two types of sources:

\begin{itemize}
 \item Data collected external to botnets. This data identifies infected machines by their telltale
behavior, such as sending spam or participating in distributed denial of service attacks;
\item  Data collected internal to botnets. Here, infected machines are identified by intercepting
communications within the botnet itself, for example by infiltrating the command and
control infrastructure through which the infected machines get their instructions.
\end{itemize}


Each type of source has its own strengths and weaknesses. The first type typically uses techniques
such as honey pots, intrusion detection systems and spam traps. It has the advantage that it is not
limited to machines in a single botnet, but can identify machines across a wide range of botnets that
all participate in the same behavior, such as the distribution of spam. The drawback is that there are
potentially issues with false positives. The second type typically intercepts botnet communications by
techniques such as redirecting traffic or infiltrating IRC channel communication. The advantage of
this approach is accuracy: a very low false positive rate. Machines connecting to the command and control server are almost always really infected with the specific type of malware that underlies that specific botnet -- the exception being sandbox or other research machines contacting the sinkhole. The downside of this accuracy is that
measurement only captures infected machines within a single botnet. Depending on how one counts, the number
of botnets can be estimated to be in the hundreds~\cite{Zhuang:2008:CBE:1387709.1387711}. That makes the data from a single sinkhole not
representative of the overall population of infected machines.

Neither type of data sources sees all infected machines, they only see certain subsets, depending on
the specific data source. In general, one could summarize the difference between the first and the
second source as a trade-off between representativeness versus accuracy. The first type captures a
more representative slice of the problem, but will also include more false positives. The second type
accurately identifies infected machines, but only for a specific botnet, which implies that it cannot
paint a representative picture.

The best available strategy to deal with these issues is to combine as many sources as possible. It is difficult to gain access to the global data captured by such sources. Data sharing is often limited to CERTS and the owners of the network, and they only get the slice of the data that pertains to their country or network. For the kind of comparisons we are making, this has very limited value.

Notwithstanding these challenges, several partners generously provided access to several data sources. We group them into two categories: global sources and sources that include only IP addresses that geo-locate to the Netherlands. Access to the latter was kindly provided by the NCSC; the former by different organization which we will acknowledge below. From these sources, only spam is categorized as ``external'' to botnets; all the other ones are internal obtained either via sinkholes or sandboxes.

\subsection{Global data sources}

\subsubsection{Spam trap dataset (Spam)}

\begin{figure}[!htbp]
        \centering
    
        \includegraphics[width=1\textwidth]{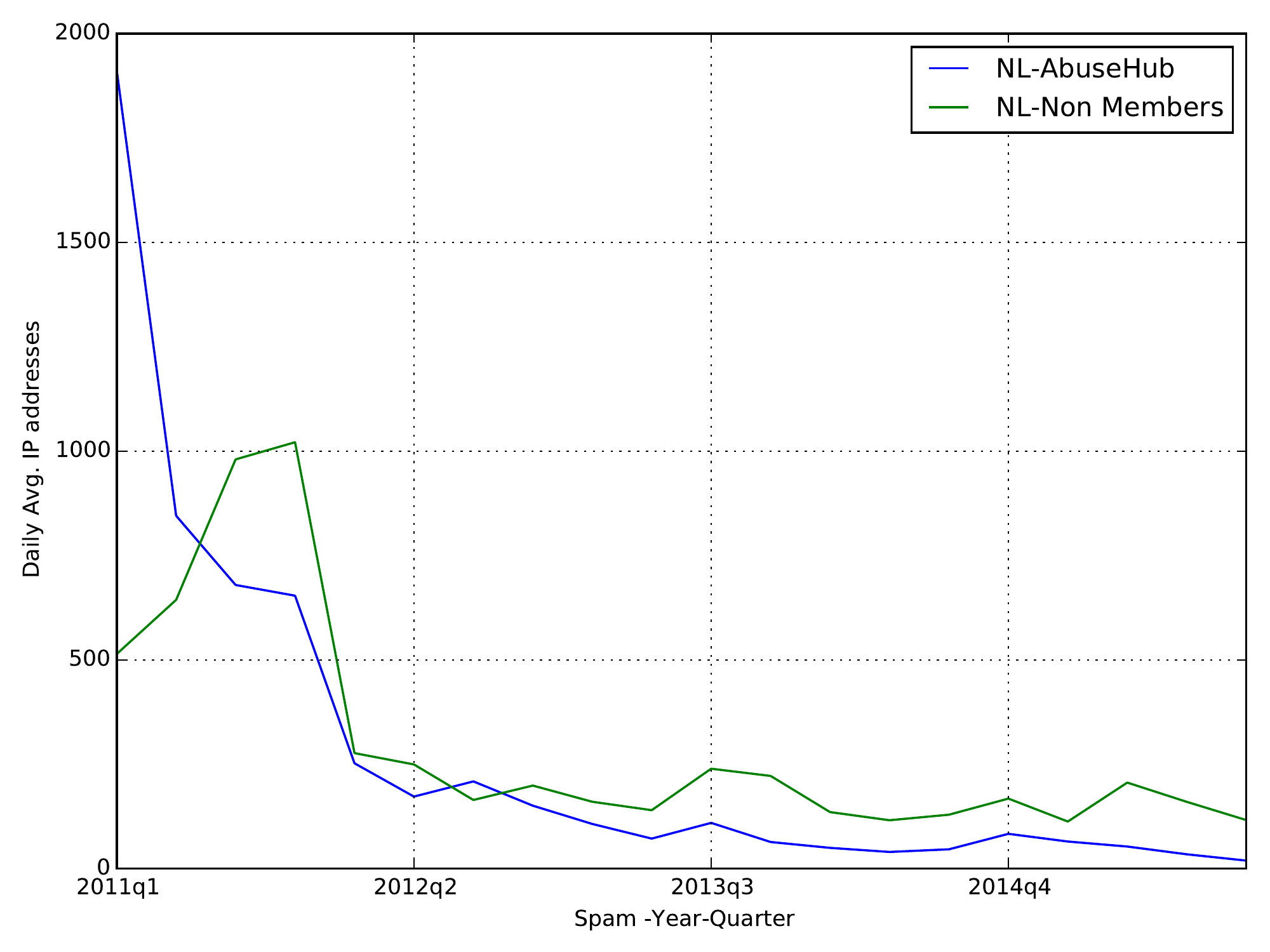}
    \caption{Spam}
    \label{spam-q1}
   \end{figure}

Spam data are obtained from a spamtrap we were generously given access to by Dave Rand of TrendMicro -- the same source as used in the baseline report. It might not be completely representative of the overall spamming trends, and also there is no guarantee that the listed spam sources are indeed originating from bots, though so far it is still a core platform for distribution. The more important limitation is that the spam has become a less important part of the botnet economy, as witnessed in the substantial drop in overall spam level. The reports of security firms seem to confirm these overall trends. Symantec reported a significant decrease in the volume of spam messages, “from highs of 6 trillion messages sent per month to just below 1 trillion”~\cite{krebs-spam} until 2012 (See Figure \ref{spam-q1}). Cisco, TrendMicro and Kaspersky show that the spam volume since that period has been fluctuating, but staying at more or less the same level (see~\cite{cisco-sender} and~\cite{trend-spam-2015}). All of this means that the source is becoming less representative of overall infection levels.

\subsubsection{Shadowserver Sinkhole Conficker data (Conficker)}

Established in 2004, the Shadowserver Foundation comprises volunteer security professionals that ``gathers intelligence on the darker side of the Internet''. They have created the Conficker working group, which provides reports and data on ``the widespread infection and propagation of Conficker bots''~\cite{shadowserver-conficker}. 

Several members of the working group run sinkholes that continuously log the IP addresses of Conficker
bots. The sinkholes work in this fashion: computers infected with Conficker frequently attempt to
connect to command and control servers to receive new payloads (i.e., instructions). In order to
protect the botnet from being shut down, Conficker attempts to connect to different C\&C domains
every day. The working group has succeeded in registering some of these domain names and logging
all connections made to them. Since these domains do not host any content, all these connections
are initiated by bots. Therefore, we can reliably identify the IP addresses of the Conficker bots.

The Conficker dataset is unique in several ways. First of all, unlike the other two datasets, it is not a
small sample of a much larger population, but rather captures the universe of its kin. This is because
of the way the bot works – most of them will eventually contact one of the sinkholes. Second, this
dataset is basically free from false positives, as, apart from bots, no other machine contacts the sinkholes.
These features make the dataset more reliable than the spam or DShield datasets. The difference,
however, is that the dataset is only indicative of the patterns applicable to one specific botnet,
namely Conficker. Although Conficker has managed to replicate very successfully, with around
several million active bots at any given moment, it has not been used for any large-scale malicious
purposes – or at least no such uses have been detected yet. This means ISPs and other market
players may have less powerful incentives to mitigate these infections, different from spam bots, for
example. These differences make the Conficker dataset complementary to the two other sets.

Overall, the Conficker dataset adds a fresh, robust and complimentary perspective to our other two
datasets and brings more insight into the population of infected machines worldwide.

 \subsubsection{Zeus GameOver Botnet (Peer and Proxy)}

Zeus botnet started making headlines in 2007, as a credential stealing botnet. The first version of Zeus was based on centralized command and control (C\&C) servers. The botnet was studied by various security researchers and multiple versions were also tracked ~\cite{zeustracker,binsalleeh2010analysis,riccardi2010framework, falliere2009zeus}. In recent years Zeus has transformed, into more robust and fault tolerant peer-to-peer (P2P) botnet, known as P2P Zeus or GameOver. It was sinkholed in the course of 2014. We were kindly provided access to the data by researchers involved in the sinkholing effort.

The botnet supports several features including RC4 encryption, multiple peers to communicate stolen information, anti-poising and auto blacklist. It also can be divided into \textit{sub-botnet},  based on BotIDs , where each sub-botnet can be used to carryout diverse task controlled by different botmasters. 

The botnet is divided into three sub-layers, which provide following functionality.

\begin{itemize}
\item \textbf{Zeus P2P Layer (Peer):}  This is the bottom most layer and contains information of infected machines.  Bots in P2P layer exchange peer list with each other in order to maintain updated information about compromised machines.
\item \textbf{Zeus Proxy Layer (Proxy) :} A subset of bots from P2P layer are assigned the status of proxy bots. This is done manually by the botmaster by sending proxy announcement message. Proxy bots are used by Peer-to-peer layer bots to fetch new commands and drop stolen information. 

\item \textbf{Domain Generation Algorithm Layer:}  DGA layer provides fall backup mechanism, if a bot cannot reach any of its  peers, or the bot cannot fetch updates for a week. Zeus algorithm  generates 1000 unique domain names per week. Bots which lose connection with all connected peers search trough these domains until they connect to live domain. 

More details about architecture and functioning of the botnet can be found in literature~\cite{shadowserverzeus,andriesse2013analysis}. 
\end{itemize} 

This dataset is sub-divided into three feeds, GameOver Peer, GameOver Proxy and GameOver DGA. The botnet is spread in around 212 countries with on average 95K unique IP addresses per day. Hence it is gives us insight of botnet infection level at global level, and compare various countries and ISPs.       

\subsubsection{ZeroAccess}
ZeroAccess is a Trojan horse, which uses a rootkit to hide itself on Microsoft Windows operating systems. The botnet is used to download more malware and open a backdoor for the botmaster to carry out various attacks including click fraud and bitcoin mining. 

The botnet is propagated and updated through various channels, including but not limited to drive-by downloads, redirecting traffic and dropping rootkits at potential host or updating the already compromised host through a P2P network. The ZeroAccess sinkhole data comes from several machines that impersonate nodes and supernodes in the network, which allows them to intercept P2P communication, including data on other infected machines in the network. These nodes see bots in around 220 countries with an average of about 12,000 unique IP addresses per day. It was generously shared with us by the operators of a ZeroAccess sinkhole, namely the team of Katsunari Yoshioka at Yokohama National University in Japan.

\subsubsection{Morto}
Morto is a worm that exploits the Remote Desktop Protocol (RDP) on Windows machines to compromise its victims. It uses a dictionary attack for passwords to connect as Windows Administrator over RDP to vulnerable machines in the network. After successfully finding a vulnerable machine, it executes a dropper and installs the payload.  

We have a time series data of Morto for the past 4 years with an average of 5,000 daily unique IP addresses distributed globally. This is relatively small, but it complements our other data sources by providing a longitudinal perspective. This data was also kindly shared with us by the team of Katsunari Yoshioka at Yokohama National University in Japan.

\subsection{Netherlands-only data sources}

In addition to the global feeds, we have obtained access from Shadowsever Foundation to botnet data pertaining only to the Netherlands. 

\subsubsection{Shadowserver's bot feed}
Shadowserver collects list of infected machines by monitoring IRC Command and Controls, IP connections to HTTP botnets, or IP's of Spam relay~\cite{shadowserverbotdrone}. This Report contains comprehensive list of IP addresses in The Netherlands of compromised machines which are infected with different malware or botnets.

This datasource enable us to compare ISPs within The Netherlands and also AbuseHUB members with non-members.

\subsubsection{Shadowserver's Microsoft Sinkhole}

Microsoft shares with Shadowserver Foundation data from botnet sinkholes\footnote{\url{https://www.shadowserver.org/wiki/pmwiki.php/Services/Microsoft-Sinkhole}}. We have also obtained this data for IP addresses located in the Netherlands.

\section{Mapping offending IP addresses to Dutch ISPs}
\label{sec:mapping}

For each unique IP address that was logged in one of our data sources, we looked up the Autonomous System Number (ASN) and the country where it was located. The ASN is relevant, because it allows us to identify what entity connects the IP address to the wider Internet --  and whether that entity is an ISP or not.

However, there are some ISPs in Table \ref{tab:isps-sig} that operate in various countries across Europe. We employ IP-geolocation databases~\cite{poese:2011:IGD:1971162.1971171} from Maxmind~\cite{Maxmind} to single out IP addresses used in The Netherlands from the other European countries when classifying the attacking IP addresses from each ISPs.


As both ASN and geoIP information change over time, we used historical records to establish the
origin for the specific moment in time when an IP address was logged in one of our data sources (e.g.,
the moment when a spam message was received or network attack was detected). This effort resulted in time series for all the variables in the datasets, both at an ASN level and at a country level.
The different variables are useful to balance some of the shortcomings of each – a point to which we
will return in a moment.

We then set out to identify which of the ASNs from which botnet IP data belonged to ISPs – that is, the actual companies who manage these networks. To the best of our knowledge, we developed the first database that reliably maps ASNs onto ISPs. This is not surprising. Estimates of the number of ISPs vary from around 4,000 – based on the number of ASNs that provide transit services – to as many as 100,000 companies that self-identify as ISPs – many of whom are virtual ISPs or resellers of other ISPs' capacity.

So we adopted a variety of strategies to connect ASNs to ISPs. First, we used historical market data
on ISPs – wired, wireless and broadband – from TeleGeography’s GlobalComms database~\cite{telegeography}. We extracted the data on all ISPs in the database listed as operating in a set of 62 countries, which were selected as follows. We started with the 34 members of the Organization for Economic Cooperation and Development (OECD), and 7 additional members of the European Union which are not part of the OECD. To this list we added 21 countries that rank high in terms of Conficker or spam bots --- cumulatively covering 80 percent of all such bots worldwide. These countries are interesting from a cybersecurity perspective. For two countries, Mexico and India, there were problems with the data on the ISPs in the country. We removed them from the ISP-level comparison. This means that the country-level comparison includes 62 countries, while the ISP-level comparison includes 60 countries.

The process of mapping ASNs to ISPs was done manually. First, using the GeoIP data, we could identify which ASNs were located in each of the 60 countries. We rank these ASNs by a number of criteria and listed 2260 ASNs for the next steps; these ASNs cumulatively cover around 80 percent of bots IPs, and approximately 75 percent of the IP addresses in their respective countries. We used historical WHOIS records to lookup the name of the entity that administers each ASN in a country. We then consulted a variety of sources – such as industry reports, market analyses and news media – to see which, if any, of the ISPs in the country it matches. In many cases, the mapping was straightforward. In other cases, additional information was needed – for example, in case of ASNs named after an ISP that had since been acquired by another ISP. In those cases, we mapped the ASN to its current parent company.

It is important to notice that ISP change their AS size over time (mergers, selling/buying blocks). To cope with this, we use historical BGP data and produce our metrics matching the timestamp of the botnet data with the BGP tables of the respective period.

\section{Compensating for known limitations}
\label{sec-dealing-with-churn}

Our approach allows us to robustly estimate the relative degree in which ISP networks harbor
infected machines. It has certain limitations, however, that need to be compensated for. The effects
of three technical issues need to be taken into account when interpreting the data: the use of
Network Address Translation (NAT), the use of dynamic IP addresses with short lease times. The key issue is to understand how these technical practices affect the number of machines that are represented by a single unique IP address.

NAT means sharing a single IP address among a number of machines. Home broadband routers often
use NAT, as do certain other networks. This potentially underrepresents the number of infected
machines, as multiple machines show up as a single address. Dynamic IP addresses with short lease
times imply that a single machine will be assigned multiple IP addresses over time. This means a
single infected machine can show up under multiple IP addresses. As such, it over-represents the number of infected machines. Both of these practices counteract each other, to some extent. This
limits the bias each of them introduces in the data, but this does not happen in a consistent way
across different networks.

This is a classic problem in the field of Internet measurement: how many machines are represented
by a single IP address? Ideally, one IP address would indicate one machine. But reality is more
complicated. Over an extended time period, a single address sometimes indicates less than one
machine, sometimes more than one. This varies across ISPs and countries. Earlier research by Stone-
Gross \textit{et al.}~\cite{stone2009your} has demonstrated that in different countries, there are different ratios of unique IP addresses to infected machines – referred to as ``DHCP churn rates''.

In this report, to control for the bias caused by churn rates, we use shorter time scales when counting the number of unique IP addresses in a network, for all the datasets.  

On shorter time scales, the potential impact of churn is very limited. Earlier research found that
churn starts to affect the accuracy of IP addresses as a proxy for machines on timescales longer than
24 hours. 12 We therefore worked with a time period of 24 hours. All our comparative analyses are
based on the daily average number of IP addresses from an ISP network. This compensates for churn,
but has a downside: in these estimates, the number of infected machines may be now grossly
undercounted, depending on the prefix, AS and/or ISP evaluated.

While the number of bots measured in a 24 hour period is the most reliable for comparisons across networks, it cannot indicate the actual infection rate of a network in absolute terms. For absolute estimates -- in other words, of the actual number of infected machines – we use larger time periods, depending on the situation: months, quarters or even the whole 18-month measurement period.
For sources we have checked both the daily average number of unique IP addresses, and the
total number of unique IP addresses for that particular metric. That way, we can compensate for the
various measurement issues. Patterns that hold across these different measurements can be said to
be robust and valid. These measurement issues are revisited in more detail in each section of the
findings chapter.

The result of this approach is time series data on the number and the location of infected machines
across countries and ISPs. We have paid special attention to whether these machines are located in
the networks of the main ISPs in the wider OECD.

\section{Generating infection metrics}

Using the data sources and infection counts outlined above, we were able to generate the reputation metrics. These basically include the average number of unique IP addresses (IPs) observed daily in each of the infection datasets per country, per ASN, and per ISP (operator). The start and end dates of datasets used in this study are summarized in Table \ref{duration}. 
The metrics include both the absolute counts, and in some instances, normalized counts – that is divided by the number of broadband subscribers in the particular country or ISP. 

In the next Chapter, we will use these metrics to evaluate infection trends in different countries and ISPs.

\begin{table}
\centering
\caption{Timeline for datasets used in this study. }
\label{duration}
\begin{tabular}{|l|l|l|}
\hline
\textbf{Data source}     & \textbf{Start date} & \textbf{End date} \\ \hline
Spam                     & 2011-01-01          & 2015-11-30        \\ \hline
Conficker                & 2011-01-12          & 2015-11-30        \\ \hline
Zeus GameOver Peer       & 2014-10-15          & 2015-11-30        \\ \hline
Zeus GameOver Proxy      & 2014-10-18          & 2015-11-30        \\ \hline
ZeroAccess               & 2014-10-12          & 2015-11-30        \\ \hline
Morto                    & 2011-07-30          & 2015-11-30        \\ \hline
Shadowserver bot feed    & 2014-04-29          & 2015-11-30        \\ \hline
Shadowserver MS sinkhole & 2014-04-29          & 2015-11-30        \\ \hline
\end{tabular}
\end{table}

\chapter{The Netherlands compared against other countries}
\label{sec:q0}

In this chapter, we use our global data sources (Section~\ref{sec:datasets}) to  rank the Netherlands against 61 other countries, and, more specifically, against a selected set of countries which we considered relevant reference points as they can be considered countries with similar development levels and botnet mitigation efforts: Germany, Great Britain, France, Finland, Italy, Spain, United States and Japan. 

The aim of this chapter is two answer the following questions:
\begin{itemize}
\item How does the Netherlands rank against other countries in terms of infection rates?
\item Is the Netherlands improving over time, compared to other countries?
\end{itemize}

The idea is that if the ranking of the Netherlands has improved, this suggests a positive impact of AbuseHUB (see Section~\ref{sec:questions} for details on why we use country-level comparisons to study the impact). 

To answer the questions, we compute the daily number of unique IP addresses from each global data feeds we have obtained, and aggregate it into quarters or weeks, depending on the time span of the data.

In Section~\ref{sec:chap2-abs}, we analyze the average performance of the Netherlands, compared to other countries, also taking into account the number of Internet users. In Section~\ref{sec:chap2-timeseries}, we present an analysis on the evolution over time of performance of the Netherlands against other countries, taking into account their respective time series.

\section{Country ranking}
\label{sec:chap2-abs}

Tables~\ref{tab:contry-abs-2014}-\ref{tab:contry-abs-2015} show the average of the daily number of  unique infected IP addresses seen in each global feed we have analyzed. We list both the top 10 most infected countries and the countries of interest we have mentioned before. Analyzing those tables, we can see that the Netherlands only has a modest share of the overall problem. Its high rank number (column \#), indicating there are many countries with more infected machines. In 2014 and 2015, the ranking of the Netherlands varies from 31 to 53 (of a total of 62, with 62 being the least infected country), for the sources we have analyzed.

\begin{itemize}
\item \textit{NOTE: the rankings for MORTO will be added in the final version of this report}
\end{itemize}

It is also relevant to assess the number of infections per user, as some countries have more users that others  (e.g., the US has a population $\sim$ 19 times of the Netherlands). To compensate for this, we have produced a ranking in which the number of unique IP addresses seen in the infection data is normalized by the subscribers of the country, obtained from the TeleGeography dataset ~\cite{telegeography}. These results can be seen in Tables~\ref{tab:contry-rel-2014}-\ref{tab:contry-rel-2015}.

In terms of it relatively infection rate, so taking the number of Internet subscribers into account, we can see that the Netherlands does quite well. In 2014 and 2015, it consistently ranks between 42 and 60 in all sources, with most sources ranking the country in the high 50s (rank 62 is the country with the least infections per subscriber for that source). Also when compared to the subset of reference countries, we can conclude that the Netherlands performs above average. It is among the least infected countries in that group. Finland is the only country that consistently performs better. Only for spam, does the Netherlands perform mediocre.

  \begin{sidewaystable}[ph!]
  \begin{center}

   \begin{tabular}{|c|c|c|c|c|c|c|c|c|c|c|c|c|}
  
    \multicolumn{13}{c}{\textit{Top 10 Most Infected Countries}}\\\hline

\textbf{\#} & \multicolumn{2}{c|}{GameOver Peer} & \multicolumn{2}{c|}{GameOver Proxy} & \multicolumn{2}{c|}{Conficker} & \multicolumn{2}{c|}{Morto} & \multicolumn{2}{c|}{ZeroAccess} & \multicolumn{2}{c|}{Spam}                       \\ \hline
            & \textbf{CC}     & \textbf{\#}      & \textbf{CC}      & \textbf{\#}      & \textbf{CC}   & \textbf{\#}    & \textbf{CC} & \textbf{\#}  & \textbf{CC}    & \textbf{\#}    & \textbf{CC} & \multicolumn{1}{c|}{\textbf{\#}}  \\ \hline
1           & JP              & 3981             & JP               & 7509             & CN            & 141120         & CN          & 711          & US             & 3180           & RU          & \multicolumn{1}{c|}{15330}        \\ \hline
2           & IT              & 3347             & IT               & 7447             & IN            & 78084          & BR          & 273          & ES             & 1841           & US          & \multicolumn{1}{c|}{5705}         \\ \hline
3           & US              & 2346             & IN               & 4617             & BR            & 66244          & US          & 245          & TR             & 1793           & UA          & \multicolumn{1}{c|}{3558}         \\ \hline
4           & UA              & 1829             & UA               & 4485             & VN            & 58578          & RU          & 116          & IT             & 1486           & IN          & \multicolumn{1}{c|}{2412}         \\ \hline
5           & GB              & 1599             & GB               & 3559             & RU            & 55222          & TR          & 113          & IN             & 1269           & VN          & \multicolumn{1}{c|}{2156}         \\ \hline
6           & IN              & 1556             & US               & 31593            & ID            & 41624          & TH          & 96           & JP             & 1207           & BY          & \multicolumn{1}{c|}{2105}         \\ \hline
7           & FR              & 1099             & GB               & 1162.03          & IT            & 32270          & IT          & 83           & BR             & 1062           & KZ          & \multicolumn{1}{c|}{1920}         \\ \hline
8           & ID              & 865              & BY               & 2856             & KR            & 32018          & IN          & 77           & VN             & 1017           & CN          & \multicolumn{1}{c|}{1745}         \\ \hline
9           & KR              & 794              & ID               & 2328             & PK            & 31726          & VN          & 66           & TH             & 903            & PL          & \multicolumn{1}{c|}{1353}         \\ \hline
10          & CN              & 726              & VN               & 1901             & US            & 29600          & DE          & 64           & FR             & 850            & AR          & \multicolumn{1}{c|}{1183}         \\ \hline
\multicolumn{13}{|c|}{Countries of Interest}                                                                                                                                                                                             \\ \hline
\textbf{\#} & \multicolumn{2}{c|}{GameOver Peer} & \multicolumn{2}{c|}{GameOver Proxy} & \multicolumn{2}{c|}{Conficker} & \multicolumn{2}{c|}{Morto} & \multicolumn{2}{c|}{ZeroAccess} & \multicolumn{2}{c|}{Spam}                       \\ \hline
\textbf{CC} & \textbf{\#}     & \textbf{IPs}     & \textbf{\#}      & \textbf{IPs}     & \textbf{\#}   & \textbf{IPs}   & \textbf{\#} & \textbf{IPs} & \textbf{\#}    & \textbf{IPs}   & \textbf{\#} & \multicolumn{1}{c|}{\textbf{IPs}} \\ \hline
\textbf{NL} & \textbf{48}     & \textbf{47}      & \textbf{52}      & \textbf{57}      & \textbf{53}   & \textbf{609}   & \textbf{36} & \textbf{9}   & \textbf{35}    & \textbf{82}    & \textbf{34} & \multicolumn{1}{c|}{\textbf{190}} \\ \hline
DE          & 16              & 349              & 14               & 1164             & 23            & 11582          & 10          & 64           & 13             & 623            & 13          & \multicolumn{1}{c|}{797}          \\ \hline
ES          & 26              & 241              & 30               & 452              & 15            & 24375          & 15          & 39           & 2              & 1841           & 18          & \multicolumn{1}{c|}{634}          \\ \hline
FI          & 62              & 1                & 62               & 6                & 62            & 39             & 60          & 0            & 60             & 4              & 59          & \multicolumn{1}{c|}{9}            \\ \hline
FR          & 7               & 1099             & 12               & 1279             & 26            & 8840           & 29          & 15           & 10             & 850            & 22          & \multicolumn{1}{c|}{416}          \\ \hline
GB          & 5               & 1599             & 5                & 3559             & 27            & 8010           & 17          & 31           & 14             & 591            & 14          & \multicolumn{1}{c|}{796}          \\ \hline
IT          & 2               & 3347             & 2                & 7447             & 7             & 32270          & 7           & 83           & 4              & 1486           & 17          & \multicolumn{1}{c|}{711}          \\ \hline
JP          & 1               & 3981             & 1                & 7509             & 22            & 14695          & 35          & 10           & 6              & 1207           & 11          & \multicolumn{1}{c|}{1144 1}       \\ \hline
US          & 3               & 2346             & 6                & 3159             & 10            & 29600          & 3           & 245          & 1              & 3180           & 2           & \multicolumn{1}{c|}{5705}         \\ \hline
\end{tabular}
   
 \caption{Average Daily Unique IP addresses Ranking (Year: 2014)}
 \label{tab:contry-abs-2014}
 \end{center}
 \end{sidewaystable}

     \begin{sidewaystable}[ph!]
   \begin{center}
   
    \begin{tabular}{|c|c|c|c|c|c|c|c|c|c|c|c|c|}
  
    \multicolumn{13}{c}{\textit{Top 10 Most Infected Countries}}\\\hline
      \textbf{\#}&\multicolumn{2}{|c|}{\textbf{GameOver Peer}} & \multicolumn{2}{|c|}{\textbf{GameOver Proxy}} & \multicolumn{2}{|c|}{\textbf{Conficker}} &\multicolumn{2}{|c|}{\textbf{Morto}} &\multicolumn{2}{|c|}{\textbf{ZeroAccess}} &\multicolumn{2}{|c|}{\textbf{Spam}} \\ \hline
  & \textbf{CC}& \textbf{\#}&    \textbf{CC}& \textbf{\#}&    \textbf{CC}& \textbf{\#}&    \textbf{CC}& \textbf{\#} & \textbf{CC}& \textbf{\#}  &\textbf{CC}& \textbf{\#}    \\ \hline
	1 & JP & 1780 & JP & 3462 & CN & 111317 & BR & 138 & US & 1600 & VN & 6965 \\ \hline
	2 & IT & 1465 & IT & 2613 & IN & 51943 & CN & 129 & ES & 1524 & US & 3205 \\ \hline
	3 & US & 900 & IN & 1449 & RU & 44718 & US & 120 & IT & 1202 & RU & 2947 \\ \hline
	4 & GB & 618 & UA & 1373 & BR & 43578 & TH & 67 & TR & 1187 & IN & 2261 \\ \hline
	5 & UA & 585 & GB & 1111 & VN & 37159 & RU & 67 & JP & 1017 & UA & 1393 \\ \hline
	6 & IN & 560 & US & 1102 & ID & 29787 & TR & 62 & IN & 898 & BR & 1391 \\ \hline
	7 & FR & 447 & BY & 969 & IT & 21740 & IN & 60 & BR & 766 & CN & 1179 \\ \hline
	8 & KR & 314 & ID & 686 & US & 21191 & VN & 57 & FR & 697 & AR & 792 \\ \hline
	9 & CN & 280 & VN & 539 & KR & 20723 & DE & 40 & AR & 677 & MX & 718 \\ \hline
	10 & ID & 273 & TR & 485 & AR & 19446 & IT & 38 & TH & 598 & KR & 578 \\ \hline
\multicolumn{13}{c}{\textit{Countries of Interest}}\\\hline
      \textbf{\#}&\multicolumn{2}{|c|}{\textbf{GameOver Peer}} & \multicolumn{2}{|c|}{\textbf{GameOver Proxy}} & \multicolumn{2}{|c|}{\textbf{Conficker}} &\multicolumn{2}{|c|}{\textbf{Morto}} &\multicolumn{2}{|c|}{\textbf{ZeroAccess}} &\multicolumn{2}{|c|}{\textbf{Spam}} \\ \hline
      
 \textbf{CC}& \textbf{\#}&    \textbf{IPs}& \textbf{\#}&    \textbf{IPs}& \textbf{\#}&    \textbf{IPs}& \textbf{\#} & \textbf{IPs}& \textbf{\#}  &\textbf{IPs}& \textbf{\#}   &\textbf{IPs} \\ \hline      
 	\textbf{NL} & \textbf{47} & \textbf{18} & \textbf{51}& \textbf{20} & \textbf{53} & \textbf{400} & \textbf{40} & \textbf{4} & \textbf{34} & \textbf{59} & \textbf{31} & \textbf{198} \\ \hline
	DE & 15 & 161 & 11 & 451 & 23 & 6998 & 9 & 40 & 13 & 538 & 11 & 571 \\ \hline
	ES & 27 & 84 & 28 & 156 & 11 & 17698 & 12 & 27 & 2 & 1524 & 18 & 469 \\ \hline
	FI & 62 & 1 & 62 & 3 & 62 & 15 & 62 & 0 & 62 & 1 & 59 & 11 \\ \hline
	FR & 7 & 447 & 13 & 438 & 25 & 6035 & 31 & 7 & 8 & 697 & 23 & 348 \\ \hline
	GB & 4 & 618 & 5 & 1111 & 29 & 4251 & 24 & 12 & 14 & 459 & 20 & 413 \\ \hline
	IT & 2 & 1465 & 2 & 2613 & 7 & 21740 & 10 & 38 & 3 & 1202 & 15 & 548 \\ \hline
	JP & 1 & 1780 & 1 & 3462 & 21 & 8899 & 39 & 4 & 5 & 1017 & 17 & 519 \\ \hline
	US & 3 & 900 & 6 & 1102 & 8 & 21191 & 3 & 120 & 1 & 1600 & 2 & 3205 \\ \hline

 \end{tabular}
 \caption{Average Daily Unique IP addresses Ranking (Year: 2015)}
 \label{tab:contry-abs-2015}
 \end{center}
 \end{sidewaystable}

    \begin{sidewaystable}[ph!]
   \begin{center}
   
    \begin{tabular}{|c|c|c|c|c|c|c|c|c|c|c|c|c|}
  
    \multicolumn{13}{c}{\textit{Top 10 Most Infected Countries}}\\\hline
      \textbf{\#}&\multicolumn{2}{|c|}{\textbf{GameOver Peer}} & \multicolumn{2}{|c|}{\textbf{GameOver Proxy}} & \multicolumn{2}{|c|}{\textbf{Conficker}} &\multicolumn{2}{|c|}{\textbf{Morto}} &\multicolumn{2}{|c|}{\textbf{ZeroAccess}} &\multicolumn{2}{|c|}{\textbf{Spam}} \\ \hline
  & \textbf{CC}& \textbf{\#}&    \textbf{CC}& \textbf{\#}&    \textbf{CC}& \textbf{\#}&    \textbf{CC}& \textbf{\#} & \textbf{CC}& \textbf{\#}  &\textbf{CC}& \textbf{\#}    \\ \hline
	1 & UA & 284 & BY & 1333 & VN & 10245 & TH & 18 & TR & 208 & BY & 982 \\ \hline
	2 & IT & 235 & UA & 696 & EG & 8966 & CY & 18 & VN & 178 & KZ & 947 \\ \hline
	3 & BY & 234 & IT & 523 & ID & 7928 & ZA & 14 & TH & 166 & RU & 600 \\ \hline
	4 & ID & 165 & ZA & 457 & PK & 7738 & IL & 14 & BG & 164 & UA & 553 \\ \hline
	5 & ZA & 165 & ID & 443 & RO & 7387 & SA & 13 & ES & 146 & VN & 377 \\ \hline
	6 & MT & 158 & PE & 340 & BG & 6859 & TR & 13 & MY & 140 & MA & 277 \\ \hline
	7 & CY & 156 & VN & 332 & RS & 5794 & BR & 12 & AR & 134 & PL & 238 \\ \hline
	8 & EE & 139 & KZ & 326 & MA & 5407 & VN & 11 & RS & 116 & CL & 237 \\ \hline
	9 & PE & 126 & MY & 299 & IN & 5021 & EG & 8 & LT & 106 & BG & 228 \\ \hline
	10 & IL & 118 & IN & 297 & MY & 4991 & PT & 8 & IT & 104 & PE & 218 \\ \hline

\multicolumn{13}{c}{\textit{Countries of Interest}}\\\hline
      \textbf{\#}&\multicolumn{2}{|c|}{\textbf{GameOver Peer}} & \multicolumn{2}{|c|}{\textbf{GameOver Proxy}} & \multicolumn{2}{|c|}{\textbf{Conficker}} &\multicolumn{2}{|c|}{\textbf{Morto}} &\multicolumn{2}{|c|}{\textbf{ZeroAccess}} &\multicolumn{2}{|c|}{\textbf{Spam}} \\ \hline
      
 \textbf{CC}& \textbf{\#}&    \textbf{IPs}& \textbf{\#}&    \textbf{IPs}& \textbf{\#}&    \textbf{IPs}& \textbf{\#} & \textbf{IPs}& \textbf{\#}  &\textbf{IPs}& \textbf{\#}   &\textbf{IPs} \\ \hline      
	\textbf{NL} & \textbf{58} & \textbf{7} & \textbf{59} & \textbf{8} & \textbf{60} & \textbf{88} & \textbf{51} & \textbf{1} & \textbf{54} & \textbf{12} & \textbf{48} & \textbf{27} \\ \hline
	DE & 54 & 12 & 47 & 40 & 50 & 397 & 43 & 2 & 42 & 21 & 49 & 27 \\ \hline
	ES & 50 & 19 & 48 & 36 & 25 & 1930 & 31 & 3 & 5 & 146 & 31 & 50 \\ \hline
	FI & 62 & 1 & 62 & 4 & 62 & 23 & 60 & 0 & 61 & 2 & 62 & 5 \\ \hline
	FR & 34 & 43 & 43 & 50 & 52 & 344 & 58 & 1 & 34 & 33 & 57 & 16 \\ \hline
	GB & 22 & 70 & 25 & 155 & 51 & 348 & 52 & 1 & 39 & 26 & 41 & 35 \\ \hline
	IT & 2 & 235 & 3 & 523 & 21 & 2265 & 20 & 6 & 10 & 104 & 32 & 50 \\ \hline
	JP & 17 & 89 & 23 & 167 & 53 & 327 & 59 & 0 & 36 & 27 & 52 & 25 \\ \hline
	US & 47 & 21 & 51 & 28 & 55 & 266 & 42 & 2 & 35 & 29 & 29 & 51 \\ \hline

 \end{tabular}
 \caption{IP addresses/Million Subscribers Ranking (Year:2014)}
 \label{tab:contry-rel-2014}
 \end{center}
 \end{sidewaystable}

 
     \begin{sidewaystable}[ph!]
    \begin{center}
    
     \begin{tabular}{|c|c|c|c|c|c|c|c|c|c|c|c|c|}
   
     \multicolumn{13}{c}{\textit{Top 10 Most Infected Countries}}\\\hline
       \textbf{\#}&\multicolumn{2}{|c|}{\textbf{GameOver Peer}} & \multicolumn{2}{|c|}{\textbf{GameOver Proxy}} & \multicolumn{2}{|c|}{\textbf{Conficker}} &\multicolumn{2}{|c|}{\textbf{Morto}} &\multicolumn{2}{|c|}{\textbf{ZeroAccess}} &\multicolumn{2}{|c|}{\textbf{Spam}} \\ \hline
   & \textbf{CC}& \textbf{\#}&    \textbf{CC}& \textbf{\#}&    \textbf{CC}& \textbf{\#}&    \textbf{CC}& \textbf{\#} & \textbf{CC}& \textbf{\#}  &\textbf{CC}& \textbf{\#}    \\ \hline
	1 & IT & 103 & BY & 452 & VN & 6499 & CY & 12 & TR & 138 & VN & 1218 \\ \hline
	2 & UA & 91 & UA & 213 & EG & 5960 & TH & 12 & ES & 121 & KZ & 277 \\ \hline
	3 & MT & 81 & IT & 183 & ID & 5674 & VN & 10 & AR & 113 & UA & 216 \\ \hline
	4 & BY & 78 & ZA & 141 & BG & 4272 & SA & 8 & BG & 113 & PE & 178 \\ \hline
	5 & EE & 69 & ID & 131 & PK & 3890 & EG & 8 & TH & 110 & IN & 145 \\ \hline
	6 & CY & 61 & KZ & 116 & MA & 3740 & IL & 7 & MY & 108 & BG & 134 \\ \hline
	7 & ZA & 58 & MY & 109 & RO & 3684 & TR & 7 & VN & 99 & AR & 132 \\ \hline
	8 & ID & 52 & IL & 107 & RS & 3483 & ZA & 6 & RS & 86 & BY & 117 \\ \hline
	9 & PE & 49 & PE & 106 & IN & 3340 & BR & 6 & IT & 84 & RU & 115 \\ \hline
	10 & IL & 46 & GR & 105 & AR & 3241 & MY & 5 & GR & 75 & CL & 111 \\ \hline
 
 \multicolumn{13}{c}{\textit{Countries of Interest}}\\\hline
       \textbf{\#}&\multicolumn{2}{|c|}{\textbf{GameOver Peer}} & \multicolumn{2}{|c|}{\textbf{GameOver Proxy}} & \multicolumn{2}{|c|}{\textbf{Conficker}} &\multicolumn{2}{|c|}{\textbf{Morto}} &\multicolumn{2}{|c|}{\textbf{ZeroAccess}} &\multicolumn{2}{|c|}{\textbf{Spam}} \\ \hline
       
  \textbf{CC}& \textbf{\#}&    \textbf{IPs}& \textbf{\#}&    \textbf{IPs}& \textbf{\#}&    \textbf{IPs}& \textbf{\#} & \textbf{IPs}& \textbf{\#}  &\textbf{IPs}& \textbf{\#}   &\textbf{IPs} \\ \hline      
	\textbf{NL} & \textbf{58} & \textbf{3} & \textbf{59} & \textbf{3} & \textbf{59} & \textbf{58} & \textbf{52} & \textbf{1} & \textbf{55} & \textbf{9} & \textbf{42} & \textbf{29} \\ \hline
	DE & 52 & 6 & 46 & 15 & 50 & 240 & 38 & 1 & 41 & 18 & 47 & 20 \\ \hline
	ES & 50 & 7 & 48 & 12 & 24 & 1401 & 30 & 2 & 2 & 121 & 34 & 37 \\ \hline
	FI & 62 & 1 & 61 & 2 & 62 & 9 & 62 & 0 & 61 & 0 & 61 & 6 \\ \hline
	FR & 35 & 17 & 44 & 17 & 51 & 235 & 60 & 0 & 32 & 27 & 56 & 14 \\ \hline
	GB & 21 & 27 & 26 & 48 & 55 & 185 & 54 & 1 & 39 & 20 & 49 & 18 \\ \hline
	IT & 1 & 103 & 3 & 183 & 23 & 1526 & 23 & 3 & 9 & 84 & 33 & 38 \\ \hline
	JP & 13 & 40 & 17 & 77 & 53 & 198 & 61 & 0 & 36 & 23 & 60 & 12 \\ \hline
	US & 46 & 8 & 50 & 10 & 54 & 190 & 43 & 1 & 44 & 14 & 41 & 29 \\ \hline
   
  \end{tabular}
  \caption{IP addresses/Million Subscribers Ranking (Year:2015)}
  \label{tab:contry-rel-2015}
  \end{center}
  \end{sidewaystable}

Our global feeds allow us to state that the Netherlands, on average, ranks quite well in the world, also in comparison to countries that are relevant points of reference and that care about botnet mitigation. We can also see, however, that Finland still outperforms the Netherlands. It has adopted effective mitigation practices before anyone else and has been a consistent top performer for years.

\section{Country performance over time}
\label{sec:chap2-timeseries}

So the Netherlands ranks quite well in the world and among countries that care about botnet mitigation. The question is to what extent we can we credit AbuseHUB with this result? In other words, to what extent did AbuseHUB reinforce the mitigation practices of ISPs and make them more effective? 

Such questions of causality are very difficult to answer under the best of circumstances, as there are always many factors that influence infection rates in addition to mitigation. However, we might gain some clues as to the impact by looking at the development over time. 

Since AbuseHUB was not fully operational in the early part of our measurements, we might gauge its impact from how infection rates developed since 2011. We can only track this development in three data sources, as those are the only ones that span multiple years: Conficker, Morto and Spam. For those sources, we looked at the ranks for the relative infection rates; so the number of infections per subscriber. 

Did the ranking of the Netherlands improve since AbuseHUB became operational? Table~\ref{tab:contries-rankings} shows the evolution of the ranking of the reference countries over the recent years. First and foremost, we can observe that most of the reference countries have remained at more or less the same position, with a few exceptions. This also holds for the Netherlands. Even compared to the reference countries, it performed well above average in the Conficker data, [... Morto ranking...], while in the spam data the ranking fluctuated more, sometimes below average, sometimes above. There is no clear trend one way or the other. The overall ranking is quite stable. In other words, here we do not find strong evidence for an impact of AbuseHUB. This does not mean, however, that this impact was missing. All reference countries are also improving their botnet mitigation, which means that relatively speaking the positions are stable.

  \begin{sidewaystable}[ph!]
  
   \begin{center}
   \begin{footnotesize}
   
    \begin{tabular}{|c|c|c|c|c|c|c|c|c|c|c|c|c|c|c|c|c|}
    
\hline

            \textbf{CC}&\multicolumn{5}{|c|}{\textbf{Conficker}} &\multicolumn{5}{|c|}{\textbf{Morto}} &\multicolumn{5}{|c|}{\textbf{Spam}} \\ \hline
      
 & 2011 & 2012 & 2013 & 2014 & 2015 & 2011 & 2012 & 2013 & 2014  & 2015& 2011 & 2012 & 2013 & 2014 &2015  \\ \hline 
  
	\textbf{NL} & \textbf{60} & \textbf{60} & \textbf{60} & \textbf{60} & \textbf{59} & 
	\textbf{52} & \textbf{49} & \textbf{55} & \textbf{51} & \textbf{52} & \textbf{44} & 
	\textbf{57} & \textbf{52} & \textbf{48} & \textbf{42} \\ \hline
	DE & 47 & 47 & 48 & 50 & 50 & 17 & 35 & 37 & 43 & 38 & 51 & 43 & 40 & 49 & 47 \\ \hline
	ES & 30 & 29 & 27 & 25 & 24 & 33 & 33 & 35 & 31 & 30 & 43 & 24 & 17 & 31 & 34 \\ \hline
	FI & 62 & 62 & 62 & 62 & 62 & 61 & 62 & 62 & 60 & 62 & 61 & 60 & 62 & 62 & 61 \\ \hline
	FR & 52 & 54 & 54 & 52 & 51 & 57 & 58 & 60 & 58 & 60 & 56 & 59 & 56 & 57 & 56 \\ \hline
	GB & 49 & 52 & 52 & 51 & 55 & 40 & 42 & 48 & 52 & 54 & 49 & 47 & 43 & 41 & 49 \\ \hline
	IT & 28 & 26 & 22 & 21 & 23 & 21 & 28 & 18 & 20 & 23 & 39 & 40 & 22 & 32 & 33 \\ \hline
	JP & 53 & 50 & 51 & 53 & 53 & 59 & 26 & 61 & 59 & 61 & 54 & 49 & 53 & 52 & 60 \\ \hline
	US & 55 & 55 & 55 & 55 & 54 & 35 & 43 & 41 & 42 & 43 & 57 & 51 & 45 & 29 & 41 \\ \hline

 \end{tabular}
 \caption{Countries Yearly Ranking (normalized by each countries' subscribers)}
 \label{tab:contries-rankings}
  \end{footnotesize}
 \end{center}

  \end{sidewaystable}

\begin{figure}[!t]
        \centering
    \includegraphics[width=0.8\textwidth]{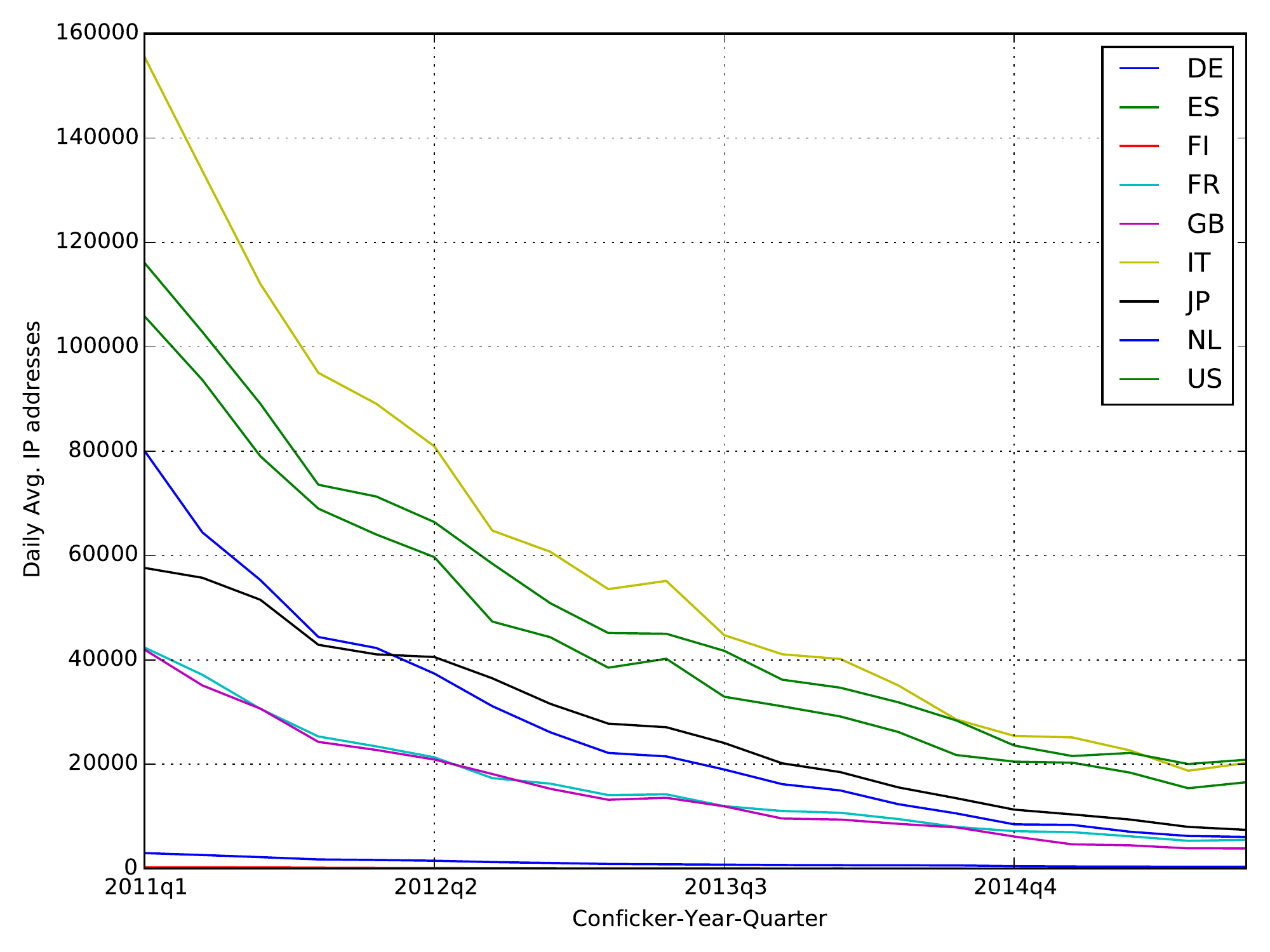}
    \caption{Conficker Countries - Daily Average}
    \label{cfk-country}
    
     \end{figure}

\begin{figure}[t]
        \centering
        
    \includegraphics[width=0.8\textwidth]{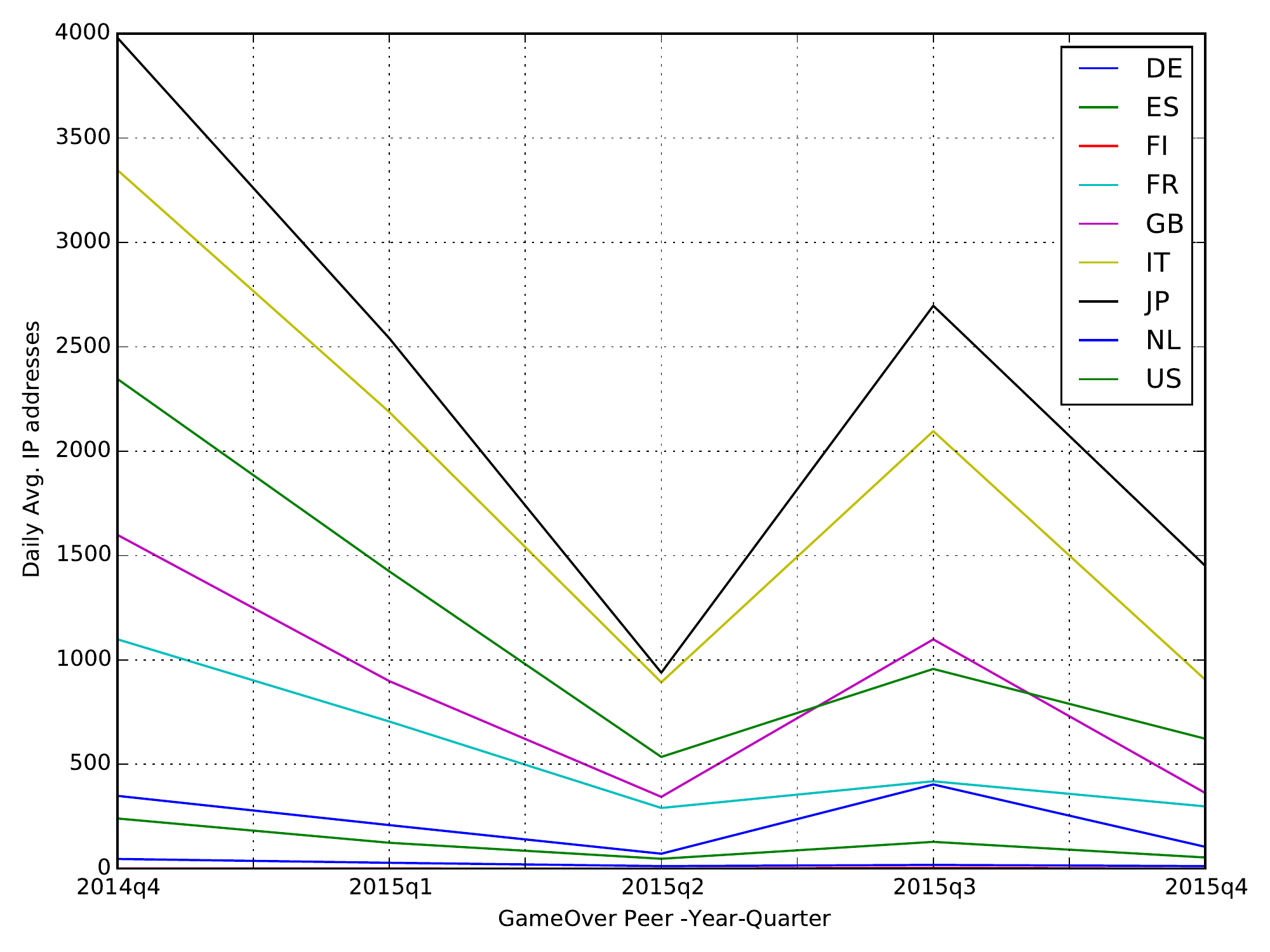}
    \caption{GameOver Peer Countries - Daily  Average}
    \label{q0-peer-country}
     \end{figure}

\begin{figure}[t]
        \centering
        \includegraphics[width=0.8\textwidth]{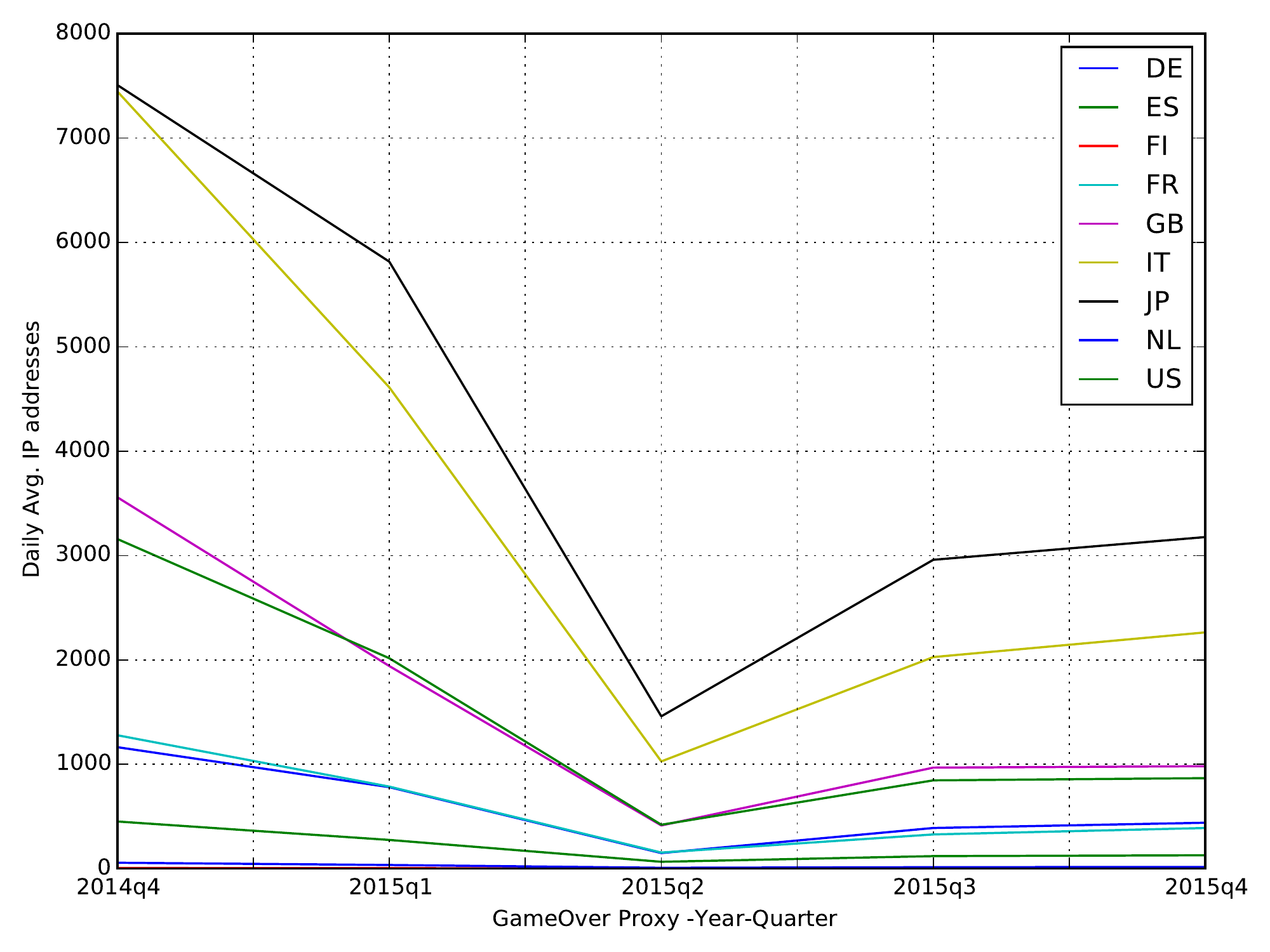}
    \caption{GameOver Proxy Countries - Daily Average}
    \label{q0-proxy-country}
    
 \end{figure}

\begin{figure}[t]
       \centering
       \includegraphics[width=0.8\textwidth]{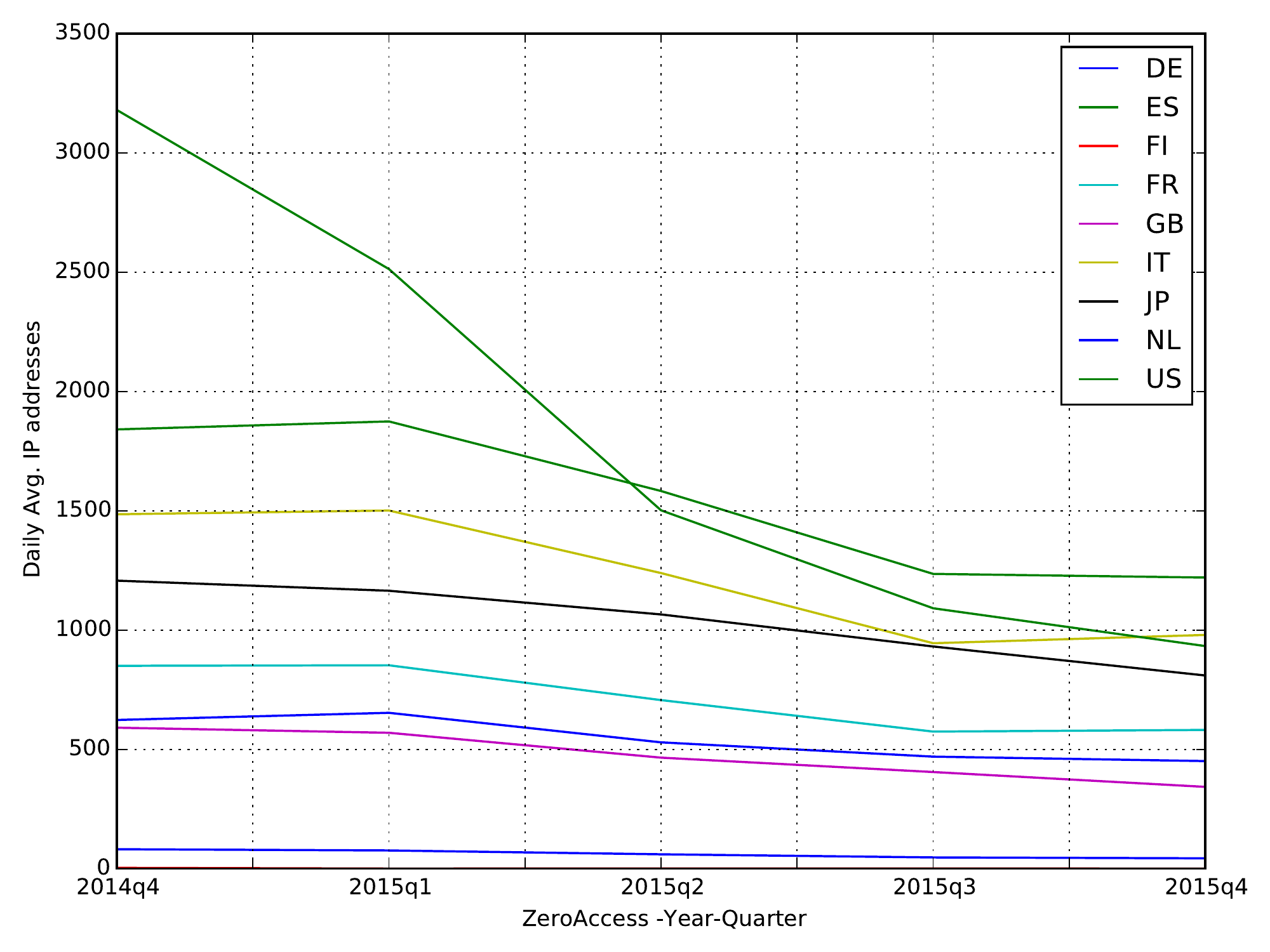}
    \caption{ZeroAccess Countries - Daily Average}
    \label{q0-zero}
 \end{figure}

\begin{figure}[t]
       \centering
       \includegraphics[width=0.8\textwidth]{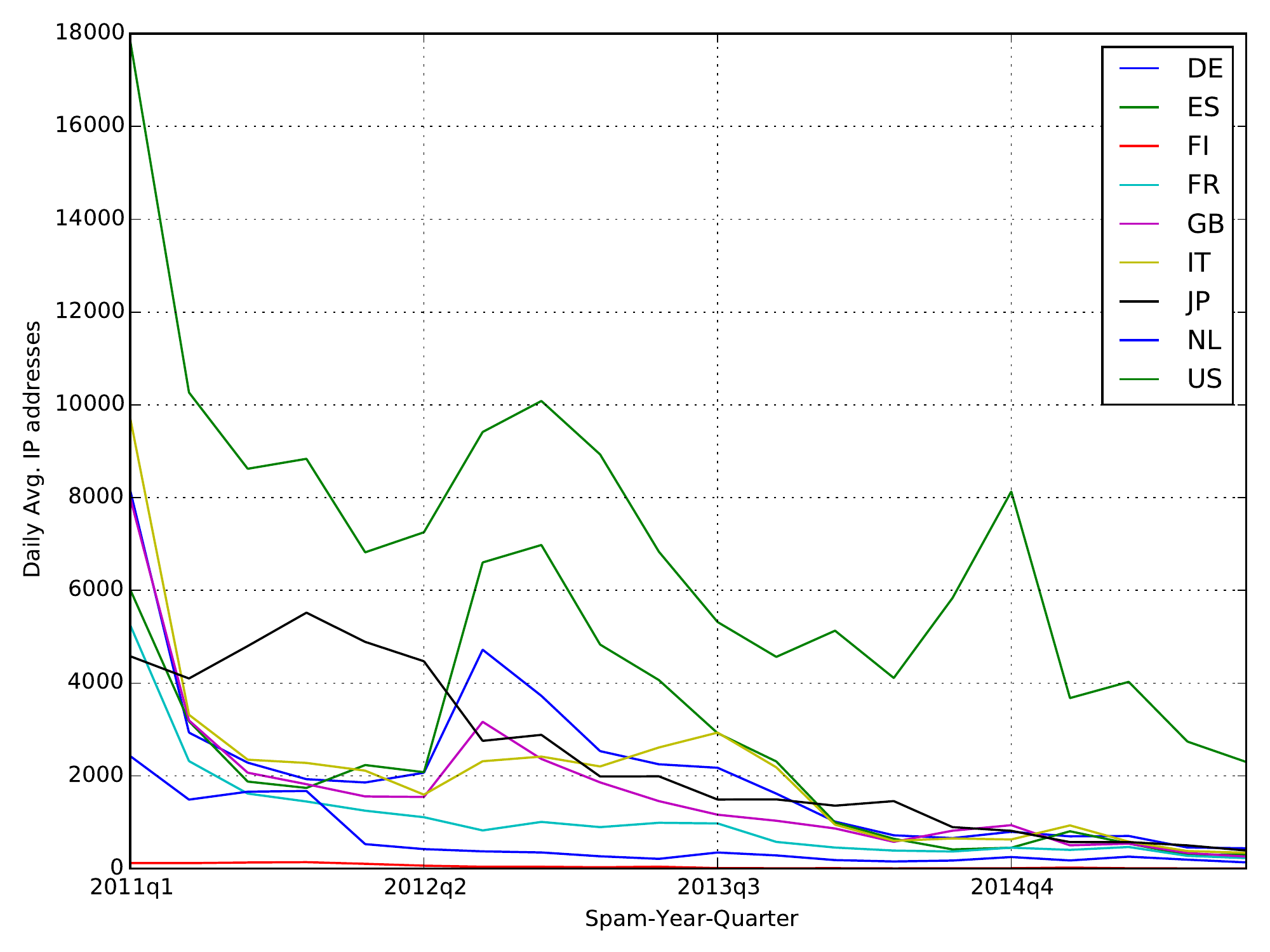}
    \caption{Spam Countries - Daily Average}
    \label{q0-spam}
 \end{figure}

\begin{figure}[t]
       \centering
       \includegraphics[width=0.8\textwidth]{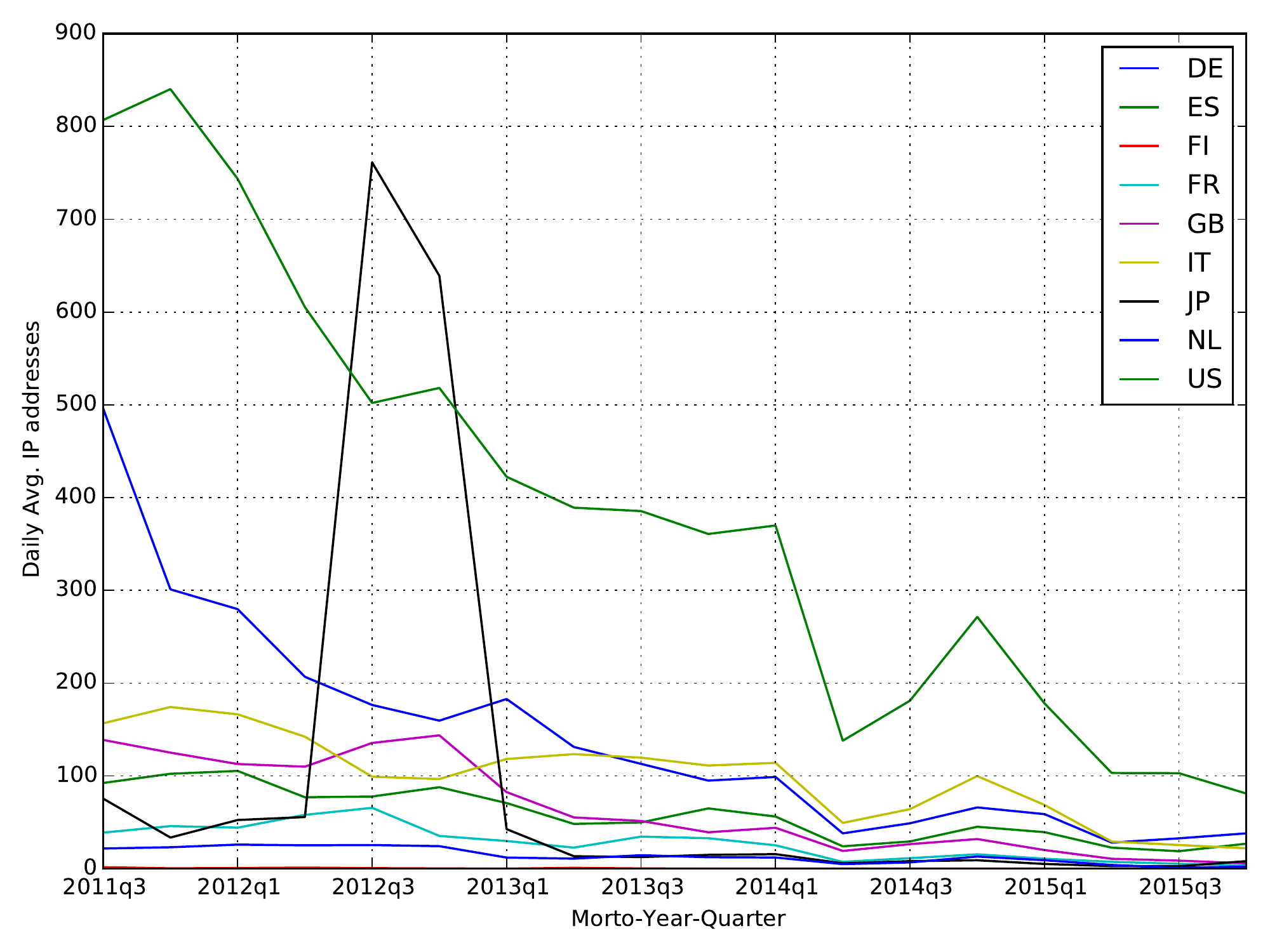}
    \caption{Morto Countries - Daily Average}
    \label{q0-morto}
 \end{figure}

We have also looked at the speed of clean-up across the reference countries. Figures~\ref{cfk-country}--\ref{q0-morto} shows the time series of daily unique IP addresses for the Netherlands and chosen countries. As can be expected, most countries move in a similar pattern, especially for the sinkhole data sources. Botnets that are successfully sinkholed are no longer under the control of the attackers. This means they slowly, but surely, shrink in size, as more and more bots are cleaned up and new infection become increasingly rare. 

\begin{figure}[t]
    \centering
    \includegraphics[width=0.8\textwidth]{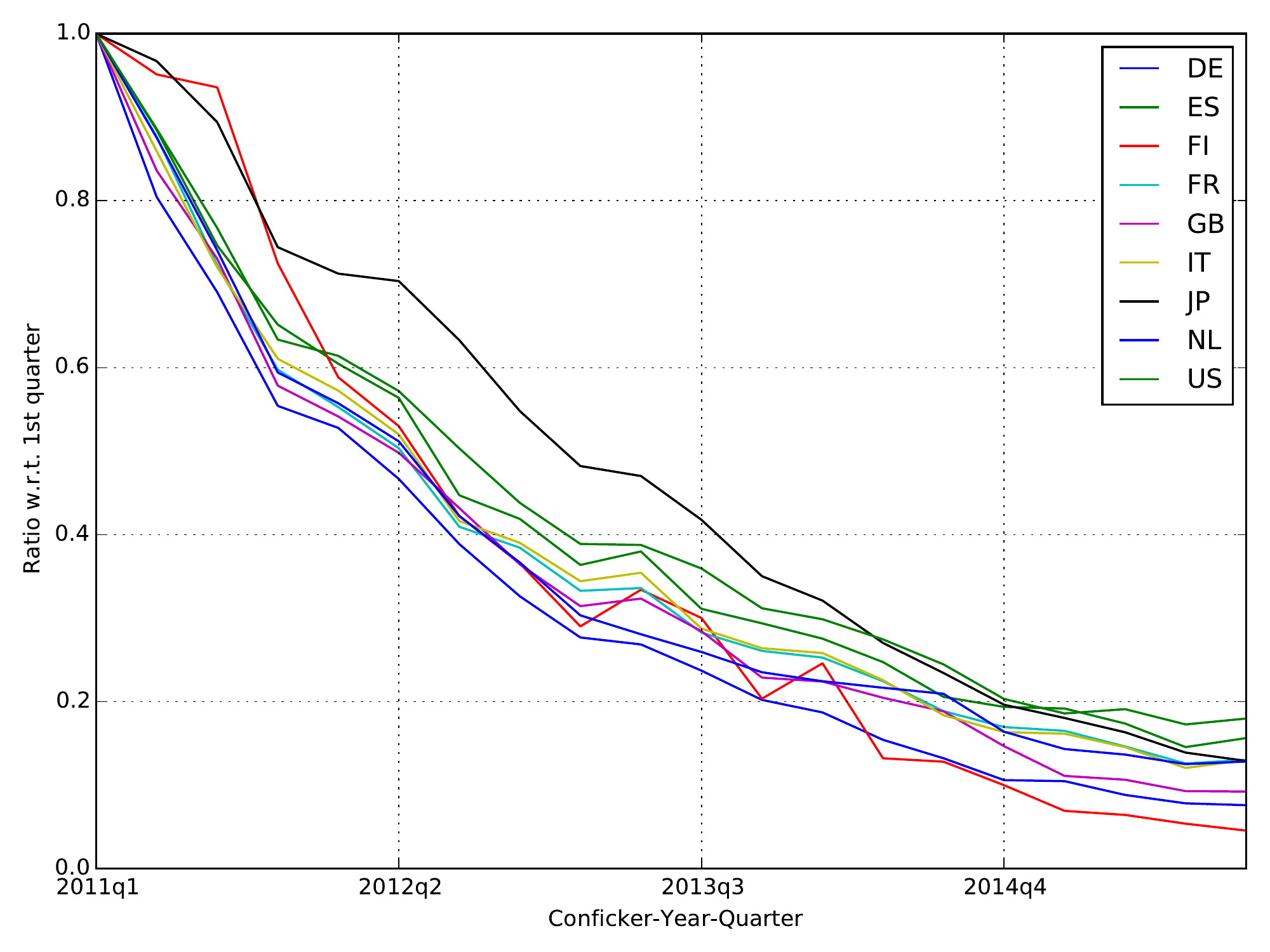}
    \caption{Conficker Countries - Indexed w.r.t. first quarter}
    \label{cfk-countries-index}
     \end{figure}

\begin{figure}[t]
    \centering
    \includegraphics[width=0.8\textwidth]{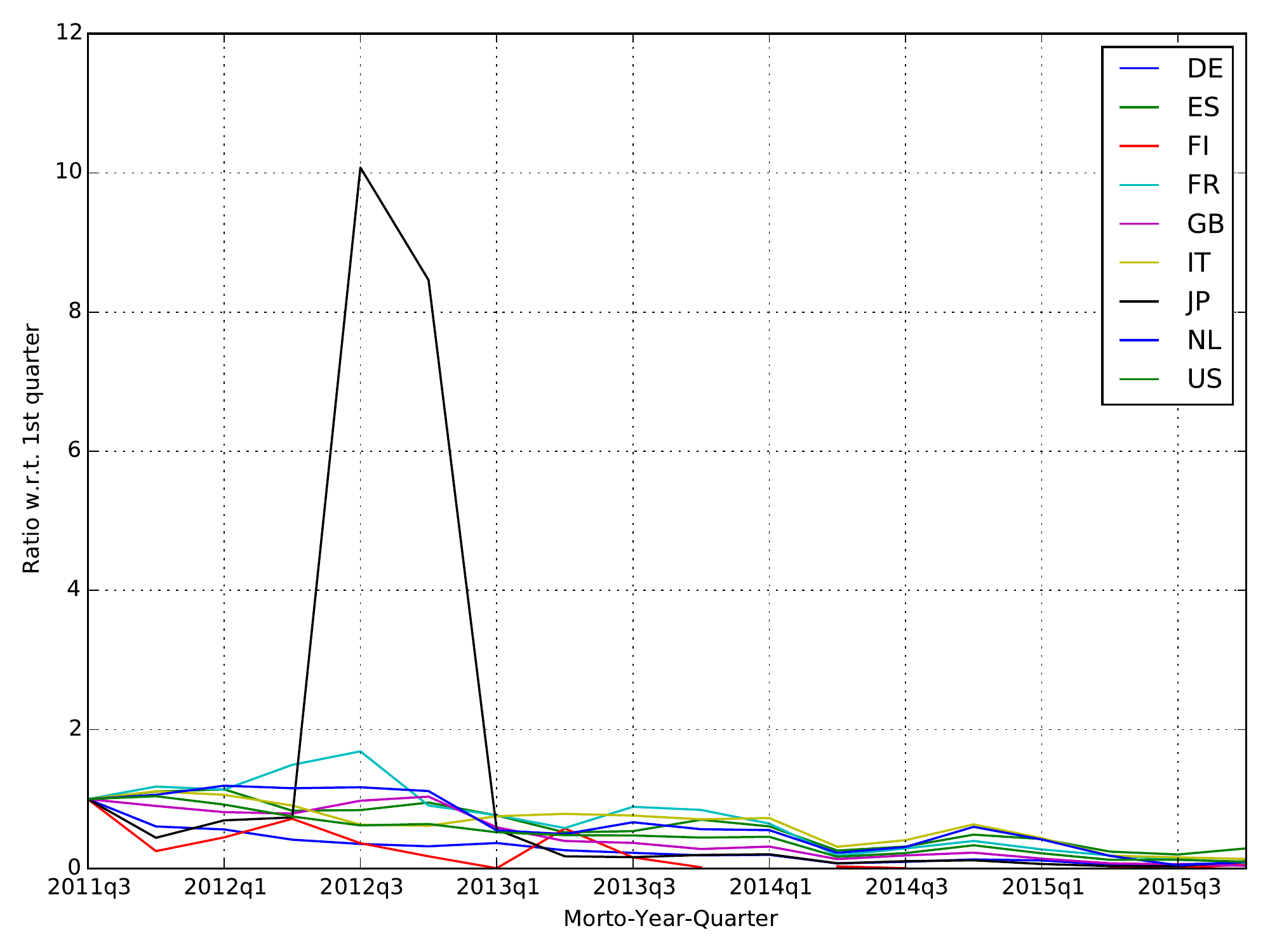}
    \caption{Morto Countries - Indexed w.r.t. first quarter}
    \label{morto-countries-index}
     \end{figure}

\begin{figure}[t]
    \centering
    \includegraphics[width=0.8\textwidth]{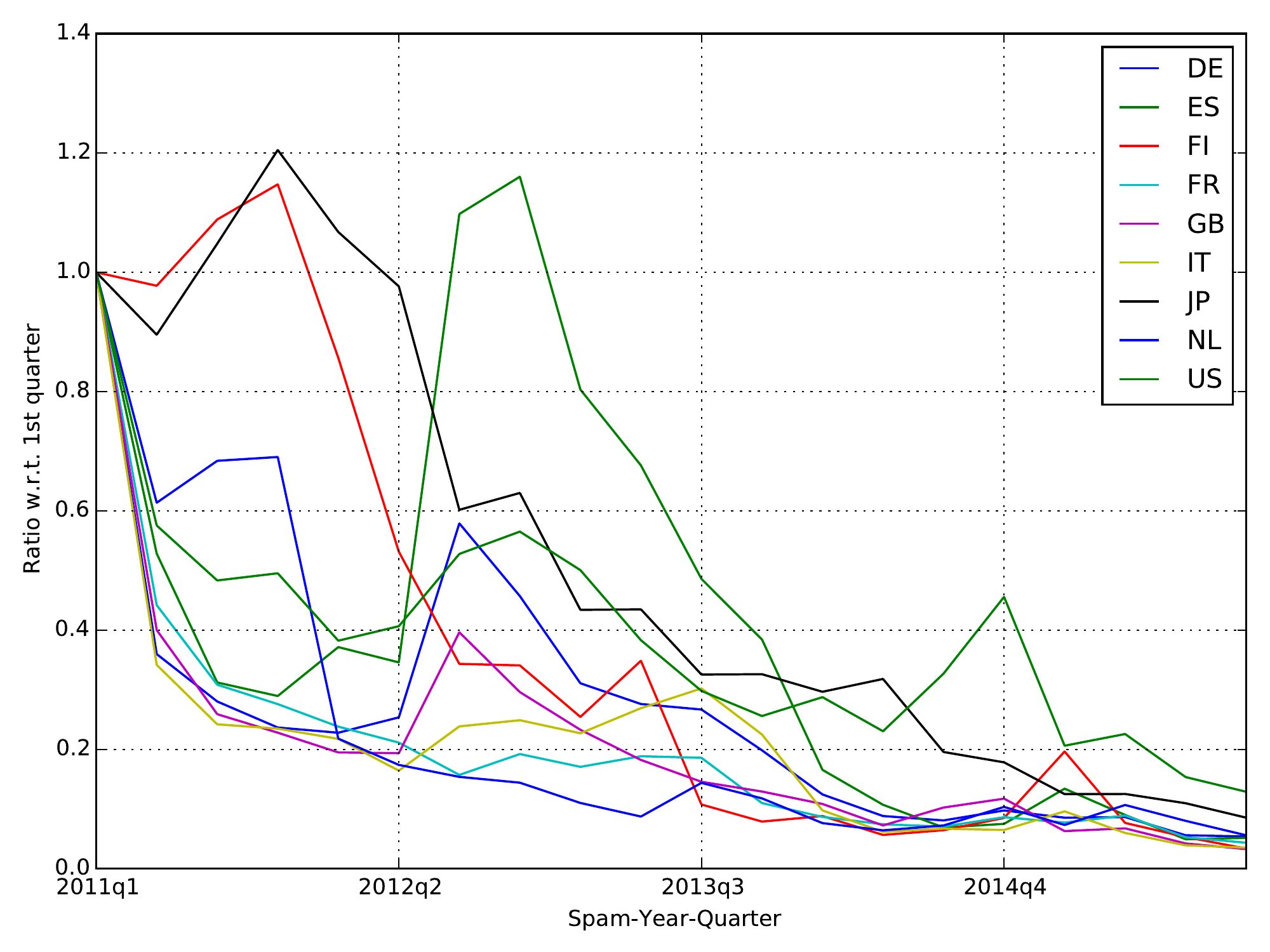}
    \caption{Spam Countries - Indexed w.r.t. first quarter}
    \label{spam-countries-index}
    \end{figure}

To get a better sense of the relative speed of clean up, we have generated indexed time series. Figures~\ref{cfk-countries-index} -- \ref{spam-countries-index} show the infection rates of the reference countries all index at 1 at start of the measurement period -- i.e., we have divided all daily averages by the first daily average of the first measurement. We only performed this for the data sources that span more than one year (Spam, and Conficker). In this way, all countries start with a value equal to 1 and their variation shows the percentage of infections that have increased or reduced.

As can be seen, the steeper curves for the Netherlands indicate better than average cleanup rates for spam and Conficker, though nothing dramatically different from the other reference countries. We should add that in many of these countries, there have also been anti-botnet efforts. In that sense, we should not expect the Netherlands to be dramatically different.

 \section{Main findings}
 
In this chapter, we have presented the evidence that the Netherlands is doing relatively well in terms of botnet mitigation, compared to the rest of the world. Also among a set of reference countries, the Netherlands performs above average, except for spam. 

It is unclear from the available evidence whether AbuseHUB itself has accelerated the process of mitigation by ISPs. The data suggest that cleanup speed in the Netherlands has been a bit higher than other countries. The evidence is inconclusive, however, whether this signals an impact of AbuseHUB, since the speed was already higher before AbuseHUB became operational. Overall, the time trends show no clear impact of AbuseHUB. This can also be explained by the fact that the Netherlands as a whole has been consistently performing above average in most respects, except spam, which makes it difficult for the impact of AbuseHUB to manifest itself in even higher rankings. Once you are on the high end of the ranking, there is not much room left to improve further. Also: the countries in that part of the ranking are also improving mitigation, so maintaining one's ranking can be consistent with improved cleanup.

\newpage

\chapter{Dutch ISPs and ABI efficacy}
\label{chap-isps-only}

In Chapter~\ref{sec:q0}, we presented an analysis of the performance of botnet infections in the Netherlands in comparison to other countries. The approached means that for each country we include all infected machines with an IP addresses that geo-locates to that country, irrespective of what network it is in.

In this chapter, however, we analyze a subset of the infected IP addresses: only those that belong to ISPs. We exclude national research networks (NRENS), hosting providers, government networks, and all non-ISP addresses, following the methodology described in Section~\ref{sec:mapping}. Compared to the country-level comparison in the previous Chapter, this focus provide further information. Botnet mitigation is predominantly an ISP effort. Indeed, AbuseHUB has first and foremost been an initiative to support ISPs. It is therefore helpful to more precisely study infection levels in ISP networks.

We investigate two research questions:

\begin{itemize}
\item How do Dutch ISPs rank in comparison to ISPs in other countries. 

\item How do Dutch ISPs compare against countries with an Anti-Botnet Initiative (ABI) similar to AbuseHUB and against peer countries without an ABI? 

\end{itemize}

We want to note that this comparison can only include those ISPs for which the TeleGeography database provides subscriber numbers in each quarter. Smaller ISPs are not covered by TeleGeography. This means that the metrics cover six Dutch ISPs: Ziggo, UPC, Vodafone, Tele2, KPN (including Telfort and XS4All), Online, CAIW. It is also important to emphasize that our ISP mappings cover 60 countries, so not the whole world. The selection of countries is explained in Section~\ref{sec:mapping}. Appendix \ref{countries} lists the complete set of included countries.

We cover the first research question in Section~\ref{sec:isponly-contries} and the second in Section~\ref{sec:isponly-abi}. Finally, in Section~\ref{sec:isps-only-findings}, we present the main findings of this chapter.

\section{Countries`s ISP rankings}
\label{sec:isponly-contries}

Tables~\ref{tab:contryisps-abs-2014}-\ref{tab:contryisps-abs-2015} show the average number of daily unique IP addresses for each global abuse feed we have analyzed. They present  the top 10 most infected countries and for the countries of reference in 2014 and 2015. This table can be seen as a subset of Tables~\ref{tab:contry-abs-2014}-\ref{tab:contry-abs-2015}, since it only includes IP addresses associated to Autonomous Systems belonging to ISPs in these countries.

Analyzing this table, we can see that the Netherlands only has a modest share of the overall problem. Its high rank number indicates there are many countries with more infected machines (see column \#; rank 1 means the highest infection rate, rank 60 the lowest). The ranking of the Netherlands varies between 32 and 55 over the different  sources we have evaluated. This is similar to what we found at the country level in the previous chapter. 

These numbers, while informative about the absolute size of the problem, cannot be compared across countries. Larger countries have more Internet subscribers and that is an important driver of the number of infections. To compensate for these differences, we count the number of infections for the ISPs of each country and normalize it by the total number of Internet subscribers of those ISPs. The subscriber data was obtained from the TelegeoGraphy GlobalComms database~\cite{telegeography}. These results can be seen in Tables~\ref{tab:contry-rel-isps-2014} -- \ref{tab:contry-rel-isps-2015}. 

In these relative metrics, which arguably give a more informative picture, we can see that the Dutch ISPs that are included display an excellent performance. They consistently rank above 50, out of the 60 countries that are included. The Dutch ISPs even outperform their peers in the reference countries. Only Finland is consistently better.

     \begin{sidewaystable}[ph!]
   \begin{center}
   
    \begin{tabular}{|c|c|c|c|c|c|c|c|c|c|c|c|c|}
  
    \multicolumn{13}{c}{\textit{Top 10 Most Infected Countries ISPs}}\\\hline
      \textbf{\#}&\multicolumn{2}{|c|}{\textbf{GameOver Peer}} & \multicolumn{2}{|c|}{\textbf{GameOver Proxy}} & \multicolumn{2}{|c|}{\textbf{Conficker}} &\multicolumn{2}{|c|}{\textbf{Morto}} &\multicolumn{2}{|c|}{\textbf{ZeroAccess}} &\multicolumn{2}{|c|}{\textbf{Spam}} \\ \hline
  & \textbf{CC}& \textbf{\#}&    \textbf{CC}& \textbf{\#}&    \textbf{CC}& \textbf{\#}&    \textbf{CC}& \textbf{\#} & \textbf{CC}& \textbf{\#}  &\textbf{CC}& \textbf{\#}    \\ \hline
	1 & IT & 2735 & IT & 6091 & CN & 126410 & CN & 616 & US & 2144 & RU & 6002 \\ \hline
	2 & JP & 2719 & JP & 4855 & BR & 48355 & BR & 233 & ES & 1691 & US & 2273 \\ \hline
	3 & US & 1584 & GB & 2848 & VN & 47844 & US & 110 & TR & 1676 & BY & 1803 \\ \hline
	4 & GB & 1178 & BY & 2555 & ID & 30302 & TR & 95 & IT & 1287 & VN & 1791 \\ \hline
	5 & FR & 925 & US & 1990 & KR & 27035 & TH & 82 & BR & 912 & KZ & 1516 \\ \hline
	6 & KR & 702 & VN & 1637 & IT & 25889 & IT & 73 & VN & 891 & CN & 1503 \\ \hline
	7 & CN & 612 & ID & 1602 & EG & 23744 & DE & 53 & FR & 800 & AR & 897 \\ \hline
	8 & TR & 563 & UA & 1551 & PK & 22222 & VN & 48 & JP & 798 & UA & 728 \\ \hline
	9 & VN & 517 & TR & 1459 & AR & 21237 & RU & 37 & TH & 744 & JP & 652 \\ \hline
	10 & ID & 473 & FR & 1089 & ES & 21178 & AR & 37 & AR & 661 & KR & 639 \\ \hline

\multicolumn{13}{c}{\textit{Countries of Interest}}\\\hline
      \textbf{\#}&\multicolumn{2}{|c|}{\textbf{GameOver Peer}} & \multicolumn{2}{|c|}{\textbf{GameOver Proxy}} & \multicolumn{2}{|c|}{\textbf{Conficker}} &\multicolumn{2}{|c|}{\textbf{Morto}} &\multicolumn{2}{|c|}{\textbf{ZeroAccess}} &\multicolumn{2}{|c|}{\textbf{Spam}} \\ \hline
      
 \textbf{CC}& \textbf{\#}&    \textbf{IPs}& \textbf{\#}&    \textbf{IPs}& \textbf{\#}&    \textbf{IPs}& \textbf{\#} & \textbf{IPs}& \textbf{\#}  &\textbf{IPs}& \textbf{\#}   &\textbf{IPs} \\ \hline      
	\textbf{NL} & \textbf{50} & \textbf{22} & \textbf{53} & \textbf{27} & \textbf{55} & \textbf{183} & \textbf{53} & \textbf{1} & \textbf{33} & \textbf{58} & \textbf{42} & \textbf{38} \\ \hline
	DE & 13 & 291 & 12 & 1017 & 19 & 9676 & 7 & 53 & 11 & 541 & 20 & 332 \\ \hline
	ES & 19 & 209 & 24 & 408 & 10 & 21178 & 11 & 35 & 2 & 1691 & 14 & 510 \\ \hline
	FI & 60 & 1 & 60 & 6 & 60 & 25 & 60 & 0 & 58 & 4 & 57 & 6 \\ \hline
	FR & 5 & 925 & 10 & 1089 & 22 & 7600 & 25 & 12 & 7 & 800 & 25 & 170 \\ \hline
	GB & 4 & 1178 & 3 & 2848 & 24 & 5437 & 19 & 16 & 12 & 508 & 13 & 525 \\ \hline
	IT & 1 & 2735 & 1 & 6091 & 6 & 25889 & 6 & 73 & 4 & 1287 & 12 & 530 \\ \hline
	JP & 2 & 2719 & 2 & 4855 & 20 & 9233 & 35 & 7 & 8 & 798 & 9 & 652 \\ \hline
	US & 3 & 1584 & 5 & 1990 & 15 & 14211 & 3 & 110 & 1 & 2144 & 2 & 2273 \\ \hline

 \end{tabular}
 \caption{Average daily unique IP addresses ranking for ISPs only (Year: 2014)}
 \label{tab:contryisps-abs-2014}
 \end{center}
 \end{sidewaystable}


     \begin{sidewaystable}[ph!]
   \begin{center}
   
    \begin{tabular}{|c|c|c|c|c|c|c|c|c|c|c|c|c|}
  
    \multicolumn{13}{c}{\textit{Top 10 Most Infected Countries ISPs}}\\\hline
      \textbf{\#}&\multicolumn{2}{|c|}{\textbf{GameOver Peer}} & \multicolumn{2}{|c|}{\textbf{GameOver Proxy}} & \multicolumn{2}{|c|}{\textbf{Conficker}} &\multicolumn{2}{|c|}{\textbf{Morto}} &\multicolumn{2}{|c|}{\textbf{ZeroAccess}} &\multicolumn{2}{|c|}{\textbf{Spam}} \\ \hline
  & \textbf{CC}& \textbf{\#}&    \textbf{CC}& \textbf{\#}&    \textbf{CC}& \textbf{\#}&    \textbf{CC}& \textbf{\#} & \textbf{CC}& \textbf{\#}  &\textbf{CC}& \textbf{\#}    \\ \hline
	1 & JP & 1216 & JP & 2254 & CN & 99479 & BR & 115 & ES & 1407 & VN & 5656 \\ \hline
	2 & IT & 1190 & IT & 2093 & VN & 30432 & CN & 100 & TR & 1100 & RU & 1290 \\ \hline
	3 & US & 606 & GB & 878 & BR & 29545 & TH & 57 & US & 1056 & US & 890 \\ \hline
	4 & GB & 439 & BY & 863 & ID & 22700 & TR & 51 & IT & 1039 & CN & 798 \\ \hline
	5 & FR & 373 & US & 681 & IT & 17431 & US & 48 & JP & 676 & BR & 602 \\ \hline
	6 & KR & 278 & VN & 462 & KR & 17337 & VN & 41 & FR & 655 & AR & 588 \\ \hline
	7 & CN & 235 & UA & 458 & EG & 16571 & DE & 30 & BR & 640 & KR & 483 \\ \hline
	8 & TR & 199 & TR & 441 & ES & 15366 & IT & 28 & AR & 550 & KZ & 442 \\ \hline
	9 & VN & 165 & ID & 434 & AR & 15107 & ES & 24 & TH & 510 & TR & 433 \\ \hline
	10 & DE & 128 & DE & 380 & RU & 12237 & SA & 22 & VN & 482 & IT & 367 \\ \hline

\multicolumn{13}{c}{\textit{Countries of Interest}}\\\hline
      \textbf{\#}&\multicolumn{2}{|c|}{\textbf{GameOver Peer}} & \multicolumn{2}{|c|}{\textbf{GameOver Proxy}} & \multicolumn{2}{|c|}{\textbf{Conficker}} &\multicolumn{2}{|c|}{\textbf{Morto}} &\multicolumn{2}{|c|}{\textbf{ZeroAccess}} &\multicolumn{2}{|c|}{\textbf{Spam}} \\ \hline
      
 \textbf{CC}& \textbf{\#}&    \textbf{IPs}& \textbf{\#}&    \textbf{IPs}& \textbf{\#}&    \textbf{IPs}& \textbf{\#} & \textbf{IPs}& \textbf{\#}  &\textbf{IPs}& \textbf{\#}   &\textbf{IPs} \\ \hline      
\textbf{NL} & \textbf{53} & \textbf{6} & \textbf{55} & \textbf{8} & \textbf{54} & \textbf{109} & \textbf{54} & \textbf{1} & \textbf{32} & \textbf{39} & \textbf{41} & \textbf{29} \\ \hline
	DE & 10 & 128 & 10 & 380 & 19 & 5748 & 7 & 30 & 11 & 451 & 20 & 206 \\ \hline
	ES & 23 & 69 & 23 & 138 & 8 & 15366 & 9 & 24 & 1 & 1407 & 11 & 354 \\ \hline
	FI & 60 & 1 & 59 & 3 & 60 & 9 & 57 & 0 & 60 & 1 & 55 & 7 \\ \hline
	FR & 5 & 373 & 11 & 369 & 21 & 5035 & 25 & 7 & 6 & 655 & 29 & 105 \\ \hline
	GB & 4 & 439 & 3 & 878 & 31 & 2610 & 30 & 5 & 12 & 391 & 22 & 189 \\ \hline
	IT & 2 & 1190 & 2 & 2093 & 5 & 17431 & 8 & 28 & 4 & 1039 & 10 & 367 \\ \hline
	JP & 1 & 1216 & 1 & 2254 & 18 & 5784 & 37 & 2 & 5 & 676 & 13 & 322 \\ \hline
	US & 3 & 606 & 5 & 681 & 13 & 10350 & 5 & 48 & 3 & 1056 & 3 & 890 \\ \hline

 \end{tabular}
 \caption{Average daily unique IP addresses ranking for ISPs only (Year: 2015)}
 \label{tab:contryisps-abs-2015}
 \end{center}
 \end{sidewaystable}

    \begin{sidewaystable}[ph!]
   \begin{center}
   
    \begin{tabular}{|c|c|c|c|c|c|c|c|c|c|c|c|c|}
  
    \multicolumn{13}{c}{\textit{Top 10 Most Infected Countries ISPs - normalized by million subscribers }}\\\hline
      \textbf{\#}&\multicolumn{2}{|c|}{\textbf{GameOver Peer}} & \multicolumn{2}{|c|}{\textbf{GameOver Proxy}} & \multicolumn{2}{|c|}{\textbf{Conficker}} &\multicolumn{2}{|c|}{\textbf{Morto}} &\multicolumn{2}{|c|}{\textbf{ZeroAccess}} &\multicolumn{2}{|c|}{\textbf{Spam}} \\ \hline
  & \textbf{CC}& \textbf{\#}&    \textbf{CC}& \textbf{\#}&    \textbf{CC}& \textbf{\#}&    \textbf{CC}& \textbf{\#} & \textbf{CC}& \textbf{\#}  &\textbf{CC}& \textbf{\#}    \\ \hline
	1 & BY & 208 & BY & 1331 & EG & 10018 & CY & 20 & TR & 197 & BY & 939 \\ \hline
	2 & EE & 133 & ID & 356 & ID & 6734 & SA & 14 & GR & 104 & KZ & 266 \\ \hline
	3 & CY & 126 & PE & 341 & VN & 5503 & ZA & 9 & VN & 102 & PE & 217 \\ \hline
	4 & IT & 123 & UA & 305 & RO & 5409 & EG & 9 & PE & 101 & VN & 206 \\ \hline
	5 & PE & 120 & IT & 274 & HU & 4013 & IL & 9 & BG & 101 & UA & 143 \\ \hline
	6 & MT & 110 & ZA & 271 & CY & 3161 & GR & 7 & RO & 96 & MA & 115 \\ \hline
	7 & ID & 105 & GR & 262 & PK & 3124 & TR & 6 & SA & 81 & IL & 106 \\ \hline
	8 & IE & 99 & SA & 246 & BG & 2990 & BR & 6 & ES & 80 & CL & 95 \\ \hline
	9 & ZA & 83 & CY & 243 & MY & 2589 & BY & 6 & RS & 76 & PL & 91 \\ \hline
	10 & UA & 71 & VN & 189 & RS & 2453 & TH & 6 & MY & 75 & AR & 90 \\ \hline
\multicolumn{13}{c}{\textit{Countries of Interest}}\\\hline
      \textbf{\#}&\multicolumn{2}{|c|}{\textbf{GameOver Peer}} & \multicolumn{2}{|c|}{\textbf{GameOver Proxy}} & \multicolumn{2}{|c|}{\textbf{Conficker}} &\multicolumn{2}{|c|}{\textbf{Morto}} &\multicolumn{2}{|c|}{\textbf{ZeroAccess}} &\multicolumn{2}{|c|}{\textbf{Spam}} \\ \hline
      
 \textbf{CC}& \textbf{\#}&    \textbf{IPs}& \textbf{\#}&    \textbf{IPs}& \textbf{\#}&    \textbf{IPs}& \textbf{\#} & \textbf{IPs}& \textbf{\#}  &\textbf{IPs}& \textbf{\#}   &\textbf{IPs} \\ \hline      
	\textbf{NL} & \textbf{57} & \textbf{1} & \textbf{58} & \textbf{2} & \textbf{59} & \textbf{11} & \textbf{56} & \textbf{0} & \textbf{54} & \textbf{4} & \textbf{57} & \textbf{2} \\ \hline
	DE & 48 & 8 & 42 & 28 & 44 & 244 & 36 & 2 & 39 & 15 & 46 & 8 \\ \hline
	ES & 39 & 16 & 40 & 31 & 25 & 919 & 23 & 3 & 8 & 80 & 27 & 24 \\ \hline
	FI & 58 & 1 & 56 & 3 & 60 & 5 & 60 & 0 & 58 & 2 & 59 & 1 \\ \hline
	FR & 41 & 15 & 49 & 17 & 48 & 111 & 51 & 0 & 38 & 15 & 55 & 2 \\ \hline
	GB & 33 & 21 & 32 & 52 & 51 & 99 & 53 & 0 & 47 & 9 & 45 & 10 \\ \hline
	IT & 4 & 123 & 5 & 274 & 19 & 1141 & 18 & 4 & 14 & 58 & 25 & 26 \\ \hline
	JP & 20 & 40 & 28 & 71 & 50 & 99 & 55 & 0 & 41 & 12 & 49 & 7 \\ \hline
	US & 54 & 4 & 55 & 5 & 58 & 21 & 50 & 1 & 53 & 5 & 53 & 3 \\ \hline

 \end{tabular}
 \caption{Daily average of unique number of IP addresses seen in data source, normalized by million subscribers in 60 countries (Year: 2014)}
 \label{tab:contry-rel-isps-2014}
 \end{center}
 \end{sidewaystable}

\begin{sidewaystable}[ph!]
   \begin{center}
   
    \begin{tabular}{|c|c|c|c|c|c|c|c|c|c|c|c|c|}
  
    \multicolumn{13}{c}{\textit{Top 10 Most Infected Countries ISPs - normalized by million subscribers }}\\\hline
      \textbf{\#}&\multicolumn{2}{|c|}{\textbf{GameOver Peer}} & \multicolumn{2}{|c|}{\textbf{GameOver Proxy}} & \multicolumn{2}{|c|}{\textbf{Conficker}} &\multicolumn{2}{|c|}{\textbf{Morto}} &\multicolumn{2}{|c|}{\textbf{ZeroAccess}} &\multicolumn{2}{|c|}{\textbf{Spam}} \\ \hline
  & \textbf{CC}& \textbf{\#}&    \textbf{CC}& \textbf{\#}&    \textbf{CC}& \textbf{\#}&    \textbf{CC}& \textbf{\#} & \textbf{CC}& \textbf{\#}  &\textbf{CC}& \textbf{\#}    \\ \hline
	1 & EE & 67 & BY & 449 & EG & 6992 & CY & 16 & TR & 129 & VN & 651 \\ \hline
	2 & BY & 66 & PE & 106 & ID & 5044 & SA & 9 & MY & 110 & PE & 175 \\ \hline
	3 & IT & 54 & GR & 105 & VN & 3500 & EG & 9 & GR & 77 & BY & 88 \\ \hline
	4 & CY & 48 & ID & 96 & RO & 2771 & ZA & 6 & BG & 69 & RO & 87 \\ \hline
	5 & PE & 47 & IT & 94 & CY & 2343 & IS & 6 & PE & 68 & KZ & 78 \\ \hline
	6 & IE & 40 & UA & 90 & HU & 2231 & TR & 6 & ES & 67 & ID & 65 \\ \hline
	7 & ZA & 37 & ZA & 85 & BG & 1889 & GR & 5 & RO & 64 & SA & 63 \\ \hline
	8 & MT & 37 & CY & 68 & PE & 1759 & VN & 5 & RS & 58 & IL & 60 \\ \hline
	9 & GR & 26 & SA & 63 & RS & 1747 & SI & 4 & AR & 55 & AR & 59 \\ \hline
	10 & ID & 26 & MY & 55 & MA & 1660 & TH & 4 & VN & 55 & BG & 58 \\ \hline

\multicolumn{13}{c}{\textit{Countries of Interest}}\\\hline
      \textbf{\#}&\multicolumn{2}{|c|}{\textbf{GameOver Peer}} & \multicolumn{2}{|c|}{\textbf{GameOver Proxy}} & \multicolumn{2}{|c|}{\textbf{Conficker}} &\multicolumn{2}{|c|}{\textbf{Morto}} &\multicolumn{2}{|c|}{\textbf{ZeroAccess}} &\multicolumn{2}{|c|}{\textbf{Spam}} \\ \hline
      
 \textbf{CC}& \textbf{\#}&    \textbf{IPs}& \textbf{\#}&    \textbf{IPs}& \textbf{\#}&    \textbf{IPs}& \textbf{\#} & \textbf{IPs}& \textbf{\#}  &\textbf{IPs}& \textbf{\#}   &\textbf{IPs} \\ \hline      
\textbf{NL} & \textbf{57} & \textbf{0} & \textbf{58} & \textbf{0} & \textbf{59} & \textbf{7} & \textbf{52} & \textbf{0} & \textbf{52} & \textbf{3} & \textbf{56} & \textbf{2} \\ \hline
	DE & 47 & 4 & 45 & 8 & 45 & 145 & 35 & 1 & 39 & 11 & 46 & 5 \\ \hline
	ES & 43 & 5 & 42 & 9 & 23 & 667 & 23 & 2 & 6 & 67 & 29 & 17 \\ \hline
	FI & 58 & 0 & 55 & 1 & 60 & 2 & 57 & 0 & 58 & 0 & 54 & 2 \\ \hline
	FR & 40 & 6 & 48 & 6 & 48 & 74 & 50 & 0 & 38 & 12 & 58 & 2 \\ \hline
	GB & 32 & 8 & 35 & 16 & 51 & 47 & 51 & 0 & 46 & 7 & 48 & 3 \\ \hline
	IT & 3 & 54 & 5 & 94 & 20 & 785 & 22 & 2 & 13 & 47 & 27 & 18 \\ \hline
	JP & 19 & 19 & 18 & 35 & 49 & 65 & 54 & 0 & 42 & 10 & 47 & 4 \\ \hline
	US & 56 & 1 & 57 & 1 & 58 & 14 & 48 & 0 & 54 & 3 & 59 & 1 \\ \hline

 \end{tabular}
 \caption{Daily average of unique number of IP addresses seen in data source, normalized by million subscribers in 60 countries (Year: 2015)}
 \label{tab:contry-rel-isps-2015}
 \end{center}
 \end{sidewaystable}

\section{Countries with an Anti-Botnet Initiative}
\label{sec:isponly-abi}

For the second research question presented in the introduction of this report, we explored how countries with a national Anti-Botnet Initiative (ABI) performed in comparison to each other and to countries without such an initiative. 

We compare three groups:

\begin{itemize}
 \item Mature ABI: countries that have ABIs initiatives for longer periods of time. 
 \item New ABI: countries that have only recently started a national ABI.
 \item No ABI: a selected group of countries without an ABI, but otherwise more or less comparable to the other groups in terms of ICT development.
\end{itemize}

Table~\ref{tab:abi-countries} shows which countries we have grouped into the three categories accordingly to the presence and status of their national ABIs.

 \begin{table}
  \begin{center}
   \begin{tabular}{|c|c|} \hline
	\textbf{Group} & \textbf{Country Codes}  \\ \hline
	Mature ABI   &AU, DE, IE, JP, KR , FI \\ \hline

	Recent ABI &BE, ES, HR, RO, IT, FR, PT, US \\ \hline

	No ABI & GB, NO, CZ, NZ, LU\\ \hline

    \end{tabular}
    \end{center}
    \caption{ABI countries group}
  \label{tab:abi-countries}
  \end{table}  
  

Tables~\ref{tab:abi-grp-cfk} through \ref{tab:abi-grp-zero} show the average ranking of each group among the 60 countries -- in other words, we averaged the rank numbers of the countries in each group to get a sense of the overall position of the group.

We can see that countries with a mature ABI rank better than those with a recent or no ABI. This holds across all sources except for the GameOver Zeus feeds. Furthermore, we can also see that the Netherlands ranks substantially better than the other groups, even than the other countries with a mature ABI.

As for improvements over time, the findings are less clear. The dominant pattern is that all groups are quite stable. For the Netherlands, the ranking remains stable, but that is to some extent to be expected, since it started already among the highest ranks. Only in terms of spam, did we witness a significant improvement over the period when Dutch ISPs increased their efforts and launched AbuseHUB. This can be interpreted as an indicator of the impact of AbuseHUB. 

\begin{figure}[!h]
        \centering
    \includegraphics[width=1.0\textwidth]{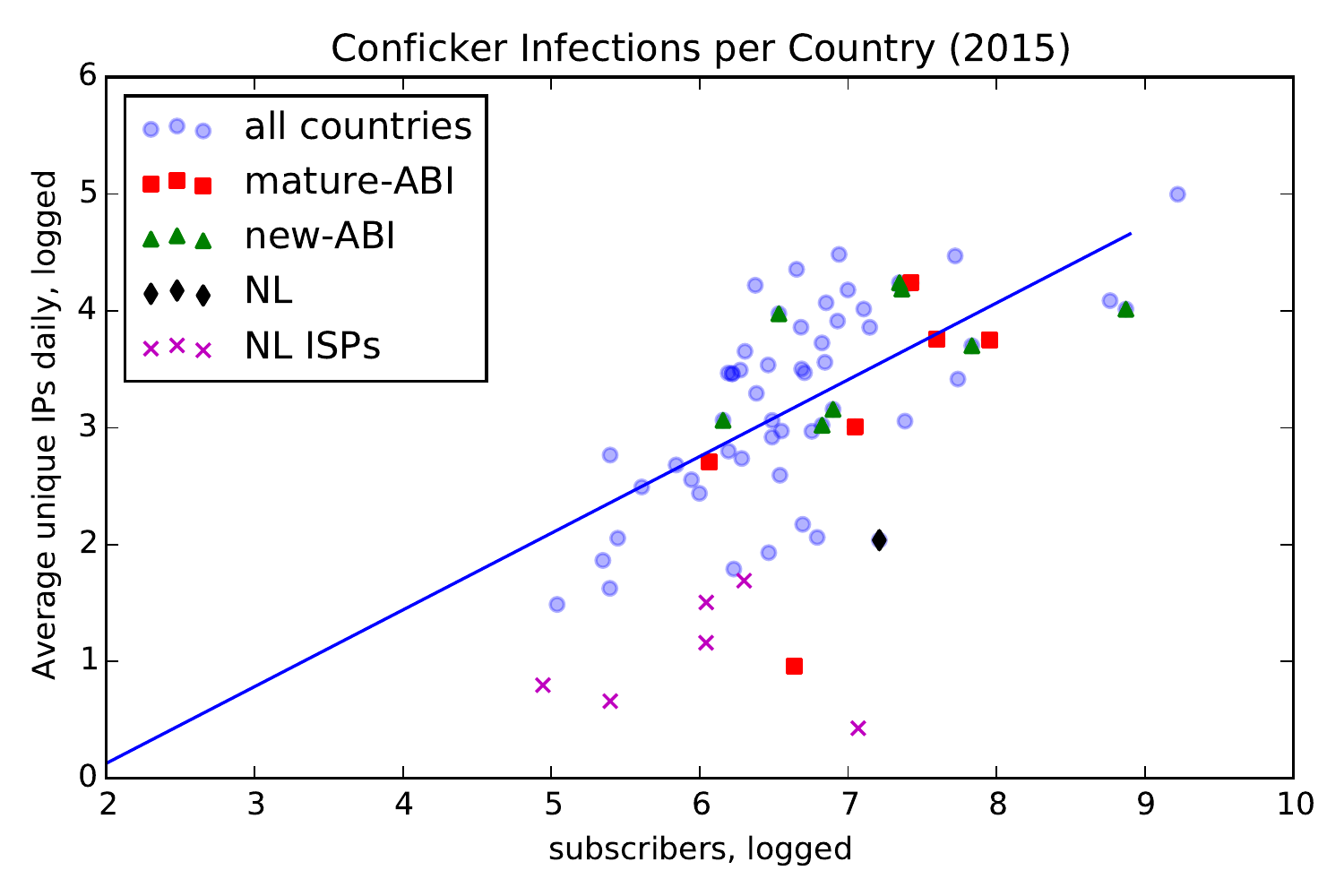}
    \caption{Conficker Countries-ABIs scatter plot }
    \label{cfk-country-isp}
    
     \end{figure}

\begin{figure}
        \centering
    \includegraphics[width=1.0\textwidth]{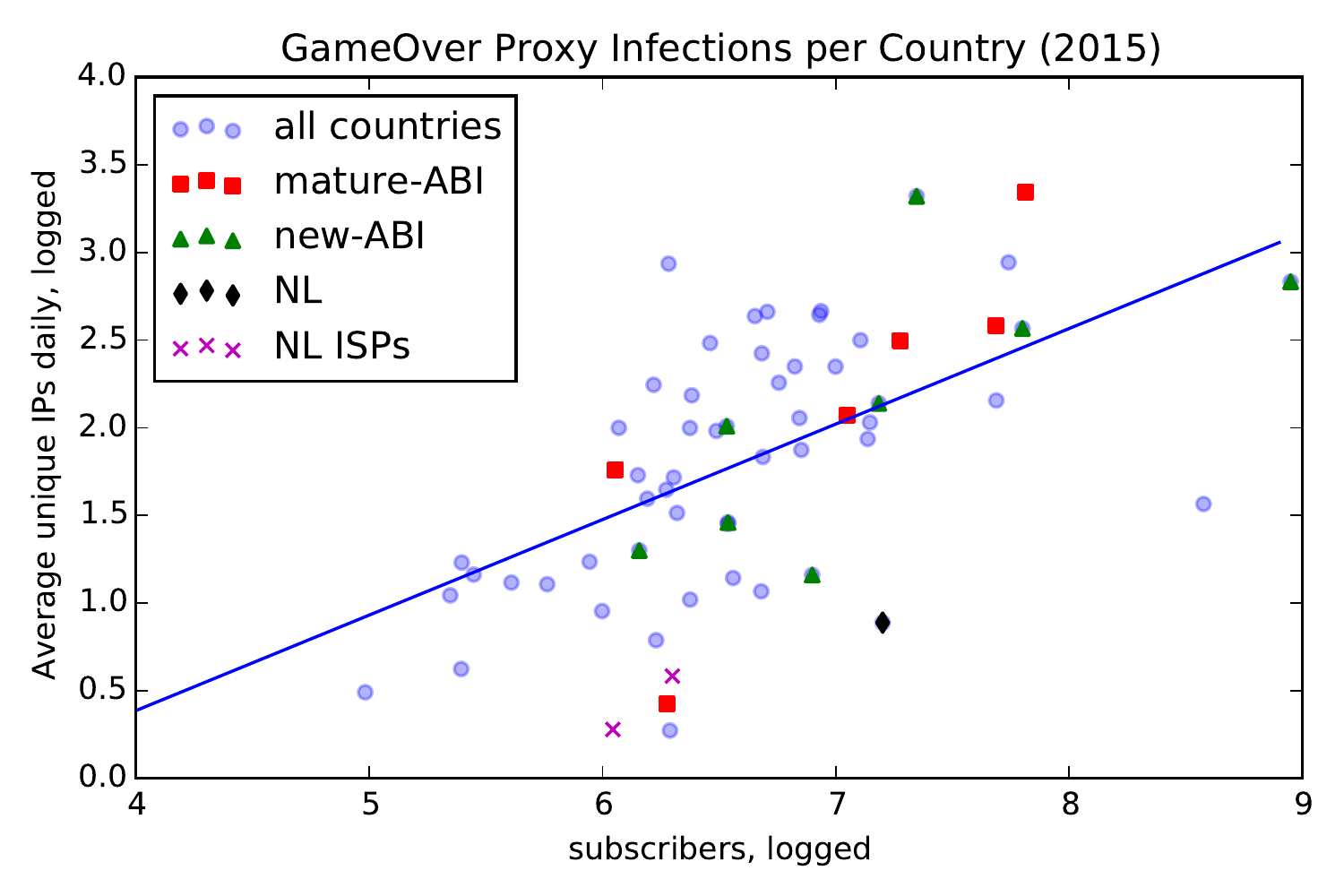}
     \caption{Game Over Proxy Countries-ABIs scatter plot }
    \label{goz_proxy-country-isp}
    
     \end{figure}

\begin{figure}
        \centering
    \includegraphics[width=1.0\textwidth]{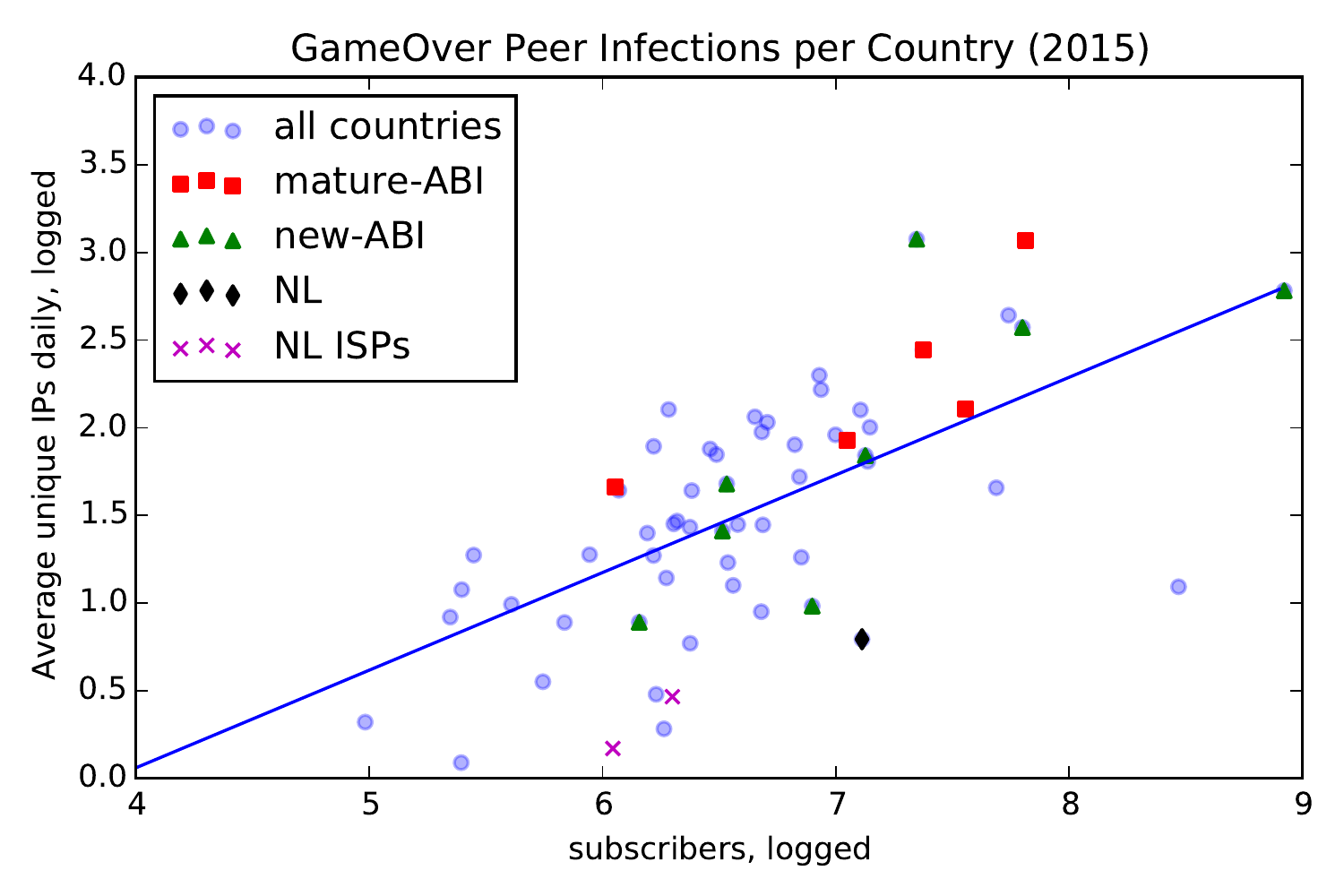}
    \caption{GameOver Peer Countries-ABIs scatter plot }
    \label{goz_peer-country-isp}
    
     \end{figure}

\begin{figure}
        \centering
    \includegraphics[width=1.0\textwidth]{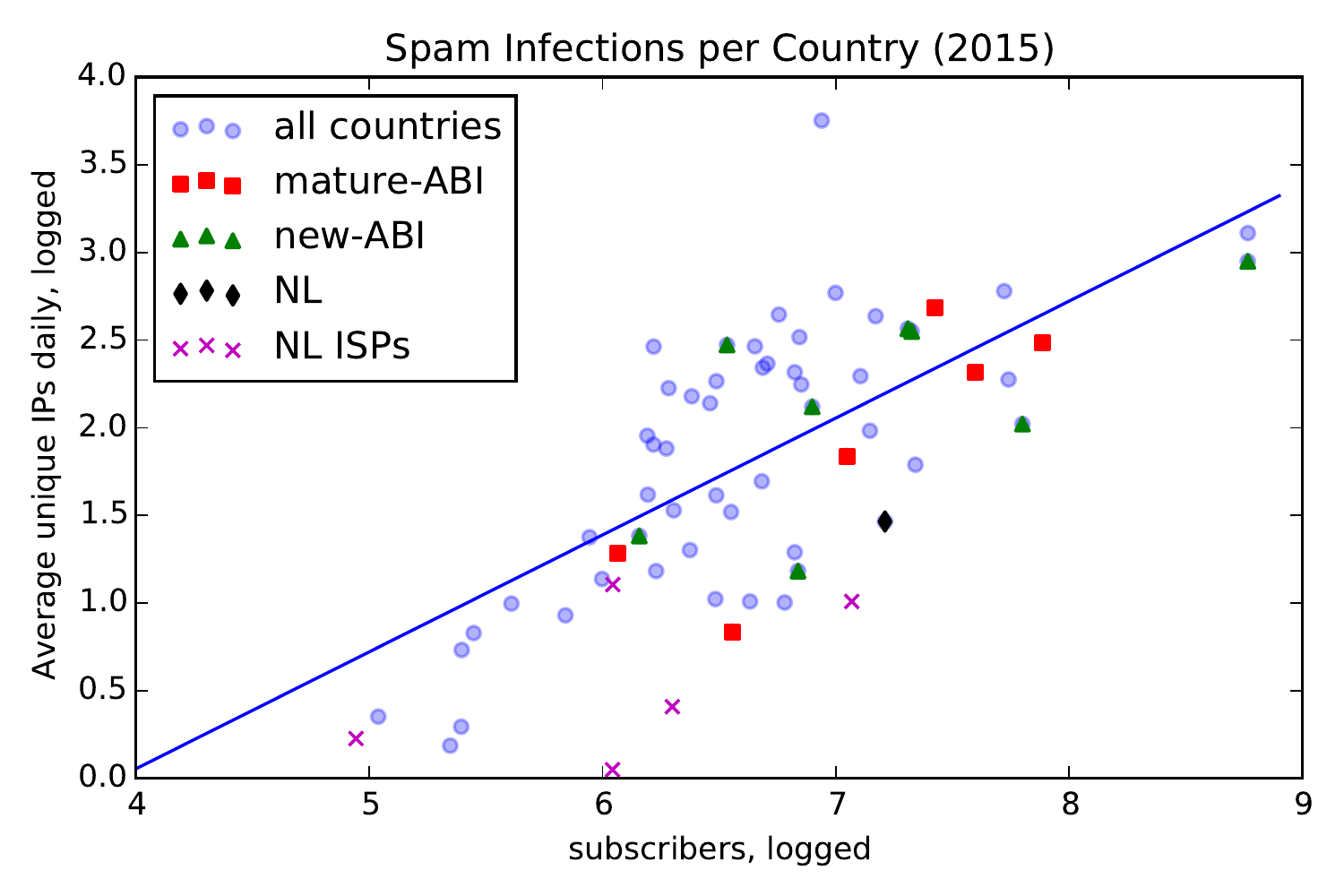}
  \caption{Spam Countries-ABIs scatter plot }
    \label{spam-country-isp}
    
     \end{figure}

\begin{figure}
       \centering
    \includegraphics[width=1.0\textwidth]{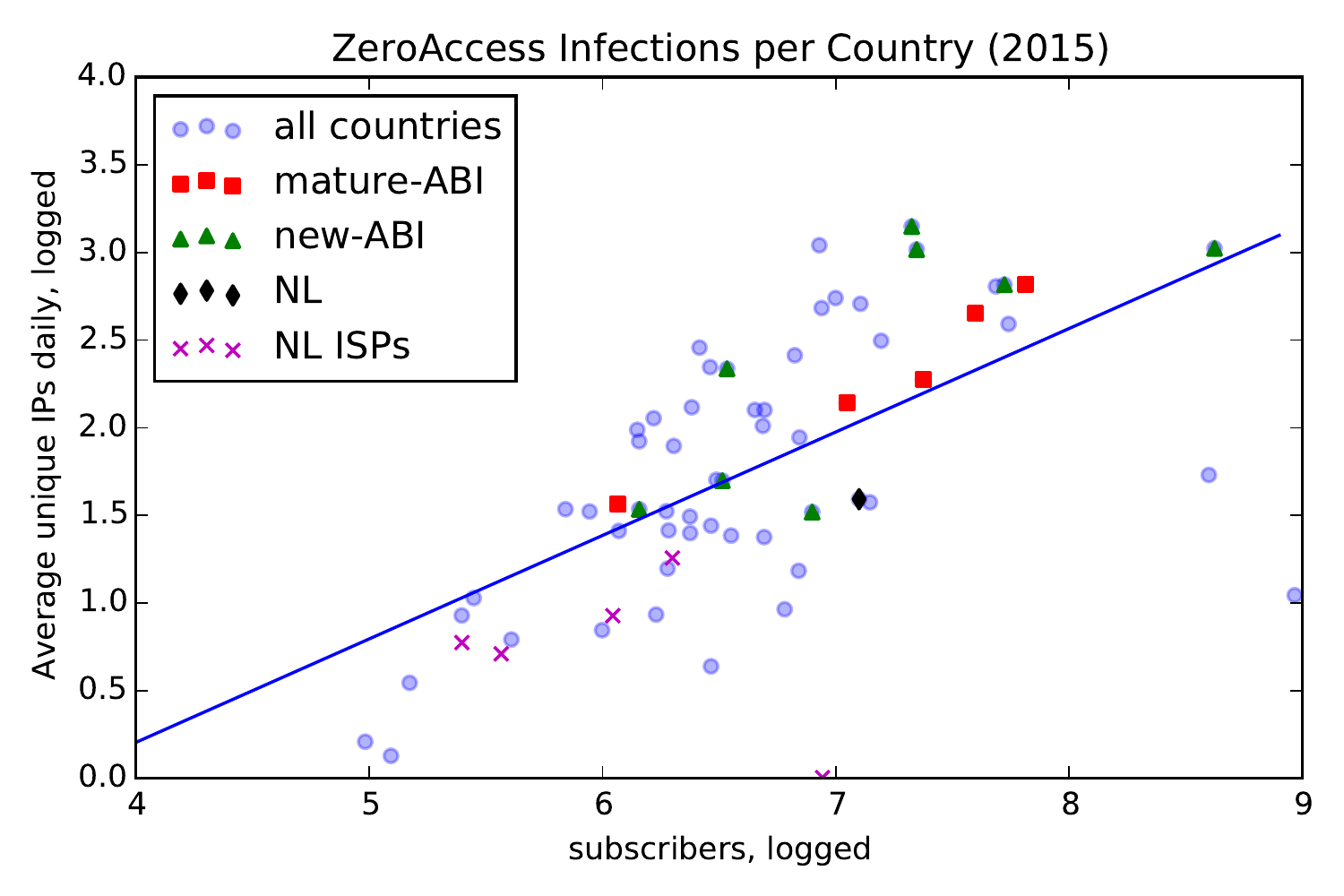}
     \caption{ZeroAccess Countries-ABIs scatter plot }
    \label{zeroaccess-country-isp}
    
     \end{figure}

In Figures~\ref{cfk-country-isp}--\ref{zeroaccess-country-isp}, we present scatter plots of the ISPs in these three groups of countries. Each dot represents all ISPs in a certain country. On the x-axis we plot the number of subscribers (summed for all mapped ISPs) and on the y-axis we plot the daily number of unique IP addresses seen in the data (summed for all ISPs). Both axes use a logarithmic scale (so each unit represents a factor of 10 difference). 

For the Netherlands, we not only include the country total, which aggregates the Dutch ISPs, but also the six individual ISPs that are included in the aggregate. The latter gives us some sense of the variance in mitigation performance among ISPs. Note that in some plots, not all six Dutch ISPs are visible. When an ISP has less than one infection per day in that data source, the metric drops below zero on the logarithmic scale, and thus outside the plotted area.

Analyzing these figures, we can observe that Dutch ISPs (marked with 'x') perform  well in comparison to other ISPs. There is variance among them, so some do better than others, but this variance seems limited. More importantly: they all perform well. Aggregate metrics can sometimes hide the poor performance of one or a few ISPs, when the rest of the group is performing well. This is not the case here. 

Also, it is clear that countries with mature ABIs, such as the Netherlands, tend to have lower infection rates per subscriber than members of the other groups, for most of the sources. That being said, these figures also clearly show that there is a lot of variance in each group. Some countries without an ABI have lower infection rates than some ISPs in countries with a mature ABI, for example. In other words, the presence of an ABI does not dictate ISP performance in botnet mitigation. Some ISPs will still perform poorly, while other ISPs do well in the absence of a national initiative.

\section{Main Findings}
\label{sec:isps-only-findings}

The main finding presented in this chapter is that Dutch ISPs perform excellently compared to ISPs from 60 other countries. Moreover, we can also see clearly in our results that the presence of mature anti-botnet initiatives correlate with lower infection rates per subscribers. However, the lack of an ABI or having a relatively recent ABI does not imply necessarily better of worse performance at the level of the individual ISP. There is a lot of variance at the level of ISPs that is sometimes larger than the variance among countries. With or without a mature ABI, some ISPs do well, others do worse. In the end, ABIs seem to nudge company policies in the right direction, but they do not dictate it. In the Dutch case, though, the six included ISPs are performing exceptionally well. Together they can rightly claim to be ''best in class''.

\begin{figure}
        \centering
    \includegraphics[width=1.0\textwidth]{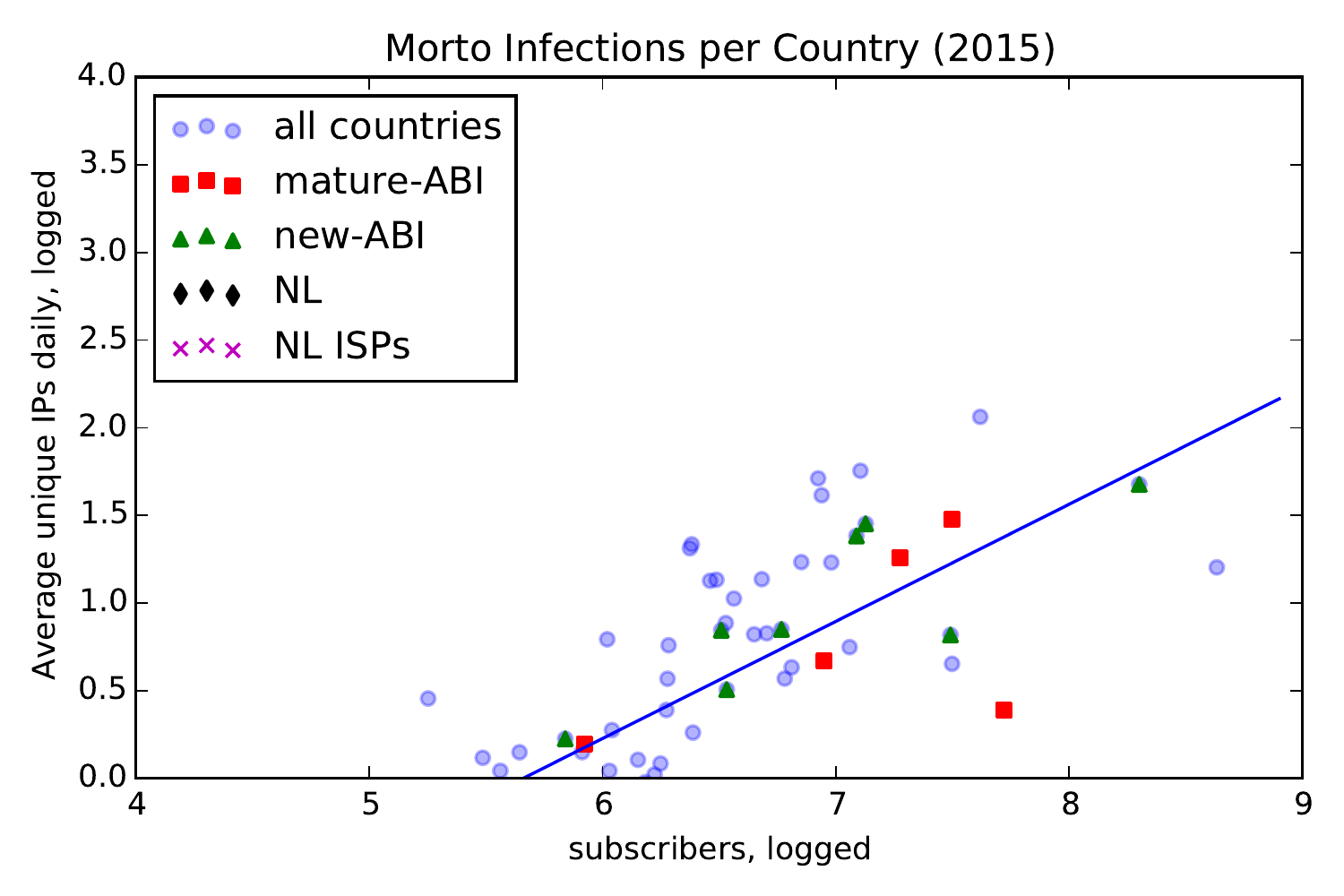}
  \caption{Morto Countries-ABIs scatter plot }
    \label{morto-country-isp}
    
     \end{figure}

 \begin{table}
  \begin{center}
   \begin{tabular}{|l|c|c|c|c|c|} \hline

Group&2011&2012&2013&2014&2015 \\ \hline
No ABI&47&46&47&47&46\\ \hline
New ABI&35&32&32&32&32\\ \hline
Mature ABI&45&44&42&42&43\\ \hline
NL&58&58&59&59&59\\ \hline

    \end{tabular}
    \end{center}
    \caption{Comparison rankings for Conficker}
  \label{tab:abi-grp-cfk}
  \end{table}

  \begin{table}
  \begin{center}
   \begin{tabular}{|l|c|c|c|c|c|} \hline

Group		&2011&	2012&	2013&	2014&	2015 \\ \hline
No ABI		&42&	45&		47&		41&		41\\ \hline
New ABI		&34&	33&		30&		28&		31\\ \hline
Mature ABI	&36&	36&		44&		42&		42\\ \hline
NL			&56&	57&		58&		56&		52\\ \hline

    \end{tabular}
    \end{center}
    \caption{Comparison rankings for Morto}
  \label{tab:abi-grp-morto}
  \end{table}

   \begin{table}
  \begin{center}
   \begin{tabular}{|c|c|c|} \hline

Group&2014 &2015\\ \hline
No ABI&39&40\\ \hline
New ABI&38&37\\ \hline
Mature ABI& 33& 32\\ \hline
NL&57&57\\ \hline
  
    \end{tabular}
    \end{center}
    \caption{Comparison rankings for GameOverPeer}
  \label{tab:abi-grp-gameoverpeer}
  \end{table}

   \begin{table}
  \begin{center}
   \begin{tabular}{|c|c|c|} \hline

Group&2014 &2015\\ \hline
No ABI& 42&41\\ \hline
New ABI&37&39\\ \hline
Mature ABI& 36&34 \\ \hline
NL& 58& 58\\ \hline

    \end{tabular}
    \end{center}
    \caption{Comparison rankings for GameOver Proxy}
  \label{tab:abi-grp-gameoverproxy}
  \end{table}

%
%

   \begin{table}
  \begin{center}
   \begin{tabular}{|l|c|c|c|c|c|} \hline

Group&2011&2012&2013&2014&2015 \\ \hline
No ABI&42&42&40&41&43   \\ \hline
New ABI& 39& 36&34&37&37  \\ \hline
Mature ABI&43&43&43&40&42    \\ \hline
NL& 45& 53& 54&57&56\\ \hline
  
    \end{tabular}
    \end{center}
    \caption{Comparison rankings for Spam}
  \label{tab:abi-grp-spam}
  \end{table}

   \begin{table}
  \begin{center}
   \begin{tabular}{|c|c|c|} \hline

Group&2014 &2015\\ \hline
No ABI&  45&  44\\ \hline
New ABI&  28& 28\\ \hline
Mature ABI& 39& 40\\ \hline
NL& 54& 52\\ \hline

    \end{tabular}
    \end{center}
    \caption{Comparison rankings for ZeroAccess}
  \label{tab:abi-grp-zero}
  \end{table}

\chapter{AbuseHUB members compared against non-members}
\label{sec:q1}

In this chapter, we take a deeper look at the infection levels within the Netherlands. In our 2011 study, we found that the bulk of all infections were located in the networks of the main broadband providers -- around 80 percent, to be precise. In the baseline report of December 2013, we found that this distribution had changed. It was unclear how these finding should be interpreted, since it was based on fewer data sources. 

The broader set of data sources we now have available allows us to revisit this issue. In addition to the global sources we have analyzed in the previous chapter, we also include two NL-only data feeds we have obtained, as described in Section~\ref{sec:datasets}.

We explore the distribution of infections in the Netherlands across AbuseHUB members versus non-members. The list of ASes belonging to the first group is provided in Table~\ref{tab:isps-sig}. The non-members comprise all other ASes that geo-locate to the Netherlands, which are too many to list here in full.

It is important to emphasize that we expect the majority of infections to be located in the networks of AbuseHUB members, for two reasons. First, AbuseHUB members cover 63.9\% of the Dutch IP address space. Second, the majority of infected computers are typically home users' machines, which are concentrated in the networks of ISPs and thus in AbuseHUB networks, since its members cover the vast majority of the Dutch retail market. In other words, the fact that most of the infections are indeed located in the networks of AbuseHUB members does not imply that the members are performing worse than non-members: it simply means that they cover the bulk of the vulnerable population.


We will look at the most relevant ASes per source. The idea is to determine:
    \begin{itemize}
    	\item How infections are distributed over AbuseHUB members versus non-members;
        \item How this distribution changes over time.
    \end{itemize}
    
The rest of this chapter is organized as follows:  Section~\ref{sec:chap3:avg} discusses the daily averages of infection for members and non-members. In Section~\ref{sec:chap3-timeseries}, we present an analysis on the distribution across of members and non-members over time. Then, in Section~\ref{sec:chap3:nonmemberAS}, we explore which non-members ASes contribute the most infections in each data source. Finally, Section~\ref{sec:chap3:summary} presents a summary of the main findings.

\section{Infections in member versus non-member networks}
\label{sec:chap3:avg}

 Table~\ref{tab:mem-non-mem-avg} shows the average daily number of unique IP addresses per group, covering the whole period for which each data source is available. As expected, AbuseHUB members are still responsible for the most of infected IP addresses -- with the exception of Morto. We analyze in more detail in which member networks these bots are located in Chapter~\ref{sec:q2}.
 
The difference among members and non-members are not negligible. In average, across all sources, we observed an average of 6,456 daily IP addresses: members are responsible for 61\% of those, while 39\% is in non-members ASes. In light of the fact that AbuseHUB members are responsible for 63,9\% of the pool of addresses, this distribution would seem expected. We know, however, that bots are not uniformly distributed over the address space. They are mostly concentrated in consumer broadband networks. In earlier research we found around 80\% of all bots in those networks. Seen in this light, the proportion of bots in AbuseHUB member networks is lower than one might expect.

In fact, the proportion of bots in member networks has actually gone down compared to non-members. The pattern confirms the shift we observed in the baseline report. In 2010, we found around 80\% of all bots resided in ISP networks. This has diminished to around 60\%. A growing proportion of the infections reside in the networks of non-members. This means that AbuseHUB members are improving faster than non-members.

      \begin{table}[t]
  \begin{center}
   \begin{small}

   \begin{tabular}{|l|c|c|c|c|} \hline
 \textbf{}&  \textbf{GameOver Peer}& \textbf{GameOver Proxy}& \textbf{Conficker}& \textbf{Morto}\\ \hline 
 \textbf{Member}  & 10.45  & 10.61 & 181.74 &  1.90  \\\hline
 \textbf{Non-member}& 7.91 &9.41 &217.79 & 2.33 \\ \hline
 
  \textbf{}& \textbf{ZeroAccess}& \textbf{Shadowserver Bot} & \textbf{Shadowserver MS} & \textbf{Spam} \\ \hline
 \textbf{Member}  &43.55         & 373.10 & 183.33 &42.04  \\\hline
 \textbf{Non-member}& 15.43 & 322.98 & 128.13 & 151.39 \\ \hline

    \end{tabular}
       \end{small}

    \end{center}
    \caption{AbuseHUB members $\times$ non-members -- average of daily IP addresses (Year: 2015)}
  \label{tab:mem-non-mem-avg}
  \end{table} 
 
 \section{Distribution over time}
\label{sec:chap3-timeseries}

\begin{figure}[!htbp]
        \centering
     \includegraphics[width=1\textwidth]{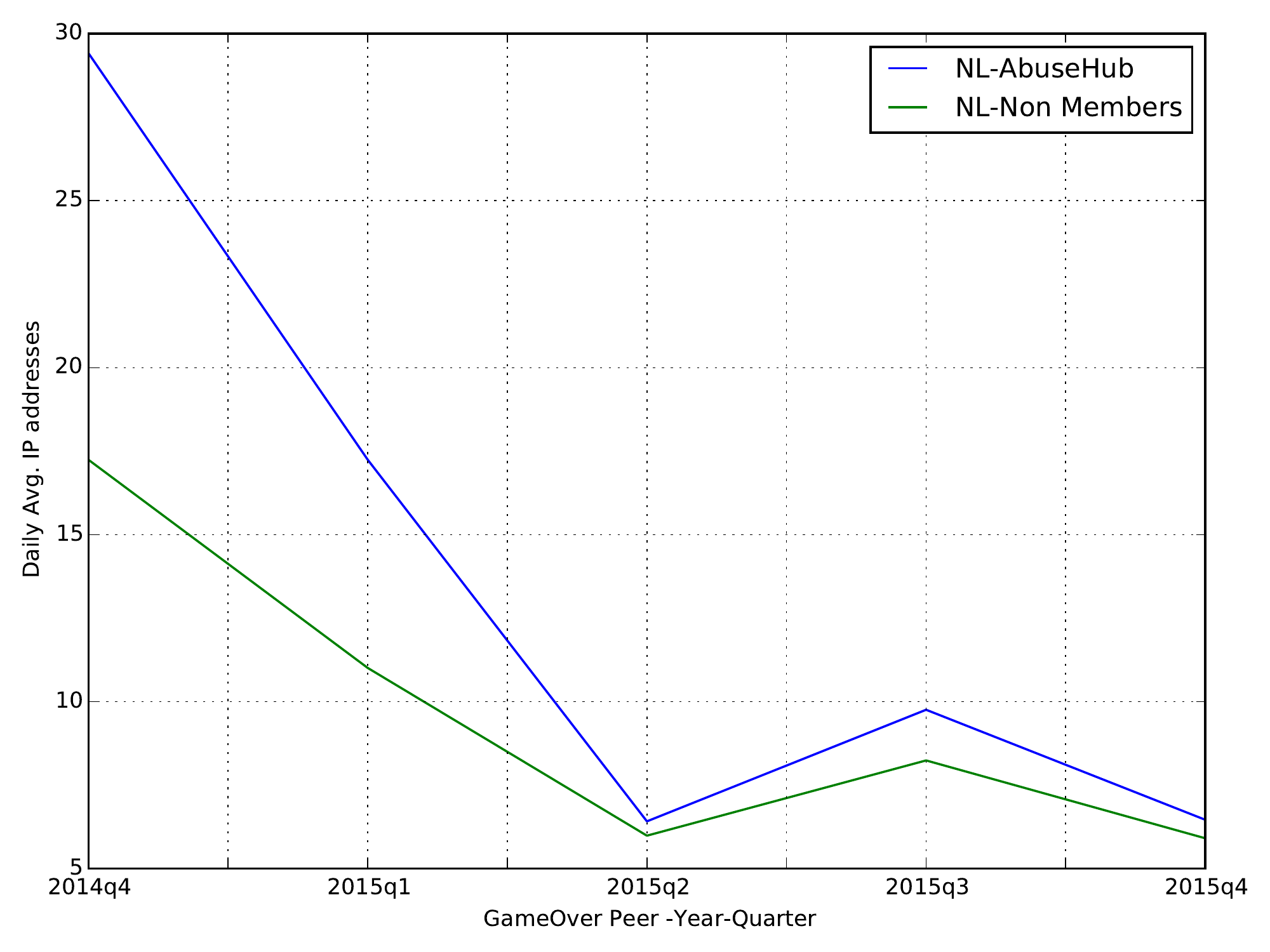}
    \caption{Zeus Peer Countries}
    \label{zeus-peer-q1}
       \end{figure}

\begin{figure}[!htbp]
        \centering
         \includegraphics[width=1\textwidth]{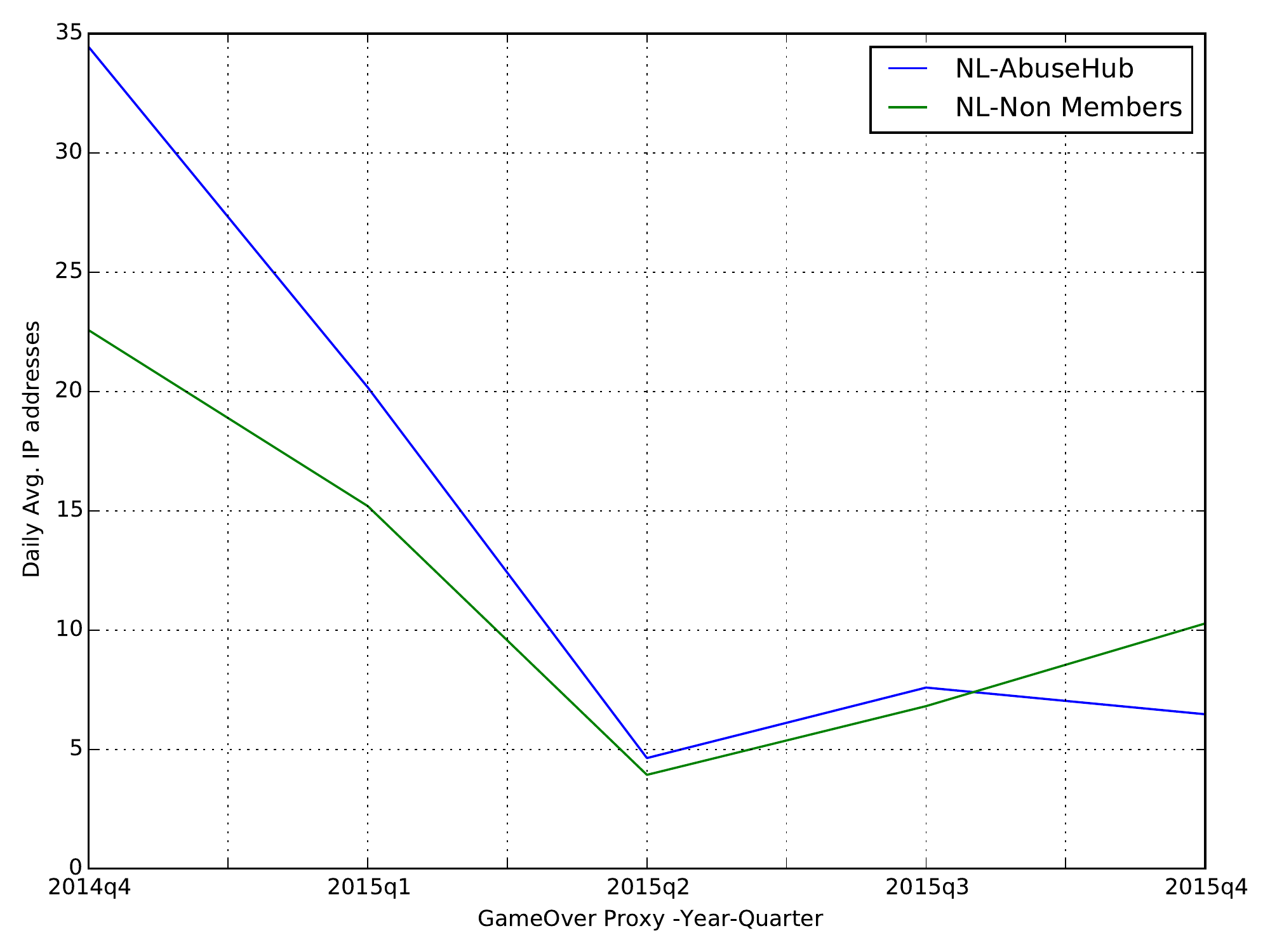}
    \caption{Zeus Proxy Countries}
    \label{zeus-proxy-q1}
      \end{figure}

\begin{figure}[!htbp]
        \centering
    \includegraphics[width=1\textwidth]{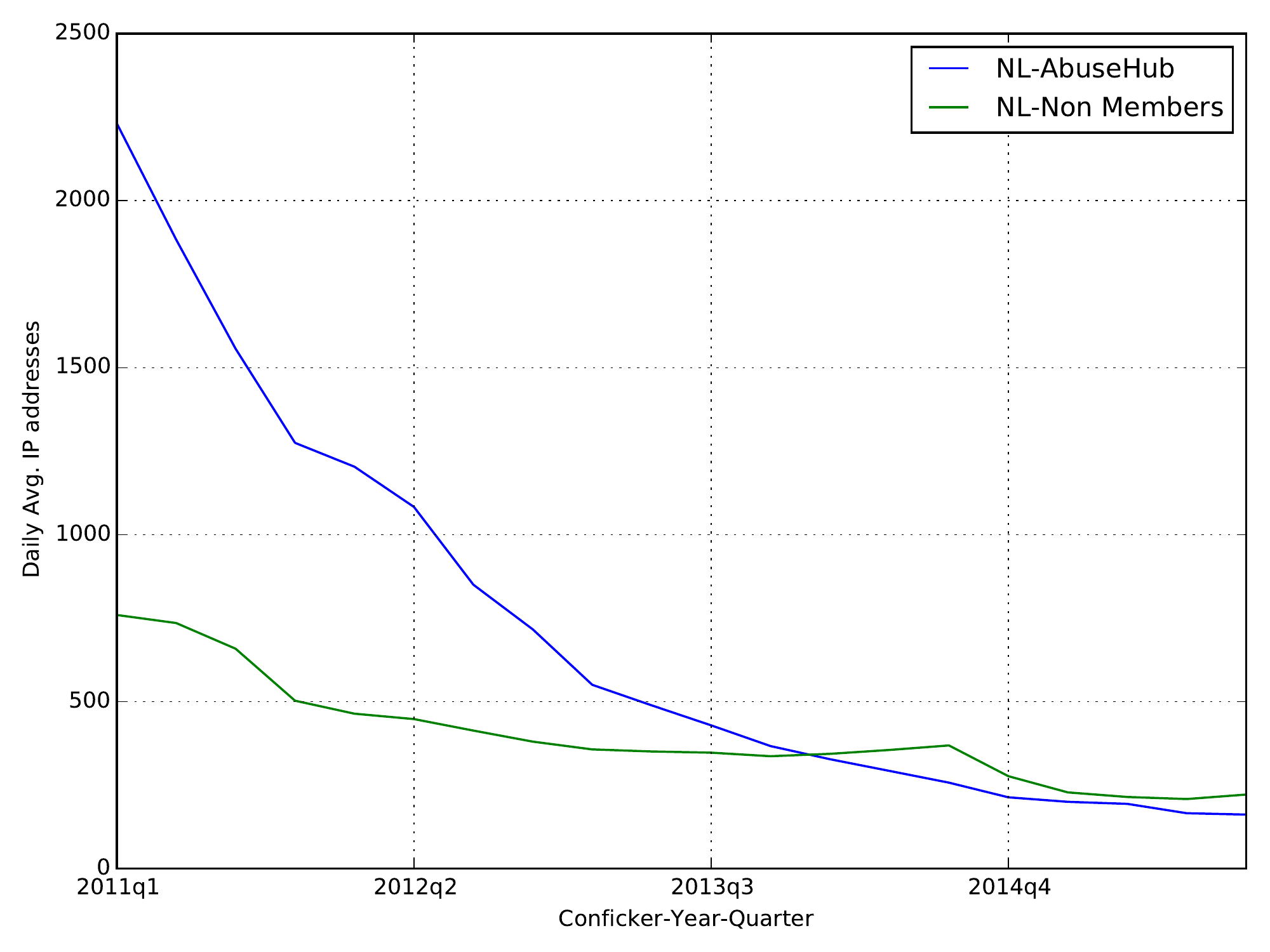}
    \caption{Conficker Countries}
    \label{cfk-q1}
      \end{figure}

\begin{figure}[!htbp]
\centering
            
         \includegraphics[width=1\textwidth]{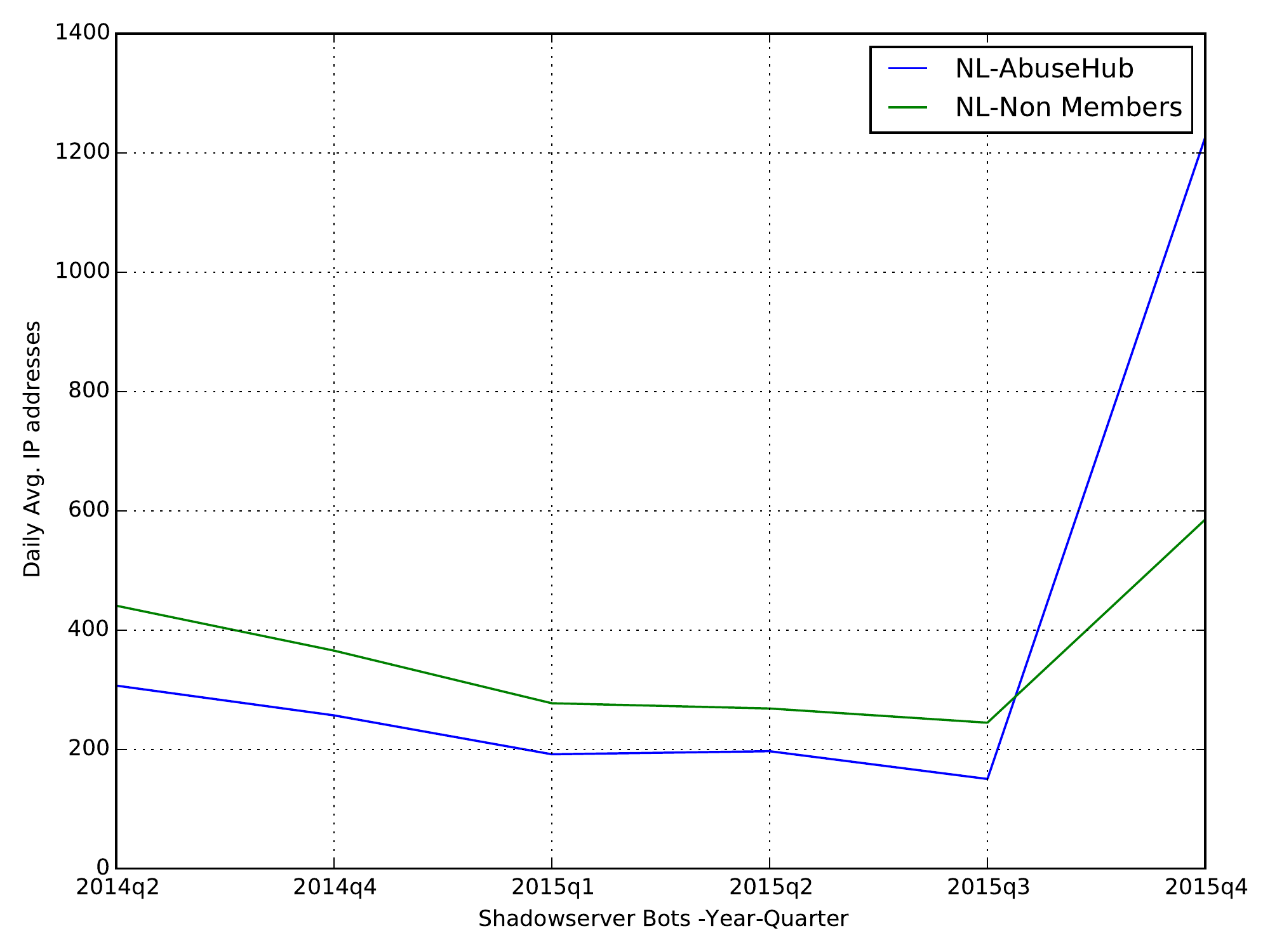}
    \caption{Shadowserver Bots }
    \label{shadowbots-q1}
       \end{figure}

\begin{figure}[!htbp]
        \centering
         
            \includegraphics[width=1\textwidth]{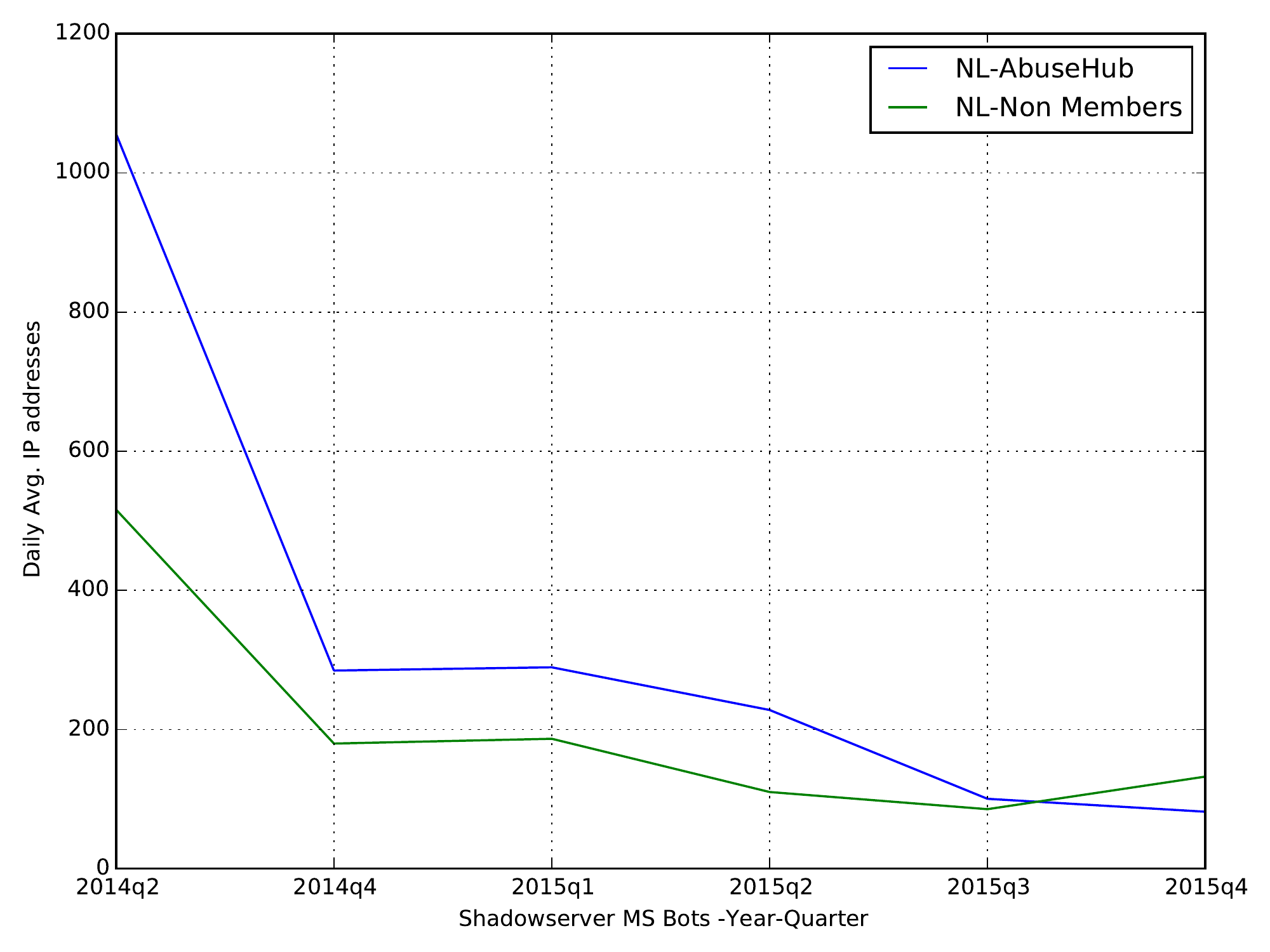}
    \caption{Shadowserver MS Bots }
    \label{microsoftbots-q1}
    
        \end{figure}

\begin{figure}[t]
        \centering
    
                \includegraphics[width=1\textwidth]{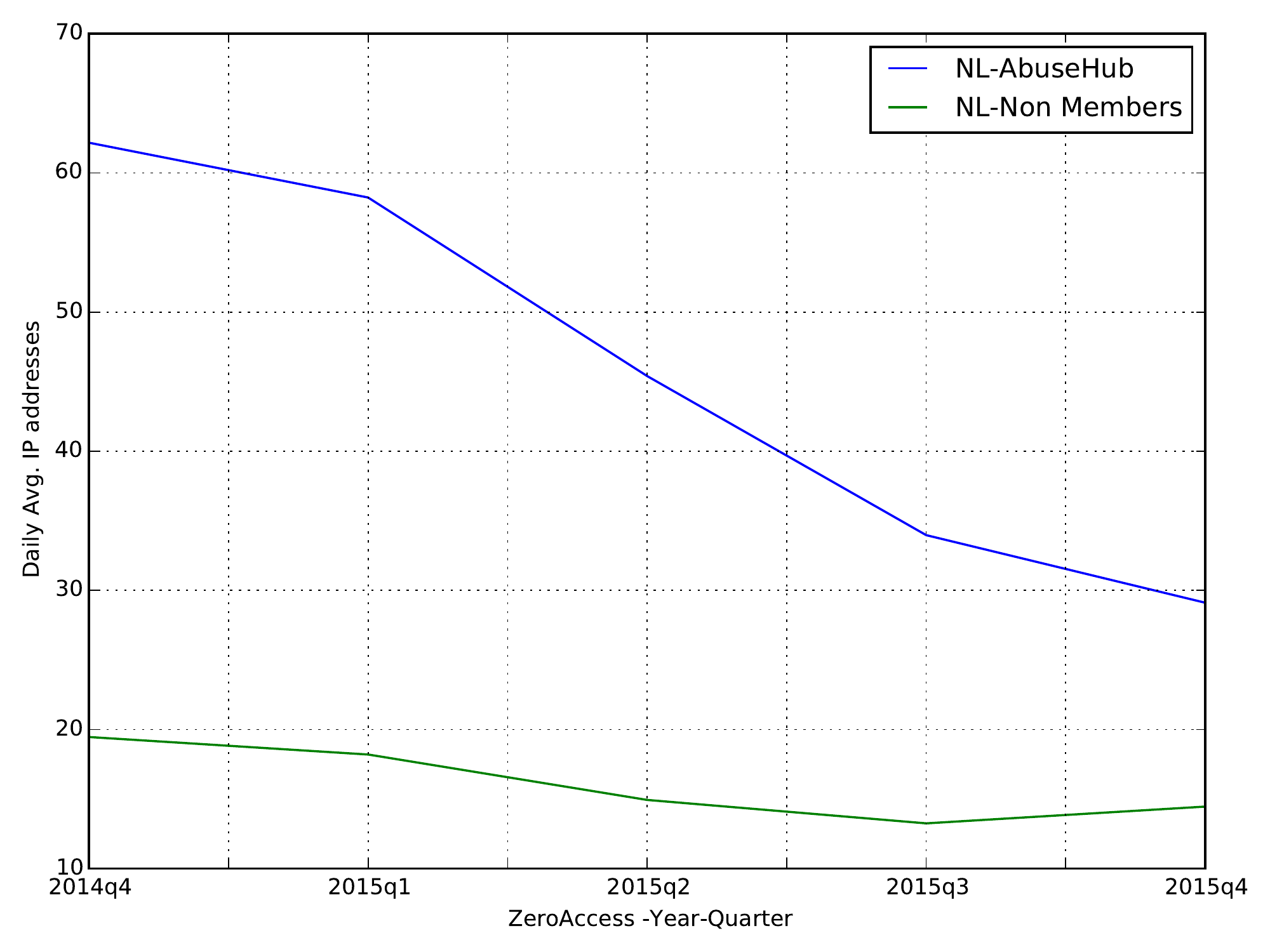}
    \caption{ZeroAccess}
    \label{zeroaccess-q1}
   \end{figure}

Figures~\ref{zeus-peer-q1} to ~\ref{zeroaccess-q1} show the temporal behavior of AbuseHUB Members versus non-members. Notice that the time period during which we track this distribution is different for each data source, depending on its availability.

The dominant pattern across the different sources is that AbuseHUB members improve over time, while non-members are relatively static or show a more modest reduction. This fits with the finding that non-members make up a larger portion of the problem than earlier studies observed. 

The shifting distribution implies that non-member ASes are becoming increasingly important in botnet mitigation in the Netherlands. To some extent, this finding is surprising. Since most infections occur in end user devices, we would expect that broadband providers would harbor most infections. AbuseHUB members cover over 90\% of the broadband market in the Netherlands, but its portion of the botnet problem is decreasing. This begs the question: what non-member networks contribute the most infections? We explore this question in the next section.
    
        \begin{figure}[t]
        \centering
      \includegraphics[width=1\textwidth]{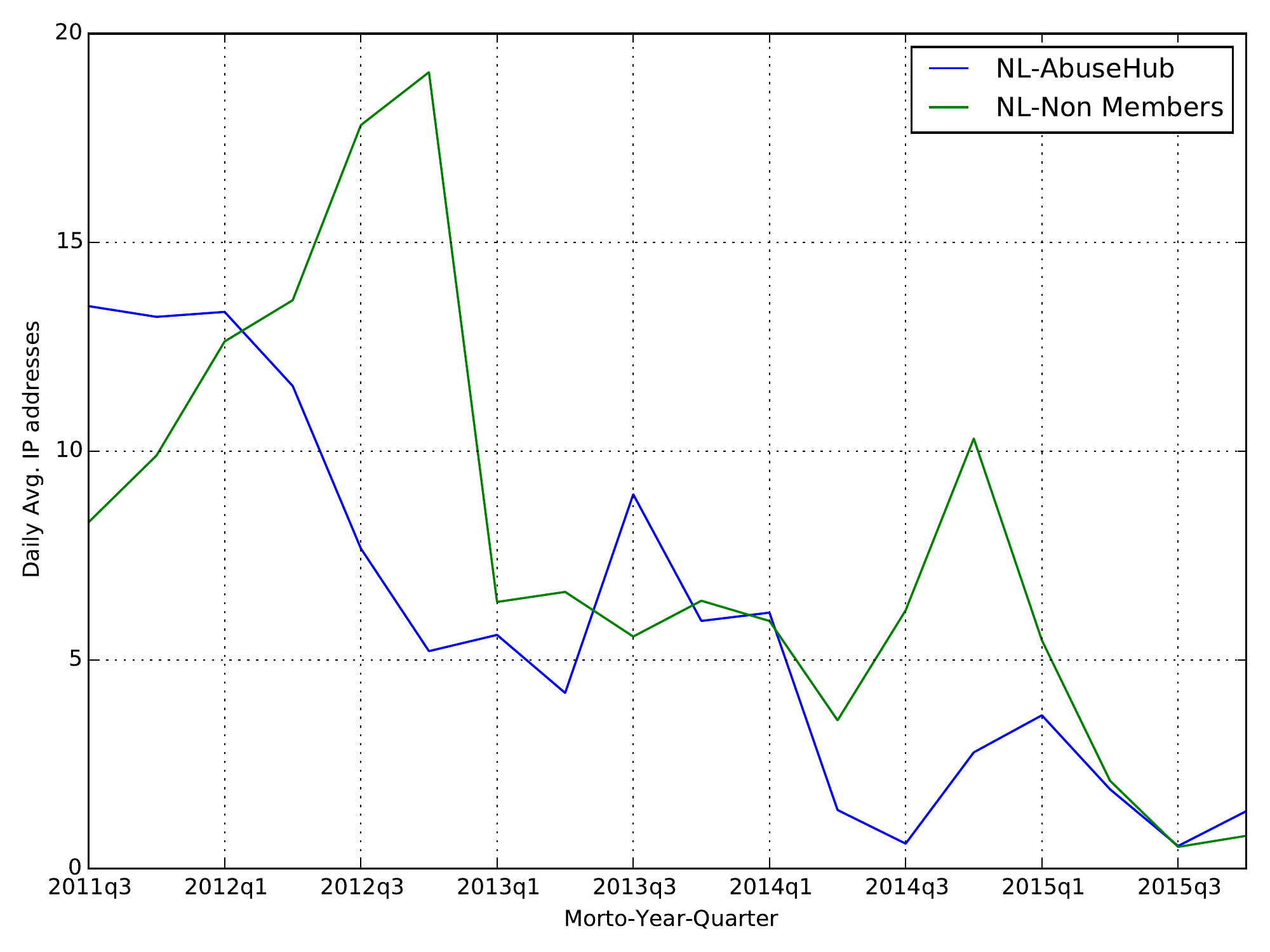}
    \caption{Morto}
    \label{morto-q1}
   \end{figure}

 \section{Most infected non-member networks}  
  \label{sec:chap3:nonmemberAS}

In the previous section we have shown that the number of infected IP addresses in AbuseHUB non-members ISPs corresponds to 39\% of the total observed in the Netherlands.  In this section, we present, for each data source, which ASes contribute the most to that infected population, based on the observed daily average number of infections.

Tables~\ref{tab:isps-gameover-peer} -- \ref{tab:isps-morto} show the results for each data source. Not all AS names may be familiar to the reader. Across all tables, four main groups of networks dominate: smaller consumer and business broadband providers that are not members of AbuseHUB (e.g., Online, CAIW, Vodafone, Xenosite, Unet),  hosting providers (e.g., NFOrce, Hostkey, Redhosting), and, to a lesser extent, mobile providers. 

The presence of hosting providers is a bit puzzling. For some sources, most notably spam, we know that hosting infrastructure is being used. However, we do know why hosting equipment would show up in botnet sinkholes, since the malware on which the botnet is based only infects Windows end user machines (home or business). We can speculatively suggest three explanations. First, hosting providers are not know as access providers, but they may nevertheless offer access services, bundled with their hosting services. Second, some of these infections may reside in hosted Virtual Machines running Windows. Third, some of the hosting servers are used as VPN nodes, which means that users connect their PC to the Internet via a VPN connection and thus an infected machine would contact the sinkhole with the IP address of the VPN exit node. Further analysis is needed to fully  clarify why we see these numbers of bots in those networks.

  \section{Main findings}
  \label{sec:chap3:summary}

Across the different data sources, we can observe that the majority of the infected machines reside in AbuseHUB member networks. This was expected, as the members cover the bulk of all the Dutch IP addresses and over 90\% of the broadband market, where bots tend to be concentrated. When compared over time, however, we see that the proportion of the infected population in AbuseHUB members is going down: from around 80\% in 2010 to around 60\% in 2015. Infections in member networks are diminishing faster than in non-members.

A substantial portion of the botnet problem (on average, 39\%) resides in non-member networks, most notably smaller broadband providers and hosting providers. Most data sources suggest this portion is growing. All of this indicates that the impact of AbuseHUB can increase by recruiting members among the remaining consumer and business broadband providers and among hosting providers -- or, to put it differently, that these parties can benefit from joining AbuseHUB in dealing with their botnet problem. 

   \begin{table}[!ht]
  \begin{center}
  \begin{small}

   \begin{tabular}{|l|c|c|} \hline
   \textbf{AS}&\textbf{Daily IPs}& \textbf{AS Name}\\ \hline
29396 & 1.2444 & Unet Network                     \\ \hline
39309 & 0.8008 & Edutel BV                        \\ \hline
15426 & 0.6466 & XenoSite B.V.                    \\ \hline
5390  & 0.6165 & Online Breedband B.V. Global AS  \\  \hline
6905  & 0.5977 & AT\&T Webhosting Amsterdam IDC   \\  \hline
25222 & 0.4962 & XANTIC B.V.                      \\ \hline
57795 & 0.4436 & Next Gen Networks B.V.           \\ \hline
8075  & 0.4173 & Microsoft Corporation            \\ \hline
8928  & 0.3496 & Interoute Communications Limited \\ \hline
51964 & 0.297  & Equant Inc.  \\\hline                   
\end{tabular}
  \end{small}
   \end{center}
    \caption{GameOver Peer -- Top 10 Most Infected Non-members (Year: 2015)}
  \label{tab:isps-gameover-peer}
  \end{table}

      \begin{table}[!ht]
  \begin{center}
  \begin{small}
   \begin{tabular}{|l|c|c|} \hline
   \textbf{AS}&\textbf{Daily IPs}& \textbf{AS Name}\\ \hline
43350 & 1.8453 & NFOrce Entertainment BV          \\ \hline
5390  & 0.7472 & Online Breedband B.V. Global AS  \\ \hline
39309 & 0.7434 & Edutel BV                        \\ \hline
15426 & 0.6302 & XenoSite B.V.                    \\ \hline
6905  & 0.6113 & AT\&T Webhosting Amsterdam IDC   \\  \hline
25222 & 0.6038 & XANTIC B.V.                      \\ \hline
29396 & 0.5736 & Unet Network                     \\ \hline
57795 & 0.5283 & Next Gen Networks B.V.           \\ \hline
8426  & 0.3472 & ClaraNET LTD                     \\ \hline
8928  & 0.3283 & Interoute Communications Limited\\\hline

\end{tabular}
\end{small}
\end{center}
    \caption{GameOver Proxy -- Top 10 Most Infected Non-members (Year: 2015)}
  \label{tab:isps-gameover-proxyx}

  \end{table}

      \begin{table}[!ht]
  \begin{center}
  \begin{small}
   \begin{tabular}{|l|c|c|} \hline
   \textbf{AS}&\textbf{Daily IPs}& \textbf{AS Name}\\ \hline
15480 & 32.7212 & Vodafone NL Autonomous System   \\ \hline
31615 & 26.1667 & T-mobile Netherlands bv.        \\ \hline
29073 & 13.2364 & Ecatel                          \\ \hline
5390  & 12.8879 & Online Breedband B.V. Global AS \\ \hline
43350 & 10.6606 & NFOrce Entertainment BV         \\ \hline
33915 & 9.7879  & Vodafone Libertel B.V.          \\ \hline
15435 & 8.6939  & CAIW Diensten B.V.              \\ \hline
29396 & 7.7545  & Unet Network                    \\ \hline
20847 & 7.6848  & Previder B.V.                   \\ \hline
39647 & 6.6576  & Redhosting B.V.                 \\ \hline

\end{tabular}
\end{small}
   \end{center}
    \caption{ShadowServer Botnet Top 10 Most Infected Non-members (Year: 2015)}
  \label{tab:isps-gameover-proxy}
  \end{table}

      \begin{table}[!ht]
  \begin{center}
  \begin{small}
   \begin{tabular}{|l|c|c|} \hline
   \textbf{AS}&\textbf{Daily IPs}& \textbf{AS Name}\\ \hline
43350 & 42.6933 & NFOrce Entertainment BV         \\ \hline
5390  & 20.9294 & Online Breedband B.V. Global AS \\ \hline
15435 & 6.7301  & CAIW Diensten B.V.              \\ \hline
12871 & 4.4172  & Concepts ICT Autonomous System  \\ \hline
57043 & 4.2638  & HOSTKEY B.V.                    \\ \hline
39309 & 3.3988  & Edutel BV                       \\ \hline
29396 & 3.0399  & Unet Network                    \\ \hline
39647 & 1.9877  & Redhosting B.V.                 \\ \hline
33915 & 1.7577  & Vodafone Libertel B.V.          \\ \hline
35470 & 1.7086  & XL Internet Services BV         \\ \hline

\end{tabular}
\end{small}
   \end{center}
    \caption{ShadowServer Microsoft --- Top 10 Most Infected Non-members (Year: 2015)}
  \label{tab:isps-microsoft}
  \end{table}

        \begin{table}[!ht]
  \begin{center}
  \begin{small}
   \begin{tabular}{|l|c|c|} \hline
   \textbf{AS}&\textbf{Daily IPs}& \textbf{AS Name}\\ \hline
15480 & 26.7638 & Vodafone NL Autonomous System \\ \hline
33915 & 9.9693  & Vodafone Libertel B.V.        \\ \hline
29073 & 9.8067  & Ecatel                        \\ \hline
32475 & 8.6319  & SingleHop                     \\ \hline
20847 & 8.3313  & Previder B.V.                 \\ \hline
29396 & 8.0215  & Unet Network                  \\ \hline
20473 & 6.8374  & Choopa                        \\ \hline
31615 & 6.5521  & T-mobile Netherlands bv.      \\ \hline
39647 & 6.5     & Redhosting B.V.               \\ \hline
15435 & 6.2822  & CAIW Diensten B.V.            \\ \hline

\end{tabular}
   \end{small}
   \end{center}
   
    \caption{Conficker -- Top 10 Most Infected Non-members (Year: 2015)}
  \label{tab:isps-cfk}
  \end{table}

 \begin{table}[!ht]
  \begin{center}
  \begin{small}
   \begin{tabular}{|l|c|c|} \hline
   \textbf{AS}&\textbf{Daily IPs}& \textbf{AS Name}\\ \hline
15426  &	0.68   & XENOSITE XenoSite B.V. \\ \hline
60781  &	0.2708 & LEASEWEB-NL LeaseWeb Netherlands B.V. \\ \hline
29073  &	0.2554 & QUASINETWORKS Quasi Networks LTD \\ \hline
39647  &	0.2123 & REDHOSTING-AS Redhosting BV  \\ \hline
32475  &	0.1508 & SINGLEHOP-LLC - SingleHop, Inc \\ \hline
50522  &	0.1508 & POCOS POCOS B.V   \\ \hline
39309  &	0.0985 & EDUTEL-AS Edutel B.V \\ \hline
199264 &	0.0954 & ESTROWEB Estro Web Services Private Limited \\ \hline
50673  &	0.0954 & SERVERIUS-AS Serverius Holding B.V \\ \hline

\end{tabular}
   \end{small}
   \end{center}
   
    \caption{Morto -- Top 10 Most Infected Non-members (Year: 2015)}
  \label{tab:isps-morto}
  \end{table}

          \begin{table}[!ht]
  \begin{center}
  \begin{small}
   \begin{tabular}{|l|c|c|} \hline
   \textbf{AS}&\textbf{Daily IPs}& \textbf{AS Name}\\ \hline
50673 & 10.253 & Serverius Holding B.V.      \\ \hline
36351 & 8.4878 & SoftLayer Technologies Inc. \\ \hline
20857 & 8      & TransIP B.V.                \\ \hline
57043 & 7.6341 & HOSTKEY B.V.                \\ \hline
49544 & 4.1402 & i3d B.V.                    \\ \hline
29073 & 4.0549 & AS29073                     \\ \hline
46652 & 3.3841 & ServerStack                 \\ \hline
43350 & 2.5274 & NFOrce Entertainment BV     \\ \hline
8315  & 2.3476 & Amsio B.V.                  \\ \hline
35017 & 2.1341 & Swiftway Sp. z o.o.         \\ \hline
\end{tabular}
\end{small}
   \end{center}
    \caption{Spam --- Top 10 Most Infected Non-members (Year: 2015)}
  \label{tab:isps-spam-non}
  \end{table}

%

%
%
%
%
%
%
%
%
%
%
%
\chapter{How do member ISPs compare among themselves?}
\label{sec:q2}

\begin{figure}[!h]
        \centering
    \includegraphics[width=1\textwidth]{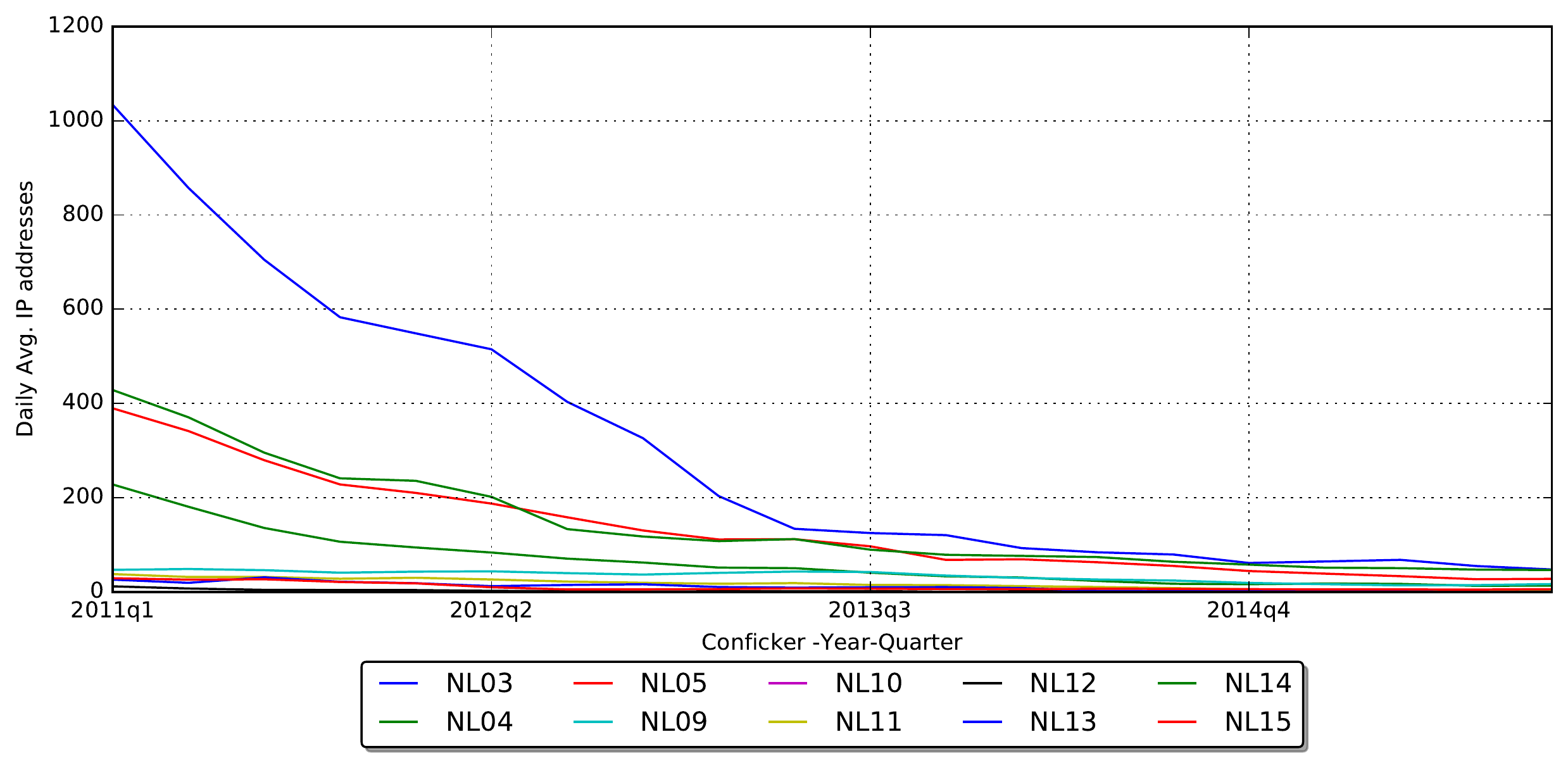}
    \caption{Conficker Members}
    \label{cfk-q2}
      \end{figure}

\begin{figure}[t]
        \includegraphics[width=1\textwidth]{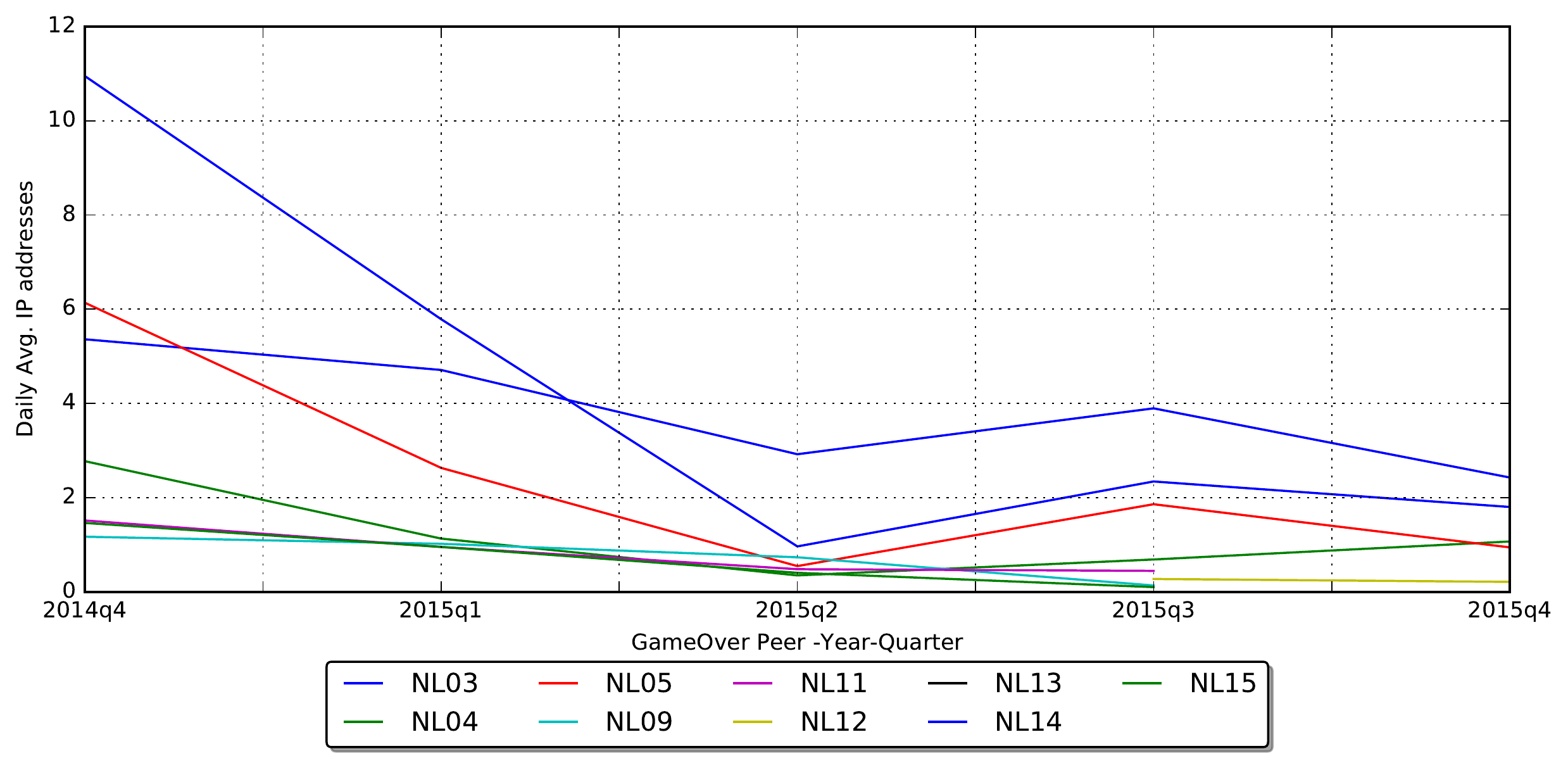}
    \caption{GameOver Peer Members}
    \label{peer-q2}
     \end{figure}

\begin{figure}[t]
        \includegraphics[width=1\textwidth]{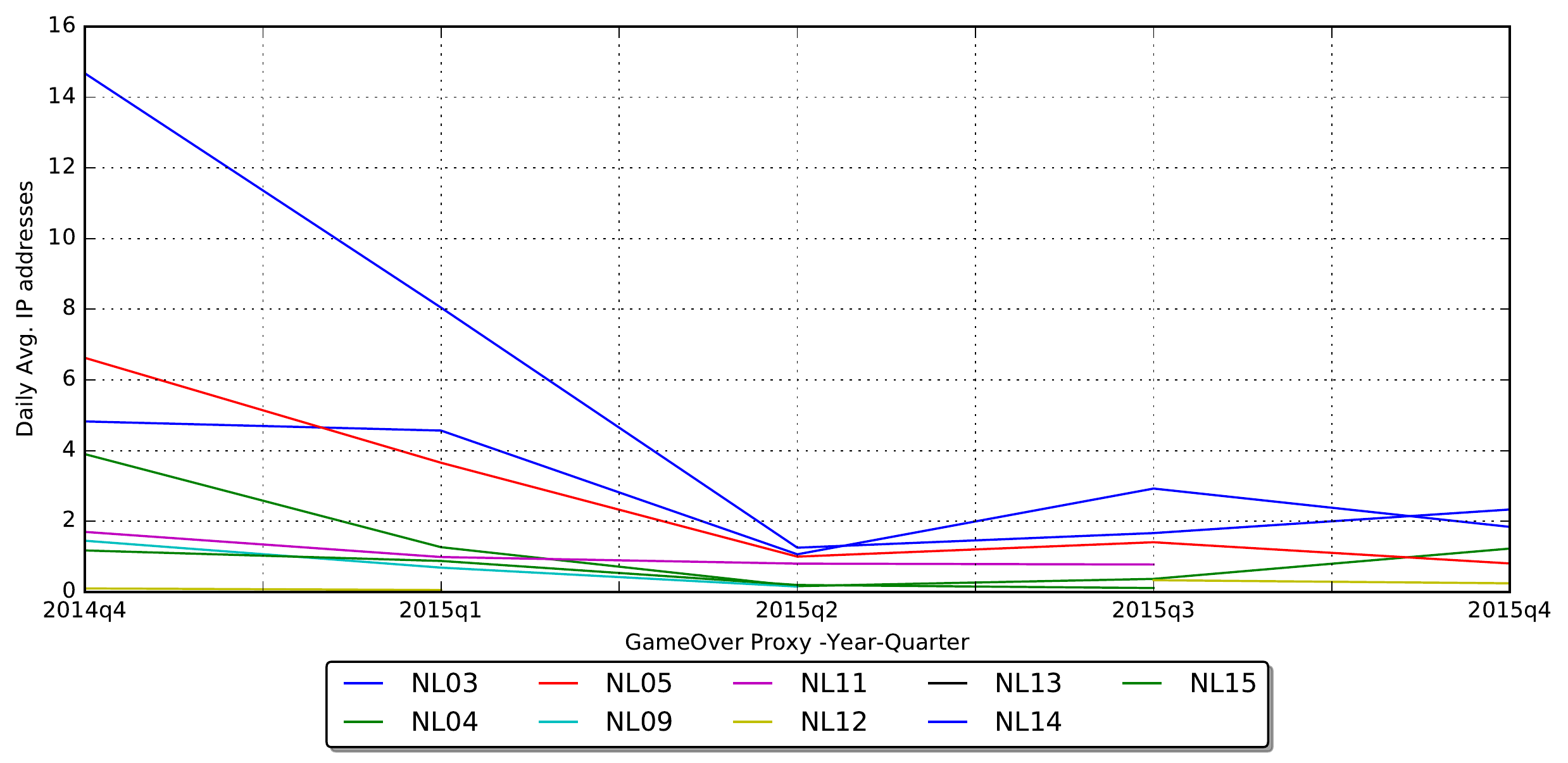}
    \caption{GameOver Proxy Members}
    \label{proxy-q2}
       \end{figure}

\begin{figure}[t]
            \includegraphics[width=1\textwidth]{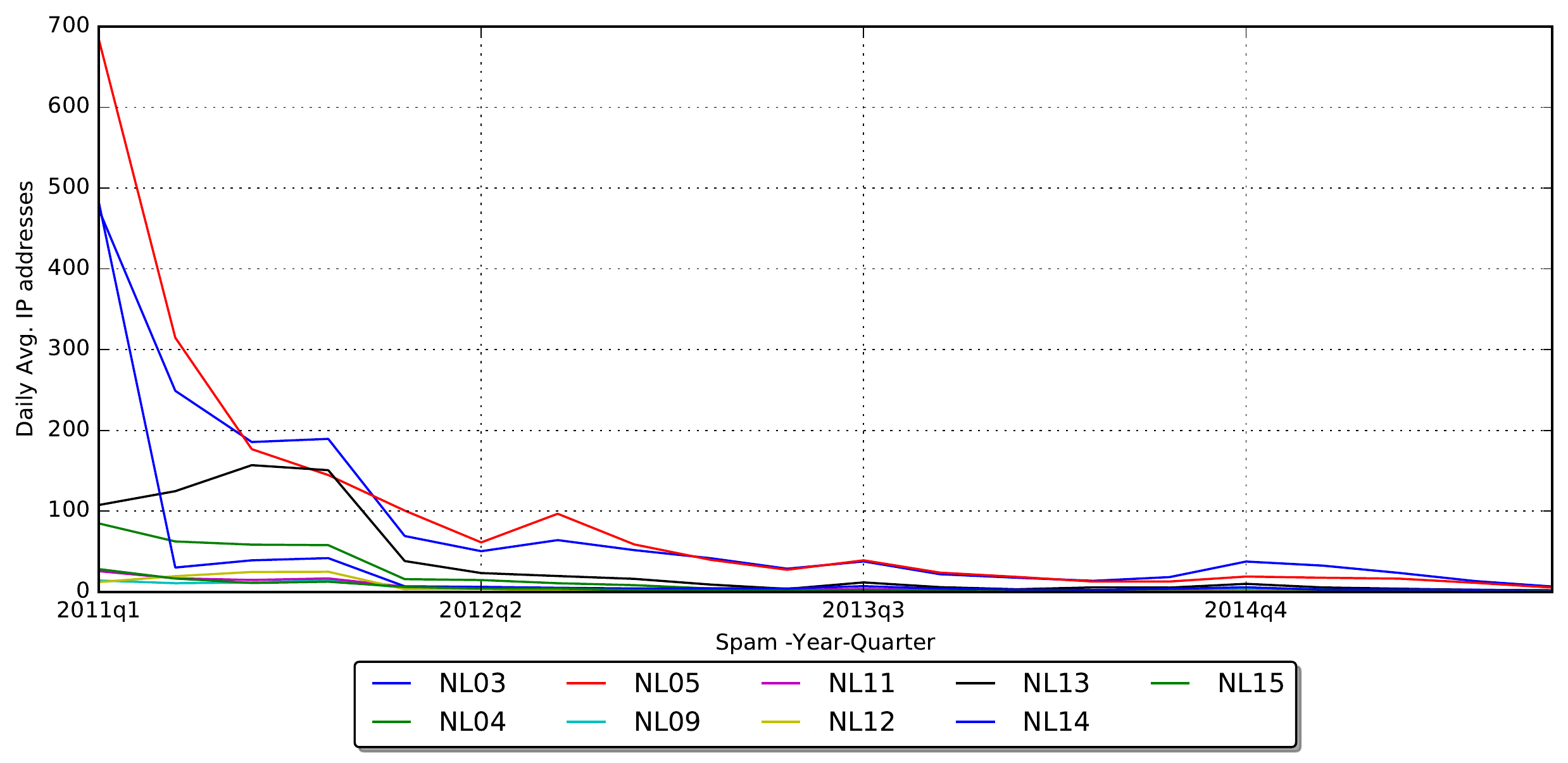}
    \caption{Spam Members}
    \label{spam-q2}
   \end{figure}

\begin{figure}[t]
            \centering
    \includegraphics[width=1\textwidth]{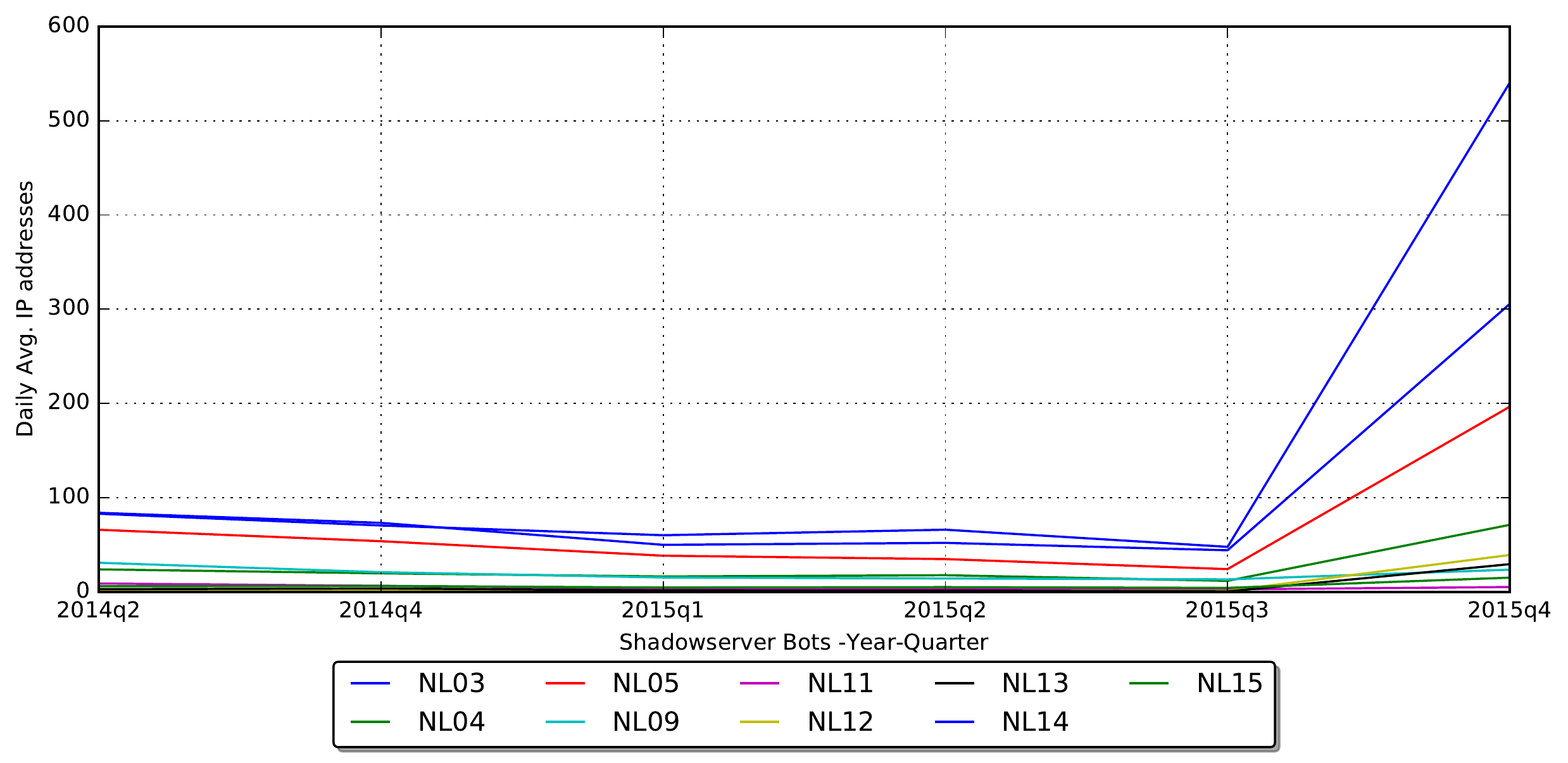}
    \caption{Shadowserver bots - Members}
    \label{drone-q2}
    
      \end{figure}

\begin{figure}[t]
            \centering
    \includegraphics[width=1\textwidth]{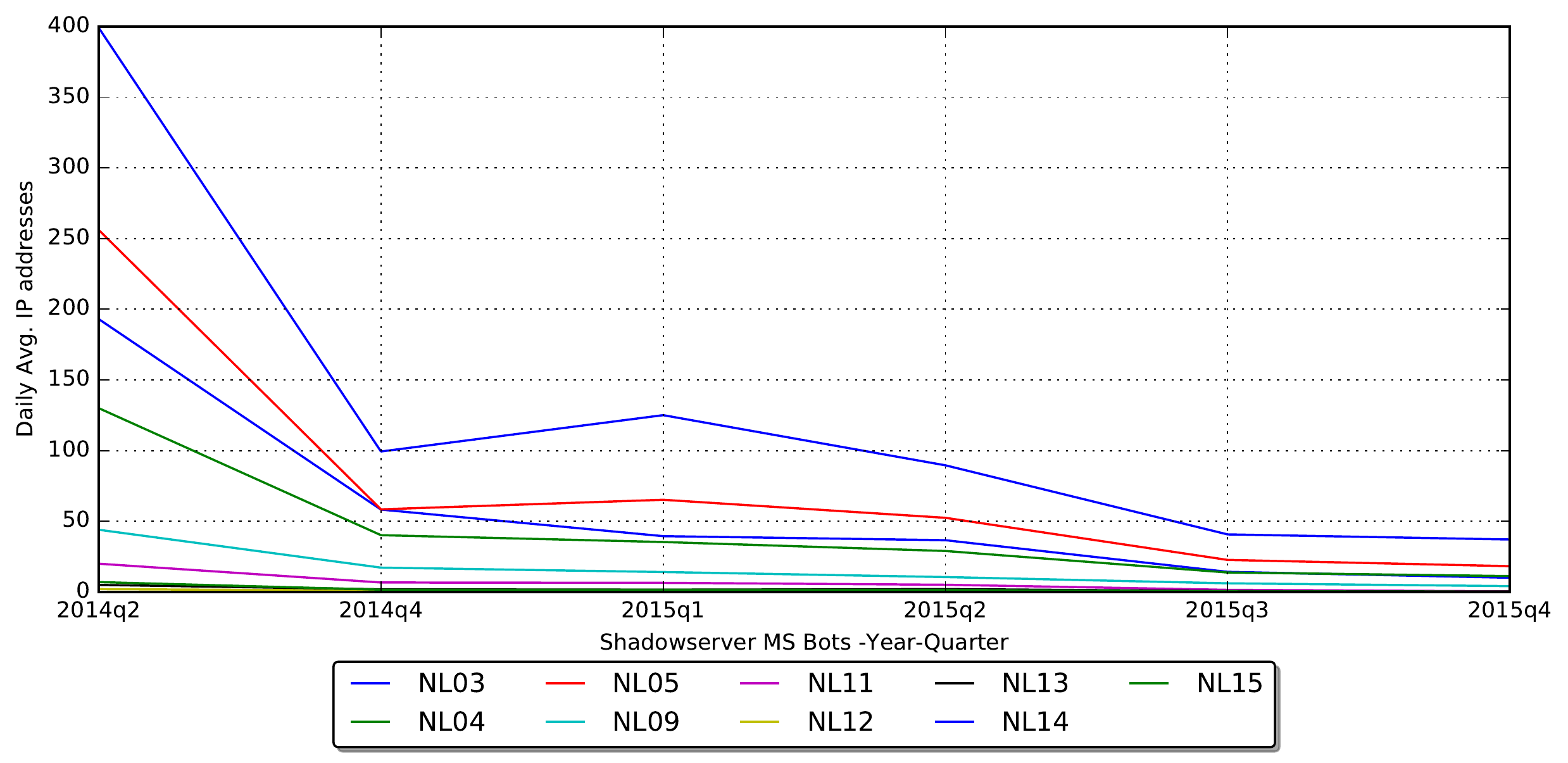}
    \caption{Shadowserver Microsoft sinkhole - Members}
    \label{ms-q2-bots}
    
      \end{figure}

\begin{figure}[t]
            \centering
    \includegraphics[width=1\textwidth]{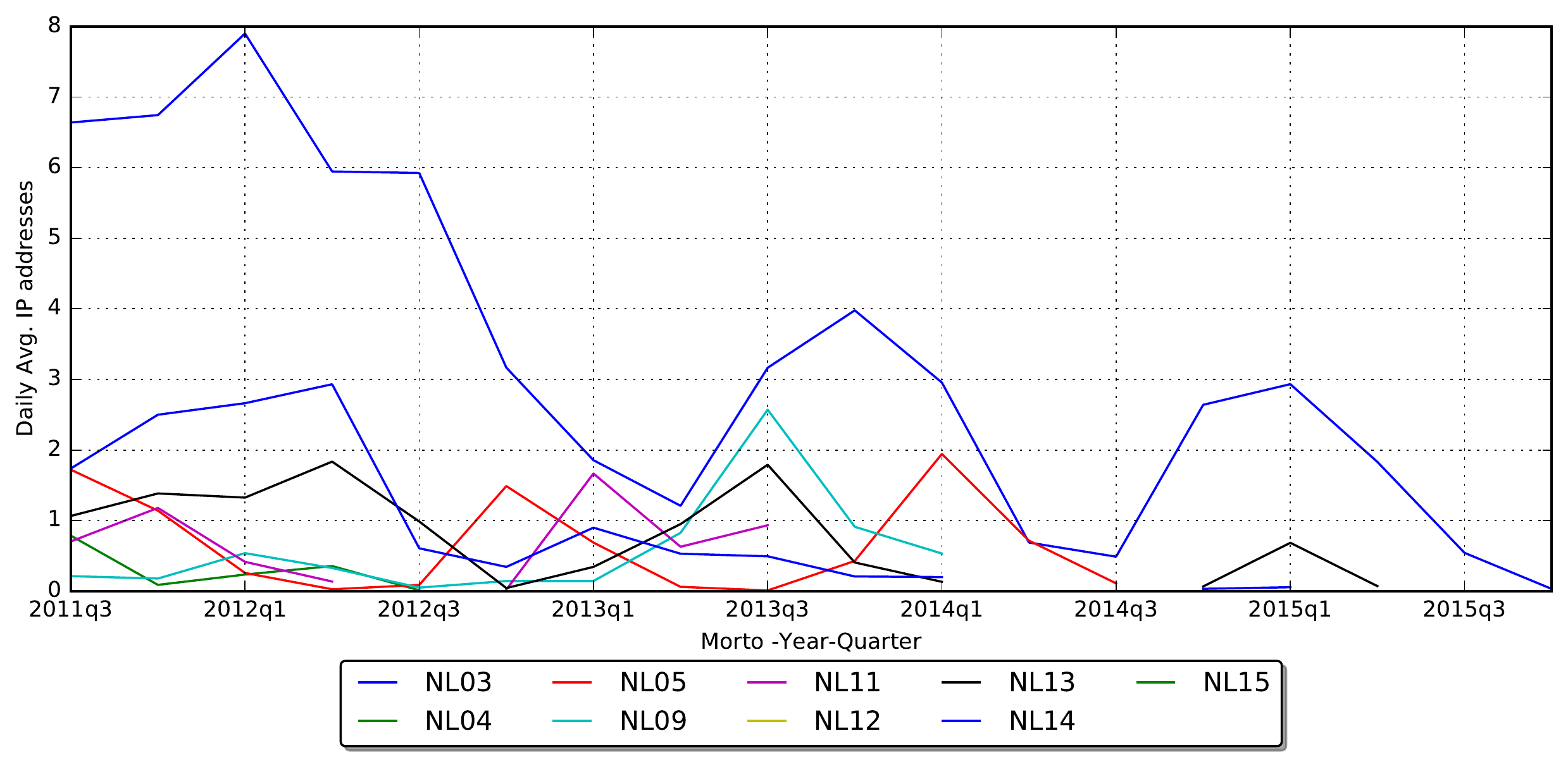}
    \caption{Morto - Members}
    \label{morto-q2}
    
      \end{figure}

\begin{figure}[t]
            \centering
    \includegraphics[width=1\textwidth]{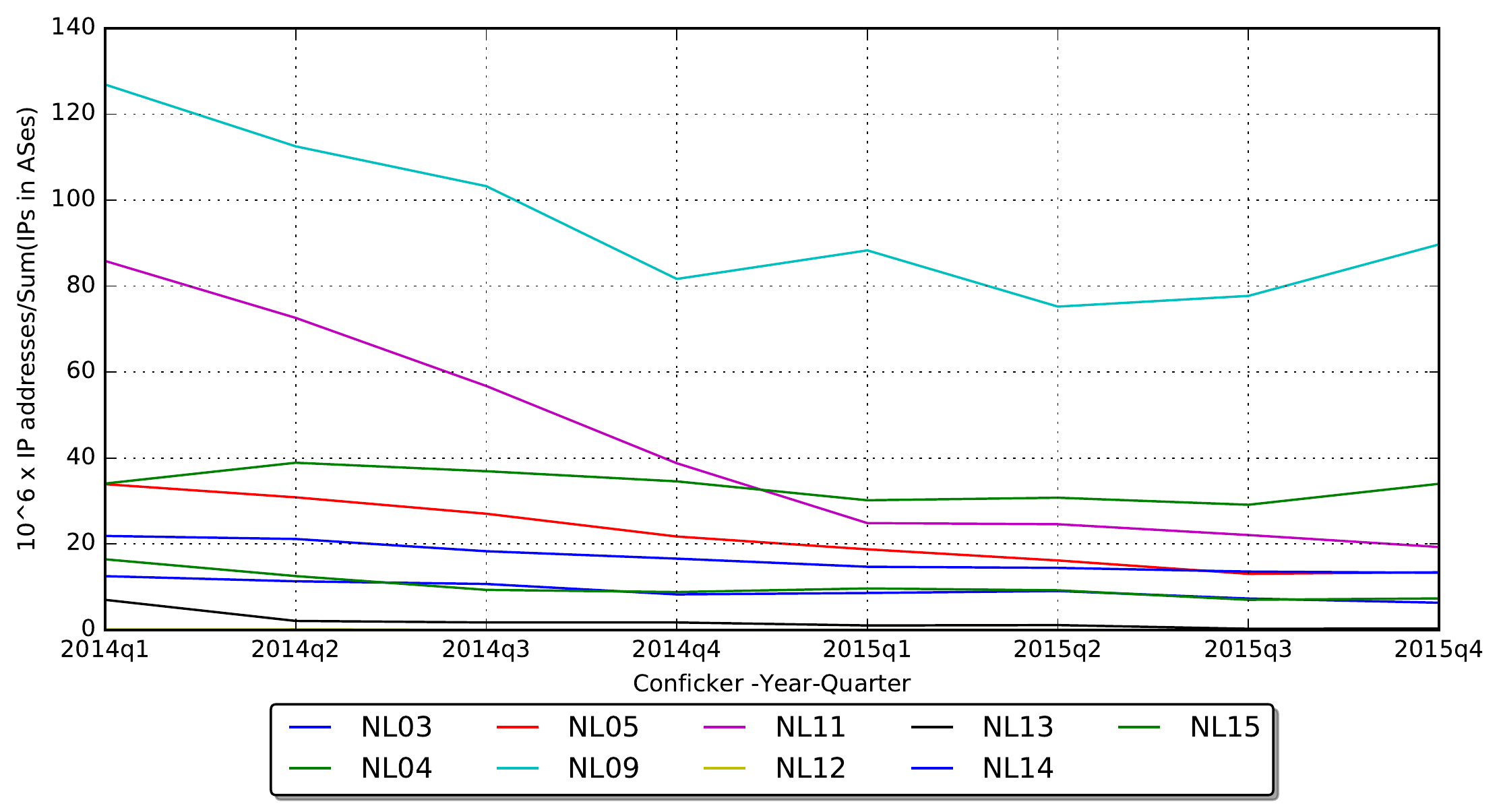}
    \caption{Conficker Members - normalized by AS size}
    \label{cfk-q2-as-size}
   \end{figure}

\begin{figure}[t]
            \centering
    \includegraphics[width=1\textwidth]{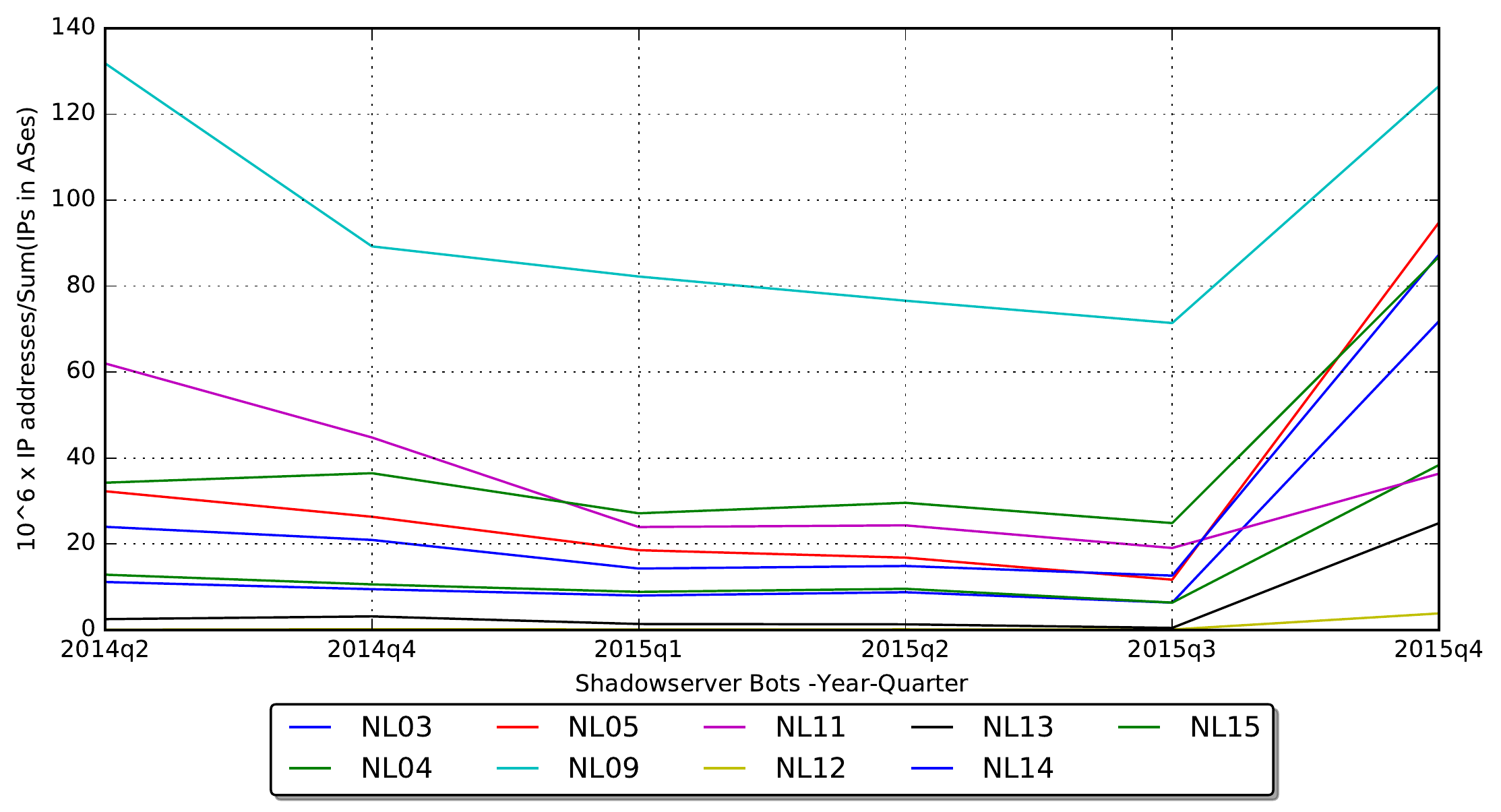}
    \caption{Shadowserver bots - normalized by AS size}
    \label{drone-norm-q2}
    
      \end{figure}

\begin{figure}[t]
            \centering
    \includegraphics[width=1\textwidth]{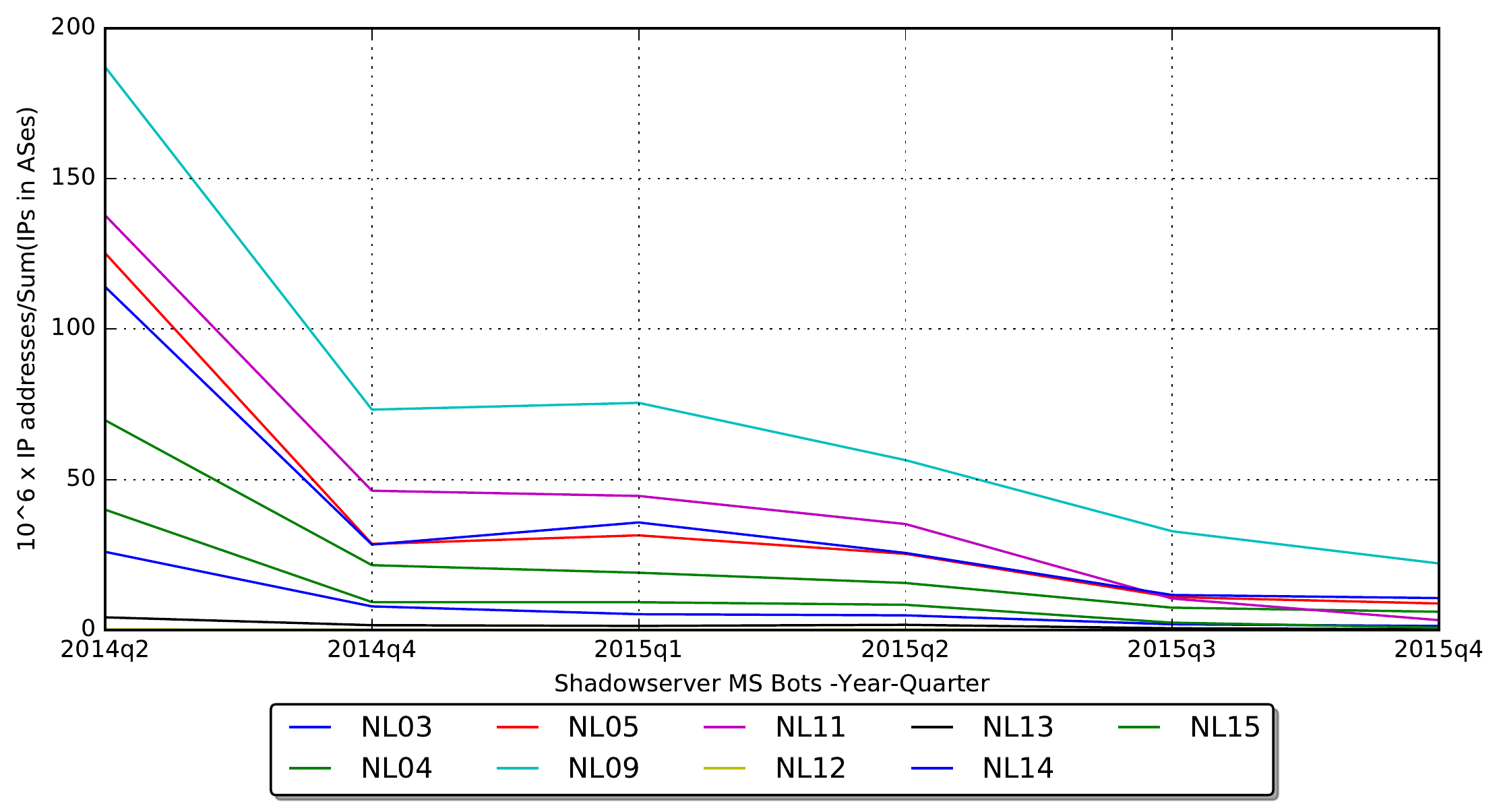}
    \caption{Shadowserver Microsoft sinkhole - normalized by AS size}
    \label{drone-norm-q2-bots}
    
      \end{figure}

\begin{figure}[t]
       \includegraphics[width=1\textwidth]{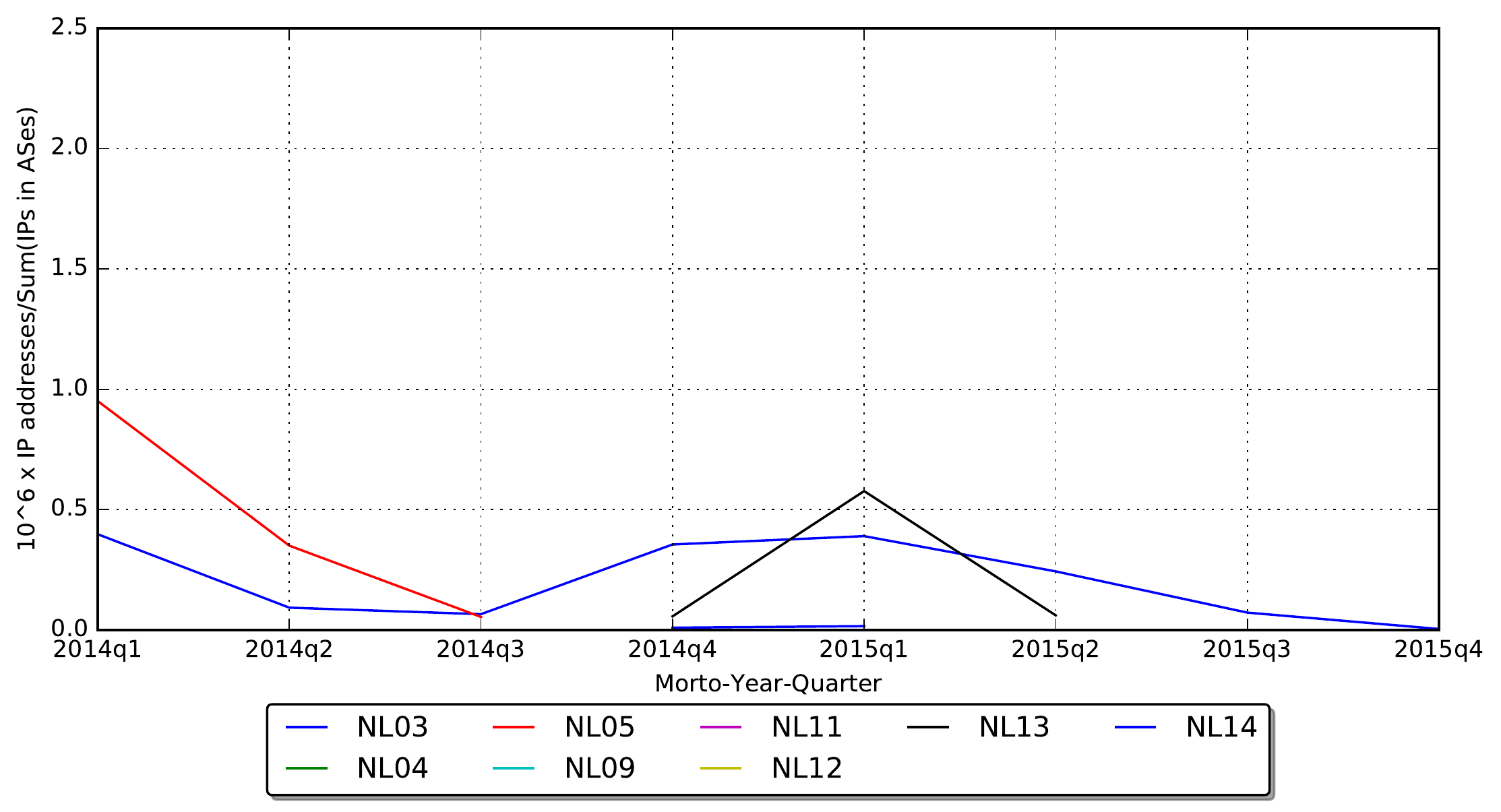}
    \caption{Morto Bots -normalized by AS size }
    \label{morto-norm}
   \end{figure}

\begin{figure}[t]
       \includegraphics[width=1\textwidth]{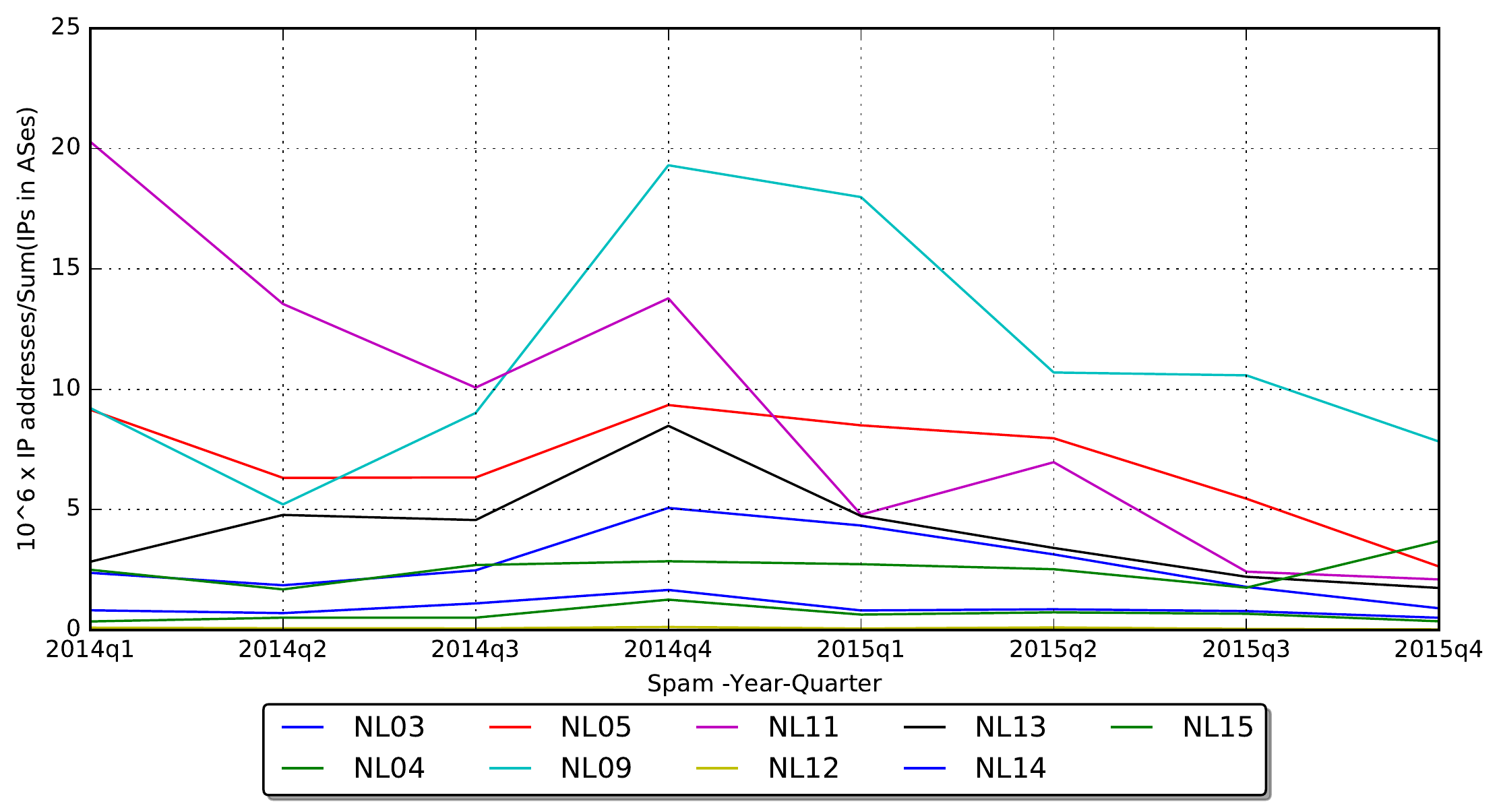}
    \caption{Spam Bots -normalized by AS size }
    \label{spam-norm}
   \end{figure}

\begin{figure}[t]
       \includegraphics[width=1\textwidth]{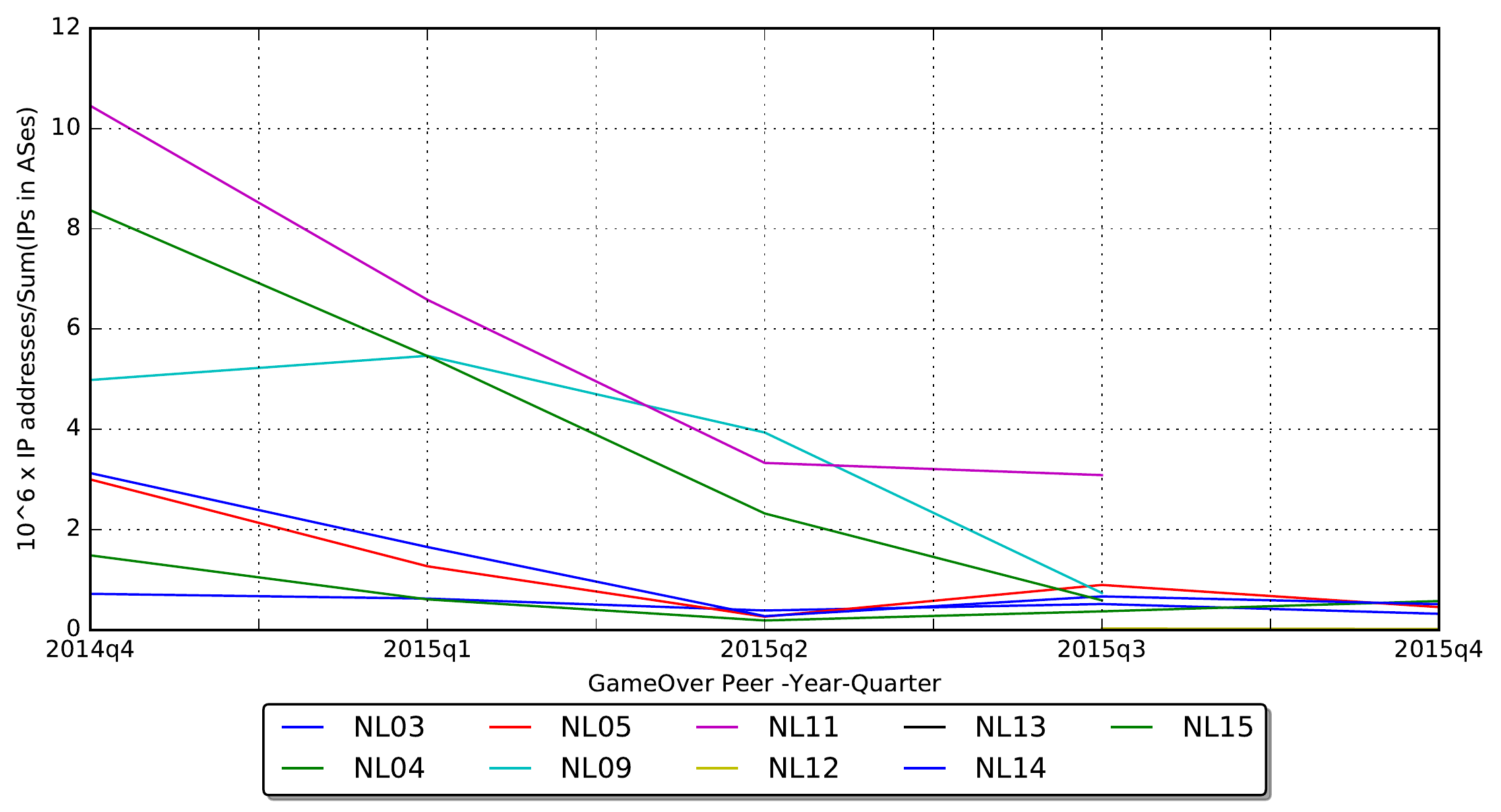}
    \caption{GameOver Zeus Peer Bots - normalized by AS size }
    \label{goz_peer-norm}
   \end{figure}

\begin{figure}[t]
       \includegraphics[width=1\textwidth]{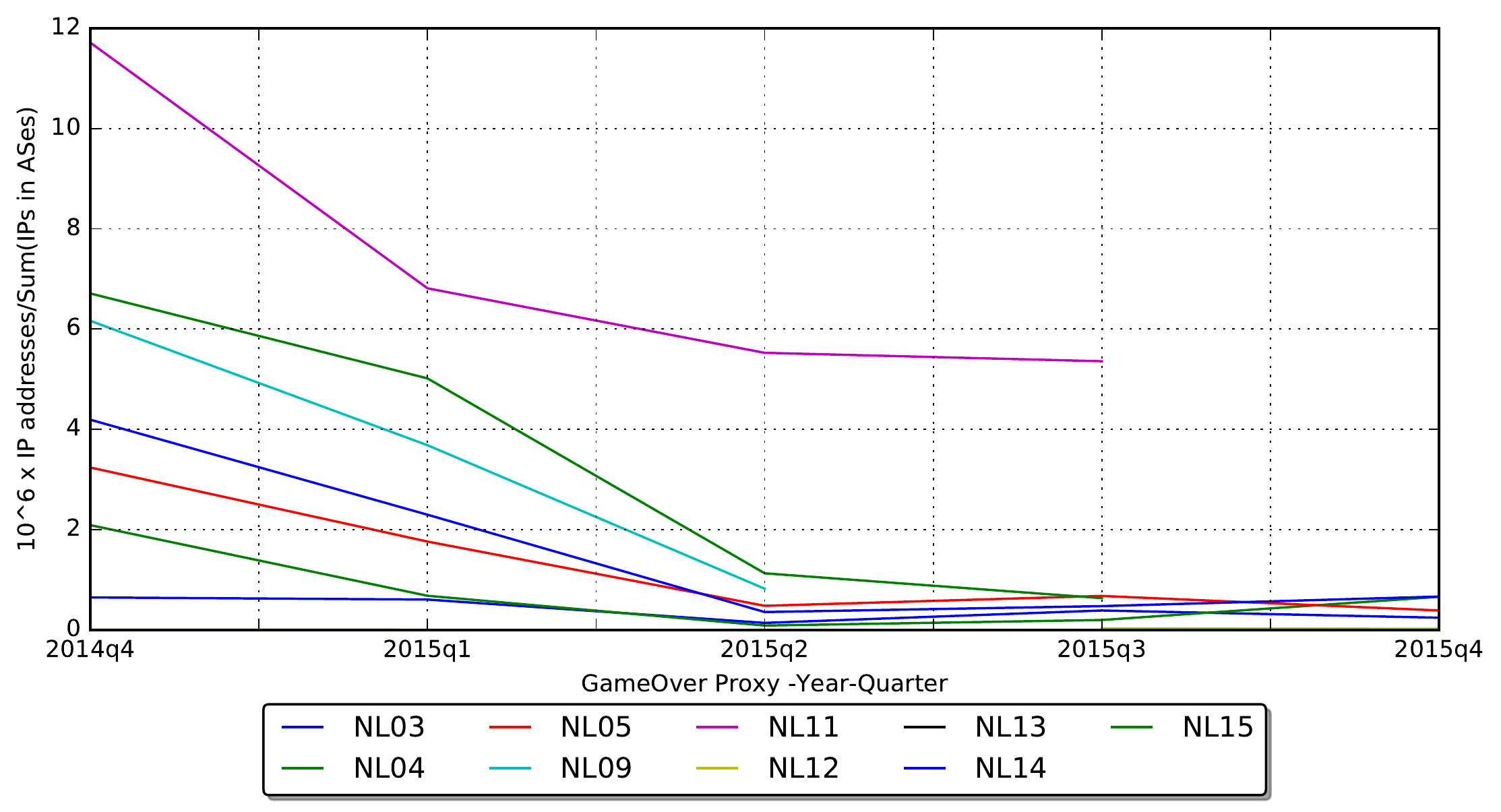}
    \caption{GameOver Zeus Proxy Bots - normalized by AS size }
    \label{goz_proxy-norm}
   \end{figure}

This chapter explores the differences in infection rates among the AbuseHUB members themselves. Figures~\ref{cfk-q2}--\ref{goz_proxy-norm} display the time series data of each data source. Please notice that Figs.~\ref{peer-q2} -- \ref{goz_proxy-norm} do not include NL10 because there no IP addresses from this network showed up in these sinkholes.


As it turns out, the differences among ISPs are not trivial. In some sources, a factor of 10 separates the best from the weakest performer. There is no completely consistent pattern of which ISPs perform better or worse, but some regularities seem to be observable. We leave these to the members themselves to evaluate. There are many factors at play here and we do not want to over-interpret the data, since we are working at a very low level of aggregation -- or to put it differently: we zoomed in on a small slice of the total dataset. This means the numbers of bots that are observed through our sources get relatively small. A further point to take into account is that we had to find a pragmatic solution that we do not have time series data for the number of subscribers for each member. So we used the size of their address space to normalize the infection rates. 

In short: these results are intended to be informative for the members themselves. They seem useful in discussing which of the ISP practices seem to have the most impact. We know that the members have their own policies for acting on the AbuseHUB data: when to act, how to act, when to quarantine, et cetera. Comparing these practices to the measurements is a good starting point for establishing best practices.

       
 \chapter{Conclusions}
\label{sec:summary}

This report provides an assessment of the impact of the AbuseHUB initiative in reducing the presence of infected computers in the networks of Dutch ISPs and other member networks. The study focuses on four research questions:
\begin{enumerate}
  \item How do member ISPs compare to non-member ISPs? 
  \item How do member ISPs compare among themselves?
  \item Do member ISPs have better data on the presence of bots in their networks?
  \item What recommendations can be identified to improve botnet mitigation in the Netherlands?
\end{enumerate}

The preceding chapters have presented and analyzed the evidence to answer the first two questions. We will summarize the main findings in the Section~\ref{sec:chap6:impact}. In Section~\ref{sec:chap6:information}, we answer the third research question. Finally, in the last section (\ref{sec:chap6:recommend}), we list the main recommendations.

  \section{Assessing the impact of AbuseHUB}
  \label{sec:chap6:impact}

Several key findings can be taken from this report:
\begin{itemize}
\item The Netherlands performs above average in botnet mitigation, not just compared to the rest of the world, but also compared to a set of reference countries (Germany, Great Britain, France, Finland, Italy, Spain, United States and Japan).

\item When we focus on the ISP networks only, which has also been the main focus of AbuseHUB, the Dutch performance is even better. The Dutch ISPs together consistently rank between position 50 and 60 across the different botnet data sources, where 60 is the country with the least infected ISP networks. This can be evaluated as ''best in class''.

\item Even compared to most ISPs in countries that also have a mature anti-botnet initiative (AU, DE, IE, JP, KR, FI), the Dutch ISPS perform significantly better. Only Finnish ISPs consistently outperform the Dutch ISPs.

\item The evidence is inconclusive whether the operational launch of AbuseHUB has significantly accelerated mitigation. The ranking of the Netherlands has stayed more or less stable between 2011 and 2014. To some extent, this is the effect of the fact that the ranking is already very high, so it cannot really go much higher. This stable ranking can still be consistent with improved mitigation. Most countries at the clean end of the ranking are improving their mitigation efforts, so the fact that the ranking of the Netherlands' ISPs stays stable at a high level within that group implies improved cleanup. The indexed metrics suggest that cleanup rates did accelerate a bit, compared to those of other countries, but this process started earlier than the operational launch of AbuseHUB. One way to interpret this is to see AbuseHUB as part of a process of improvement that was already ongoing. 

\item Within the Netherlands, AbuseHUB members still harbor the majority of botnet infections, as can be expected since most bots are concentrated in consumer broadband networks. That being said, the distribution has been shifting towards non-members. The latter have contributed, on average, 39\% of the infections. This proportion was around 20\% in 2010. This means that the AbuseHUB members have improved faster than non-members and that the latter are becoming more important to further reduce botnets.

\item The most infected non-members are smaller consumer and business ISPs, hosting providers and, to a lesser extent, mobile providers. These operators may benefit from connecting with AbuseHUB and vice versa.

\item There is variance in infection rates among AbuseHUB members and there appear to be some regularities as to which ISPs perform better or worse. There are limitations, however, of the analyzing the datasets at such a granular level. We therefore do not want to over-interpret the results and suggest that they be used by the members themselves as a starting point for a discussion on best practices.

\end{itemize}

  \section{Improved data for mitigation}
  \label{sec:chap6:information}

In an earlier study on botnet mitigation in the Netherlands~\cite{michel-dutch-market}, we found that ISPs were using only a limited set of data feeds to identify infected customers in their networks. This led to the recommendation to increase the intake and use of available feeds of botnet data. A year or two later, this would become one of the core objectives of the AbuseHUB initiative.

Has this objective been realized? Are members now receiving more comprehensive botnet data? The short answer is unequivocally: Yes, they do. A more informative answer distinguishes several factors that determine the quality of data that is available for mitigation: 
\begin{itemize}
\item Coverage: how many feeds are included and how well do these feeds cover the botnet ecosystem?
\item Continuity: how well is the set of available feeds maintained and kept relevant?
\item Quality: is the quality of received data evaluated and enriched to enable follow-up by abuse teams?
\item Transaction cost: does AbuseHUB reduce the transaction cost for external parties, such as researchers, CERTS or law enforcement, to distribute ad hoc botnet data to ISPs?  
\end{itemize}

Starting with the first factor, coverage, we can unequivocally state that it has improved. AbuseHUB collects and processes an increasing number of feeds that go well beyond what each of the ISPs was using at the time of our first report on botnet mitigation in the Netherlands. They also go beyond what we were able to use in this report. Many data feeds are not accessible to outside researchers. They are only (partially) shared with network owners, who can request the subset of data that pertains to their IP address space. A non-exhaustive list of feeds currently received by AbuseHUB: all available Shadowserver feeds, XBL bot blocklist feed from Spamhaus, and spam feedback loops from Microsoft (Hotmail, Live, Outlook.com) and AOL. Other useful feeds are being sought, so that AbuseHUB staff can reach out to the relevant data provider. All this does not negate the fact that botnet data feeds are always limited. Many bots go undetected by the specific methods underlying these feeds.  

Continuity is an important aspect of collecting and distributing botnet data. It might seem that once the process is set up to receive, parse and distribute feeds, the work is done. In practice, however, this process is dynamic and needs monitoring, updating and constant effort. First, the feeds change and automated processes need to identify these changes and adapt the relevant parsing code, otherwise this data will be lost and not used for mitigation. Second, and arguably more important, is the need to continually work on the set of feeds. Many botnet feeds are based on sinkholes, for example, and they become obsolete over time. Sinkholed botnets are out of reach of their attackers, so they slowly decline over time. Other types of feeds might also become obsolote, because of changes in the botnet ecosystem. Spam feeds, for example, are becoming have become a lot less informative than a few years ago, as the spam ecosystem itself has partially collapsed. Spammers, therefore, are purchasing less botnet capacity to distribute spam. In other words, only a small slice of all infected machines will engage in spamming and thus become visible via spam feeds. All of this means constant effort is needed to maintain the continuity of the botnet data that can be used for mitigation.

Quality of data is also a key factor in mitigation. The abuse teams that receive the data from AbuseHUB have to decide when and how to act on it. In all feeds, there is a chance of false positives: an IP address that shows up in the data does not harbor an infected machine. This is intrinsic to the methods via which the data are collected, though some methods are more prone to false positives than others (see Section \ref{sec:datasets} for more details about these methods). The presence of false positives presents a dilemma: if they contact the customer, it might be a false alarm and the customer might have a negative experience that is unnecessary. If they do not contact the customer, a potential infection goes untreated to the detriment of both the customer and the provider. For this reason, it is important to be able to evaluate the quality of the data. What is the chance of a false positive? How accurate is the time stamp associated with the event? Is it user-reported, as in the spam feedback loops, or is it machine generated? How trustworthy and reliable is the provider of the data? AbuseHUB does not merely forward the incoming data to the abuse teams of the ISPs, but it has also invested in different indicators of quality to accompany the data. This improves the decision making process of abuse handling staff.   

One final factor is often overlooked: transaction cost. So far we have talked about abuse feeds that are disseminated by organizations like Shadowserver. These are dedicated to collecting abuse data. There are other organizations, however, that also uncover infection data on a more ad hoc basis. Law enforcement sometimes discovers infection data during investigations, scientists undertake research projects that temporarily track certain botnet activity, CERTS that get notified e-commerce providers (think of banks, for example) might discover account takeover or suspicious activity on the machines of their users suggesting they are infected. This kind of data is typically not used for mitigation, because these entities do not have the time, resources or mandate to contact ISPS, educate them about the data, put it in a format that the ISPs can process and then consistently share it with the ISPS in a timely fashion. Data that arrives just a few weeks after is was collect is already useless for mitigation purposes. The fact that AbuseHUB is now a hub in the Dutch landscape for infection data drastically reduces the transaction cost of making this data available for mitigation. This can be of great benefit for the future of botnet mitigation in the Netherlands.

All in all, for members of AbuseHUB, the quality of data available for mitigation has been substantially improved.

  \section{Recommendations}
  \label{sec:chap6:recommend}

In the final part of this report we explore a number of recommendations to improve botnet mitigation in the Netherlands. They flow from the observations we made at different points in this report. We first present the recommendations for the members of AbuseHUB. Next, we identify recommendations for the Ministry of Economic Affairs, who has co-funded the development of AbuseHUB. 

We have identified the following recommendations for AbuseHUB members: 
\begin{enumerate}

\item Expand the membership of AbuseHUB to include the remaining consumer and business ISPS and the hosting providers that account for most of the remaining bots. As discussed in Chapter \ref{sec:q1}, these providers are harboring a growing portion of the problem. Their mitigation efforts can be jump started more efficiently via AbuseHUB than on their own.

\item Sustain the effort to acquire new feeds and other sources of botnet data. We have discussed the necessity of this work already in the previous section, under the issue of continuity. Some data sources are hard to acquire for ISPs. In those cases, the NCSC might be able to mediate the acquisition, as it holds a trusted position as the national CERT.

\item Consider setting up own bot detection methods. In light of the limitations of the available data feeds, it makes sense to increase detection through non-intrusive means that are locally operated, such as a darknet sensor, monitoring botnet-related DNS queries, and other tools. Finnish ISPs, among others, have long been running such tools and might be able to share them in a way that are easily deployable by AbuseHUB itself or by the members. These are relatively cheap to operate and can complement the feeds received from third parties, which will always suffer from under-reporting of bots.   

\item Critically evaluate when botnet data is deemed good enough to actually initiate a customer notification. In earlier stages, ISPs waiting for two independent feeds to identify a customer as being infected. This is understandable in light of the issue of false positives, which we discussed above. But it errs too far on the side of caution. Research has shown time and again that different abuse feeds share almost no overlap \cite{CERTCC, Noroozian}. An infected machine has typically a less than one percent chance to show up in two feeds at once. So this threshold means more than 90\% of all data points would not lead to cleanup \cite{CERTCC}. Analysis of the issue of false positives is needed to establish better rules on when to initiate cleanup. 

\item Establish an easy-to-find mechanism to submit data for entities that have ad-hoc sources of botnet data. We suspect that ad-hoc sources will become increasingly important over time, as they tend to be closer to the cutting edge of botnet development. Research, law enforcement, CERTs, they tend to be involved in cases or incidents that generate data well before such botnets are sinkholed and added to the existing feeds of organizations like Shadowserver. Many of the entities that capture ad-hoc sources will not invest into building relationships with network operators or anti-botnet initiatives, as this is not their job and they typically lack the resources to do so. While AbuseHUB provides a single point of contact for Dutch providers, which is an improvement of the earlier situation, there is still considerable 'friction' (to borrow a term from e-commerce) in the process. AbuseHUB is not easily discoverable for international abuse reporting entities. Furthermore, entities cannot send data for mitigation without first going through a process of contacting AbuseHUB and communicating with them. One way to reduce friction is to register AbuseHUB as member of the Trusted Introducer network, so that AbuseHUB is discovered automatically by international entities.\footnote{For more details, see https://www.trusted-introducer.org/} Another option is to provide an API or other mechanism on the website for abuse reporting entities to submit data, without first going through manual email contact. 

\item Incentivize current entities that posses relevant infection data to provide it to AbuseHUB. This is the complement of the previous recommendation. In some cases, AbuseHUB wants to do mitigation on the data, but it is still not receiving it. An example that deserves a special mention here is the banks. Their monitoring generates reliable data on customers who are suspected to be infected with financial malware. So far, the banks have interested in sharing this data, but they have been unable to overcome their own reluctance related to potential privacy and public relations issues associated with providing the data to AbuseHUB. These hurdles are currently being slowly overcome, but until this happens, it is loss for everyone.

\item Use experimental methods to develop evidence-based best practices for customer notification and cleanup. Research in other areas has shown that even minor variations in the notification message can have substantially different results \cite{179521}. Current practices are based on common sense and anecdotal evidence. When should customers be contacted? What type of contact is the most effective in getting them to cleanup? How can this be executed efficiently? A/B testing can be a very powerful way to get evidence as to the true effectiveness of abuse policies and cleanup. This also implies better tracking of what customers have done in terms of cleanup. There are tools available that can close this feedback loop, thus allowing direct monitoring of the effectiveness of the notification. Evidence-based practices would have great relevance also beyond the Netherlands. Working groups at industry organizations like M3AAWG have been grappling with these issues for years. 

\item Push towards scaling up the number of cleanup actions. The process of acquiring botnet data and notifying customers is a necessary, but not sufficient condition to overcome botnets. We have undertaken other studies, on Conficker cleanup in 60 countries and on the national support centers in Germany and Spain \cite{190882, ACDC}. In both we have found a rather limited impact of such processes. The explanation for this somewhat dissapointing result seems to be that true impact requires doing notifications and cleanup at a much higher scale than is currently taking place. The best approximation for the number of bots comes from Microsoft's Malicious Software Removal Tool. It puts the infected population at roughly one percent of all Windows PCs are infected at any moment in time. This number is two orders of magnitude larger than the number of customer notifications that the German and Spanish national centers have done. Scaling up to the actual infection level seems necessary to have a substantial impact. This is also relevant for the Netherlands and AbuseHUB: one percent means tens of thousands of infections per month. Many of these are resolved through automatic cleanup, without ISP intervention. Still, a significant amount will remain. 

\end{enumerate}

In addition to these eight recommendations for AbuseHUB members, we have also identified four recommendations for government, most notably the Ministry of Economic Affairs. The are the following:

\begin{enumerate}

\item Incentivize other providers to join. Now that AbuseHUB members are investing resources into better cleanup, it becomes more important to incentivize others in the market to join this effort or undertake similar efforts. The idea is to prevent so-called free riders: providers that do not invest in abuse management but who do reap the benefits of the improvements of the sector as a whole. One way in which the government can do this is to reach out to the most infected non-members directly. Another mechanism is to start working towards publicly available metrics of abuse, so that the market become more transparent. This way, good providers can distinguish themselves from lax providers.

\item Incentivize current entities that posses relevant infection data to provide it to AbuseHUB.  One of the main reasons why existing infection data is not being submitted to AbuseHUB is because of legal risk avoidance by the entities that posses the data. Issues around privacy, existing legal agreements and reputation effects are hurdles that are hard to overcome for the operational people that typically set up data sharing. To illustrate: banks and even the government's own NCSC has data that is valuable for mitigation, but it has taken a very long time for this data to be submitted to AbuseHUB. Apparently, the moment were data will actually be shared is close at hand, but it still has not happened. Moreover, the fact that it has taken so long is hard to stomach. The government can engage the leadership of these organizations and get them to commit to the data sharing. Engaging the leadership is probably more efficient in terms of overcoming the risk-avoidant behavior of legal departments than the attempts of abuse handling staff.  

\item Push towards scaling up the number of cleanup actions. This is the governmental complement to the same recommendation for AbuseHUB. As was explained there (see recommendation 8), scale is very important to achieve impact. The current scale is probably at least one order of magnitude too small. Putting the process in place has been a very labor-intensive effort. Now that this has been achieved, it should be feasible to scale up the process, so the impact on infection rates will be more substantial. 

\item Facilitate making cleanup cheaper. If cleanup is to scale successfully, the cost of each action is a major factor. The main cost is customer support. While ISPs give their customers advice about what to do, these instruction are still not trivial to execute for many users. One reason for the relative complexity is liability: the more directive and straightforward the advice (''click here to run this tool''), the more the ISP might be held liable for when it doesn't work or disrupts the machine. This could be overcome by a central or public institution offering a single cleanup tool, so the ISPs could point their customers there. The German experience with the botfrei.de center and website is the example of this. Their cost for customer support turned out to be much lower than anticipated. This centralized cleanup tool has another advantage; it can report back to the support center that it has run and what malware, if any, it has encountered. This is very valuable to close the feedback loop mentioned in recommendation \#7.

\end{enumerate}

\begin{appendices}

\label{sec:appendix}

\begin{table}
\centering
\caption{List of countries used in the report }
\label{countries}
\begin{tabular}{llllllll}
\textbf{CC} & \textbf{Country  name}  & \textbf{CC}  & \textbf{Country name}  & \textbf{CC}  & \textbf{Country  name} &\textbf{CC}  & \textbf{Country  name}   \\
AR & Argentina      & LV & Latvia       & DK & Denmark      & RU & Russia         \\
AU & Australia      & LT & Lithuania    & EG & Egypt        & SA & Saudi Arabia   \\
AT & Austria        & LU & Luxembourg   & EE & Estonia      & RS & Serbia         \\
BY & Belarus        & MY & Malaysia     & FI & Finland      & SK & Slovakia       \\
BE & Belgium        & MA & Morocco      & FR & France       & SI & Slovenia       \\
BR & Brazil         & NL & Netherlands  & DE & Germany      & ZA & South Africa   \\
BG & Bulgaria       & NZ & New Zealand  & GR & Greece       & KR & South Korea    \\
CA & Canada         & NO & Norway       & HU & Hungary      & ES & Spain          \\
CL & Chile          & PK & Pakistan     & IS & Iceland      & SE & Sweden         \\
CN & China          & PE & Peru         & ID & Indonesia    & CH & Switzerland    \\
CO & Colombia       & PH & Philippines  & IE & Ireland      & TH & Thailand       \\
HR & Croatia        & PL & Poland       & IL & Israel       & TR & Turkey         \\
CY & Cyprus         & PT & Portugal     & IT & Italy        & UA & Ukraine        \\
CZ & Czech Republic & RO & Romania      & JP & Japan        & GB & United Kingdom \\
MT & Malta          & VN & Vietnam      & KZ & Kazakhstan   & US & United States  \\
\footnote{Used in Chapter 2 only}MX & Mexico         &\footnote{Used in Chapter 2 only} IN & India        &    &              &    &               
\end{tabular}
\end{table}
\end{appendices}

\bibliographystyle{IEEEtran}

\bibliography{report}
\end{document}